\documentclass[11pt]{article} 
\usepackage{jheppub}

\usepackage{amsmath,amsfonts,amsbsy,amssymb,array,accents,dsfont}
\usepackage{enumerate}
\usepackage[shortlabels]{enumitem}
\usepackage{graphicx} 
\usepackage{subcaption} 
\usepackage{upgreek}
\usepackage{bm}
\usepackage{slashed}

\let\includefigures=\iftrue
\let\useblackboard==\iftrue
\newfam\black

\usepackage[dvipsames]{xcolor}

\definecolor{myblue}{RGB}{85,130,255}
\definecolor{myred}{RGB}{200, 45, 40}

\PassOptionsToPackage{ 
     colorlinks=true,
     linkcolor=myBlue,
     urlcolor=blue,
     filecolor=black,
     citecolor=myRed,
}{hyperref}

\usepackage{xparse}
\usepackage{xspace}

\NewDocumentCommand\eqn{om}{%
  \IfNoValueTF{#1}
     {\[ #2 \]}
     {\begin{equation}\label{#1} #2  \end{equation} \expandafter\newcommand\csname #1\endcsname{\eqref{#1}\xspace}\ignorespaces}
}
\NewDocumentCommand\eqna{om}{%
  \IfNoValueTF{#1}
    {\begin{align*} #2 \end{align*}}
    {\begin{equation}\label{#1}\begin{split} #2  \end{split}\end{equation} \expandafter\def\csname #1\endcsname{\eqref{#1}\xspace}\ignorespaces}
}

\newcommand{\rcite}{\cite}

\def\sl{\text{sl}}

\def\tight#1{\! #1 \!}  

\def\({\left(}
\def\){\right)}
\def\[{\left[}
\def\]{\right]}

\def\eg{{e.g.}}

\def\etc{{etc}}

\def\nfive{{n_5}}

\def\uhat{{\sfu}}
\def\vhat{{\sfv}}

\def\A{{\mathsf A}}
\def\B{{\mathsf B}}

\def\sfA{{\mathsf A}}
\def\sfB{{\mathsf B}}

\def\sfb{{\mathsf b}}

\def\sfu{{\mathsf u}}
\def\sfv{{\mathsf v}}

\DeclareMathSymbol{\medhatsym}{\mathord}{largesymbols}{"62} 

\DeclareMathSymbol{\medtildesym}{\mathord}{largesymbols}{"65}

\makeatletter
\newcommand*\rel@kern[1]{\kern#1\dimexpr\macc@kerna}
\newcommand*\widebar[1]{%
  \begingroup
  \def\mathaccent##1##2{%
    \rel@kern{0.8}%
    \overline{\rel@kern{-0.8}\macc@nucleus\rel@kern{0.2}}%
    \rel@kern{-0.2}%
  }%
  \macc@depth\@ne
  \let\math@bgroup\@empty \let\math@egroup\macc@set@skewchar
  \mathsurround\z@ \frozen@everymath{\mathgroup\macc@group\relax}%
  \macc@set@skewchar\relax
  \let\mathaccentV\macc@nested@a
  \macc@nested@a\relax111{#1}%
  \endgroup
}
\makeatother

\def\One{{\hbox{1\kern-1mm l}}}

\def\barray{\begin{array}}
\def\earray{\end{array}}
\def\be{\begin{equation}}
\def\ee{\end{equation}}
\def\bea{\begin{align}}
\def\eea{\end{align}}
\def\bal{\begin{align}}
\def\eal{\end{align}}
\def\nn{\nonumber}

\newcommand{\bR}{{\mathbb R}}
\newcommand{\bS}{{\mathbb S}}
\newcommand{\bT}{{\mathbb T}}

\definecolor{cardinal}{rgb}{0.6,0,0}
\definecolor{darkgreen}{rgb}{0,0.4,0}
\definecolor{green}{rgb}{0,0.4,0}
\definecolor{golden}{rgb}{0.92, 0.7, 0}
\definecolor{midnight}{rgb}{0, 0, 0.5}
\definecolor{darkblue}{rgb}{0, 0, 0.7}

\numberwithin{equation}{section}

\mathchardef\mhyphen="2D

\def\cD{\mathcal {D}}

\def\one{{\hbox{\kern+.5mm 1\kern-.8mm l}}}
\def\zero{{\hbox{0\kern-1.5mm 0}}}

\def\id{\textrm{id}}

\def\id{{1 \kern-.28em {\rm l}}}

\def\journal#1&#2(#3){\unskip, \sl #1\ \bf #2 \rm(19#3) }
\def\andjournal#1&#2(#3){\sl #1~\bf #2 \rm (19#3) }

\def\eg{{\it e.g.}}

\def\etc{{\it etc}}

\def\One{{1\hskip -3pt {\rm l}}}

\catcode`\@=11
\def\slash#1{\mathord{\mathpalette\c@ncel{#1}}}
\def\underrel#1\over#2{\mathrel{\mathop{\kern\z@#1}\limits_{#2}}}

\catcode`\@=12

\def\exp{{\rm exp}}

\def\O{{\cal O}}

\def\eg{{\it e.g.}}

\title{
{
BPS Fivebrane Stars I:
Expectation Values of Observables
}}

\author{Emil J. Martinec$^a$, Yoav Zigdon$^{b}$\\}

\affiliation[a]{
Kadanoff Center for Theoretical Physics, Enrico Fermi Institute, and Department of Physics\\ 
University of Chicago\\ 
5640 S. Ellis Ave.\\
Chicago IL 60637\\ 
}

\affiliation[b]{
Department of Applied Mathematics and Theoretical Physics,
University of Cambridge, Cambridge, CB3 0WA, United Kingdom \\
}

 \emailAdd{e-martinec@uchicago.edu}
 \emailAdd{ yz910@cam.ac.uk}

\abstract{%
We study ensembles of 1/2-BPS bound states of fundamental strings and NS-fivebranes (NS5-F1 states) in the AdS decoupling limit. We revisit a solution corresponding to an ensemble average of these bound states, and find that the appropriate duality frame for describing the near-source structure is the T-dual NS5-P frame, where the bound state is a collection of momentum waves on the fivebranes.  We find that the fivebranes are generically well-separated; this property results in the applicability of perturbative string theory.
The geometry sourced by the typical microstate is not close to that of an extremal non-rotating black hole; instead the fivebranes occupy a ball whose radius is parametrically much larger than the ``stretched horizon'' scale of the corresponding black hole.  These microstates are thus better characterized as BPS fivebrane stars than as small black holes.

When members of the ensemble spin with two fixed angular potentials about two orthogonal planes, we find that the spherical ball of the non-rotating ensemble average geometry deforms into an ellipsoid. This contrasts with ring structures obtained when fixing the angular momenta instead of the angular potentials; we trace this difference of ensembles to large fluctuations of the angular momentum in the ensemble of fixed angular potential.
}
\begin{document}
\maketitle
\hypersetup{pageanchor=false}
\hypersetup{pageanchor=true}
\pagenumbering{arabic}








\section{Introduction and summary} 
\label{sec:intro}

The study of bound states of fundamental strings and NS5-branes has been particularly fertile ground for the investigation of non-perturbative aspects of string theory \cite{Strominger:1996sh,Dijkgraaf:1997nb}.

The 1/2-BPS states of $n_5$ Neveu-Schwarz fivebranes wrapping $\bT^4 \times \bS^1_y$, bound to $n_1$ fundamental strings that wind $\bS^1 _y$, comprise a large ensemble of degenerate microstates. The Lunin-Mathur construction~\rcite{Lunin:2001fv,Kanitscheider:2007wq} provides supergravity descriptions of the vast majority of these states in terms of a set of profile functions ${f^I}(v)$, including those which specify the shape of the bound state in transverse space, where $v=t+y$, $t$ is time and $y$ parametrizes the circle.  The generic element of this ensemble has a deep throat with a large redshift to the core of the bound state, but no horizons, closed timelike curves, or other pathologies; this class of backgrounds has thus been considered as a prototype of the {\it fuzzball paradigm}~\rcite{Mathur:2005zp}~-- the idea that in string theory, the black hole interior is supplanted by some horizon-scale structure having itself no horizon.

An important fact about the Lunin-Mathur solutions is that they are parametrized by harmonic forms and functions that can be superposed to generate new solutions.%
\footnote{Note that the metric is not always linear in harmonic functions (for instance in the NS5-F1 frame it depends on the inverse of the onebrane harmonic function $H_1$, see eq.~\eqref{LMgeom}), so the quantum superposition of sources has a rather nonlinear effect on geometry; indeed, if one were to simply average 1/2-BPS metrics, the result would not be BPS.  The Hilbert space of 1/2-BPS states is embedded in the supergravity configuration space in a highly nontrivial manner.}
It is possible to average these forms and functions by putting the system in a grand canonical ensemble where the chemical potential conjugate to the string charge is fixed, and calculate analytically a corresponding solution~\rcite{Alday:2006nd,Balasubramanian:2008da,Raju:2018xue}. 
In this case, the resulting ensemble average solution forms a non-spinning, spherically symmetric blob of radius $\tilde{r}_b$, in which the redshift saturates at a large value.  
One can interpret the solution as describing typical states in the ensemble.
While there is also a classical extremal black hole solution carrying the same charges which has a horizon of microscopic size corresponding to the number of microstates, the suggestion is that in string theory this solution is to be replaced in the near-source region by the ensemble of horizonless objects, having the horizonless ``ensemble geometry'' given by the spherical blob of radius~$\tilde{r}_b$.

More recently, this proposal has been criticized on several grounds~\rcite{Raju:2018xue}: First, that expectation values of observables in typical microstates differ from those of the black hole solution by only exponentially tiny amounts, while the ``ensemble geometry'' of the two-charge state differs substantially from the extremal black hole geometry; second, that the supergravity approximation is not valid near the source, because some parts of the geometry become sub-Planckian in size; and third, that the quantum fluctuations remain large even well away from the source, rendering classical supergravity inapplicable.

One of the aims of the present work is to revisit these issues.  The reason that the supergravity approximation breaks down in the D1-D5 and NS5-F1 duality frames, so that they are inappropriate for describing the near-source structure, is that the circle the branes are wrapping is much smaller than the string scale already far away from the source. 
If we locally T-dualize the NS5-F1 geometry to the NS5-P duality frame, the geometry is sourced by momentum waves propagating along the fivebranes, whose stress-energy is a pressure causing the T-dual circle to expand to a large size near the source.  Thus an effective supergravity analysis in this frame is appropriate for describing this region of spacetime.  We conclude that while classical supergravity in the NS5-F1 frame breaks down, classical string theory does not (at least, not on this account).

When we consider the near-source geometry in the appropriate duality frame, we find that a black hole background is not the right description~-- the fivebrane source is spread out over a region much larger than the stretched horizon of the extremal black hole background.  The ``ensemble geometry'' is a description of an explicit, smeared brane source which is not close to collapsing behind a horizon.  It differs substantially from the black hole background near the source, where it describes a ``fivebrane star''.

As to the third issue, the analysis of~\rcite{Raju:2018xue} calculated the fluctuations of the harmonic functions in the grand canonical ensemble, and pointed out that they are order one over a region extending out to the blob radius $\tilde{r}_b$ and beyond; we address this critique in a companion paper \cite{PaperB}. 

The horizonless spherical blob solution appears to be weakly coupled throughout space, implying that strands of the fivebranes are well-separated from each other in target space (since it is coinciding fivebranes that leads to strong coupling~\rcite{Giveon:1999px,Martinec:2019wzw,Martinec:2020gkv}). We verify this by explicitly calculating the probability that the fivebrane strands are separated by coordinate distance $d$
\be
\Big\langle \delta\big(d^2-|\vec{f}(v_1) \tight- \vec{f}(v_2)|^2\big)\Big\rangle~.
\ee
We will see that when the fivebranes are highly excited, this quantity peaks at a value of $d$ corresponding to an invariant distance of a few $\sqrt{Q_5}$ length scales, and actually vanishes quadratically in $d$ at small separation, 
ensuring that weakly coupled string theory is a reliable description. This result is of interest because the analysis of~\rcite{Martinec:2019wzw,Martinec:2020gkv,Martinec:2022okx} argued that near-intersections of the fivebranes lead to the appearance of ``W-strings'' of little string theory which are lighter than fundamental strings when the fivebranes are sufficiently close, making strong coupling physics necessary. This phenomenon does not take place in typical 1/2-BPS states, suggesting that such self-intersections are relatively rare in the 1/2-BPS configuration space.

\begin{figure}
\centering
\includegraphics[scale=0.75]{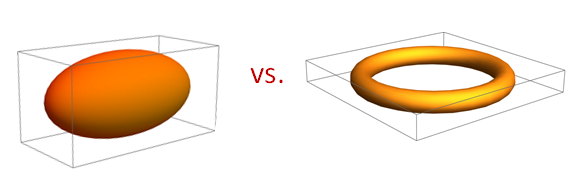}
\caption{On the left: The ellipsoidal structure of the ensemble average bound state when the angular potentials are fixed. On the right: A ring shape of the bound state with fixed angular momenta. }
\label{TwoEnsembles}
\end{figure}

In a separate thread of investigation of ``ensemble geometries'', we consider ensembles of rotating brane sources. Known examples with fixed angular momentum are characterized by a ring transverse to the compactification manifold~\rcite{Lunin:2001fv}. One can again ask: What structures arise in the string theory of typical rotating bound states of fivebranes and strings?  We address this question both in the ensemble of fixed angular momentum, and the conjugate ensemble of fixed angular potential.

The four dimensions transverse to the fivebranes permit two independent angular potentials about two orthogonal planes. We first calculate the solution corresponding to an ensemble average where the chemical potential conjugate to the string charge and the two angular potentials are fixed.  A priori, one might expect that the structure of the resulting ensemble average would exhibit two phases: First, for low angular potentials, the blob structure should arise as one sees in the absence of rotation. Then for large angular potentials, one might expect that an angular momentum barrier would cause the ensemble average solution to be localized about a ring, as one sees in known rotating NS5-F1 solutions~\rcite{Lunin:2002qf}.  Somewhat surprisingly, the solution instead displays an ellipsoidal rather than toroidal structure across all possible values of the angular potentials. 

We trace this phenomenon to order one relative fluctuations of the angular momentum in the grand canonical ensemble of fixed angular potential. For small angular potentials, the average angular momentum scales linearly with the angular potential while the fluctuations are constant, making the relative fluctuations large. For large values of the angular potential, a mode that winds once around the spatial circle begins to fluctuate wildly.  The relative fluctuations of the angular momentum are set by the relative fluctuations of the occupation number in this mode, which become of order unity. 

A calculation demonstrates that relative fluctuations of the angular momentum surpass 90\%.  In this case, predictions from the two conjugate ensembles are allowed to disagree. Having the result for fixed angular potential, one can apply a Fourier transform to switch to ensembles of fixed angular momentum, and reproduce explicitly the effect of repulsion from the rotation axis leading to ring shapes.  In the ensemble of fixed angular potential, one is averaging over rings of varying size, and because the fluctuations are large, the result smears the ensemble of ring geometries into the observed ellipsoid.

The outline of the paper runs as follows.  In section \ref{sec:GCSolution} we review the Lunin-Mathur solutions and the solution corresponding to an ensemble average in the grand canonical ensemble. We make a transition to the microcanonical ensemble of fixed charge, and recall that the circumference of the $y$-circle is sub-Planckian near the bound state \rcite{Raju:2018xue}. 
This section concludes with an analysis of the solution written in the NS5-P duality frame where this particular issue is resolved. 

In section \ref{sec:SeparationStrands}, we show that the fivebranes in the grand canonical ensemble with small chemical potential are typically separated by a few $\sqrt{Q_5}$ length scales. Section \ref{sec:RotatingSolution} is devoted to computing ensemble averages of the harmonic functions in the rotating ensemble with fixed angular potentials, finding an ellipsoid spinning in transverse space.  Section~\ref{sec:FixedAngularMomenta} performs the transformation to an ensemble of fixed angular momentum, upon which repulsion from the rotation axis emerges. We draw conclusions in section \ref{sec:Conclusions}.


\section{Review of the non-rotating solution}
\label{sec:GCSolution}
Consider an ensemble of NS5-F1 bound states, with each member making excursions in transverse space. Suppose one fixes the charges of the state with $n_1,n_5\gg 1$. One can ask what is the bulk structure of any member of this microcanonical ensemble? It is possible to address this question by first weighting configurations with different charges using a grand canonical distribution, through fixing a chemical potential conjugate to the string excitation level. We denote this chemical potential by $2\pi \tau_2$ because we will utilize a formalism that resembles string theory on a worldsheet torus. The limit of $\tau_2 \to 0^+$ corresponds to nearly equal weighting of the microstates. Then as we review below, the harmonic functions describing the grand canonical ensemble average admit simple analytical expressions.  Later we make a transition to the microcanonical ensemble of fixed charges, which is the maximally mixed state consisting of all states of the effective string at a fixed total oscillator level $L_0=n_1n_5$.

\subsection{The Lunin-Mathur geometries}
\label{subsec:LMgeoms}

This subsection reviews the NS5-F1 Lunin-Mathur geometries \rcite{Lunin:2001fv} which constitute a starting point for many of the subsequent calculations in this paper.

We adopt the following notations: $g_s$ is the asymptotic string coupling, $(2\pi)^4V_4$ is the asymptotic volume of the four-torus and $2\pi R_y$ is the asymptotic circumference of the spatial $y$-circle. The charges carried by each member of the ensemble are expressible in terms of the number of fundamental strings and fivebranes:
\begin{equation}
	Q_1=\frac{g_s^2 n_1 (\alpha')^3}{V_4} ~~,~~~~ Q_5=n_5\alpha'  ~.
\end{equation}
The string frame metric and B-field are given by:
\begin{align}
\label{LMgeom}
\begin{split}
 ds^2 &= \frac{1}{H_1}\left( -(dt+\A)^2 + (dy+\B)^2\right)+H_5\, d\vec{x}^2 + |d\vec{z}|^2 
\\[.2cm]
B &= \frac{1}{H_1}(dt+\sfA)\wedge(dy+\sfB) + \sfb
~~,~~~~ d\sfb = *_\perp dH_5  ~.
\end{split}
\end{align}
The spatial coordinates $\vec{x}$ are transverse to the compactification manifold $\bT^4\times \bS^1 _y$. One can work with spherical coordinates where $d\vec{x}^2 = dr^2 + r^2 d\Omega_3 ^2$. The coordinates of the $\bT^4$ factor are denoted by $\vec{z}$. 
The dilaton is given by
\begin{equation}
\label{LMdil}
	e^{2\phi} = g_s ^2\frac{H_5}{H_1} ~.
\end{equation}
The harmonic forms and functions $H_1,H_5$ \etc. are given by the convolution of the four-dimensional Euclidean Green's function with one-dimensional sources located along $\vec x=\vec f(v)$   
\begin{align}
\label{A}
		H_5 (\vec{x}) &= \frac{Q_5}{L}\int_0 ^{L} \frac{1}{|\vec{x}-\vec{f}(v)|^2}dv
\nn\\[.2cm]
		H_1 (\vec{x} ) &= \frac{Q_5}{L}\int_0 ^{L} \frac{|\partial_v \vec{f}(v)|^2}{|\vec{x}-\vec{f}(v)|^2}dv
\\[.2cm]
		\vec{A}(\vec{x}) &= -\frac{Q_5}{L}\int_0 ^{L} \frac{\partial_v \vec{f}(v)}{|\vec{x}-\vec{f}(v)|^2}dv
  ~~,~~~~   	dB = -*_{\perp} dA ~.
\nn\end{align}
where $*_{\perp}$ denotes the Hodge dual operation with respect to flat four-dimensional transverse space, and we have dropped constant terms in the harmonic functions in order to take the AdS$_3$ limit.
The parameter $L$ is obtained from successive applications of U-duality operations starting from the duality frame where the conserved charges are F1-P (see \eg~\rcite{Mathur:2005zp}): 
\begin{equation}
	L = \frac{2\pi Q_5}{R_y} ~.
\end{equation}
Far away from the sources, 
\begin{equation}
		H_1 (\vec{x}) \approx \frac{Q_1}{|\vec{x}|^2 }~~,~~~~ H_5(\vec{x}) \approx \frac{Q_5}{|\vec{x}|^2} \qquad (|\vec{x}|\to \infty)~.
\end{equation}
By rescaling
\begin{equation}
\label{scaledcoords}
r= \frac{\sqrt{Q_1 Q_5}}{R_y} \tilde{r}~~,~~~~ y= R_y Y ~~,~~~~ t= R_y T  ~,
\end{equation}
the asymptotic line element can be written as
\begin{equation}\label{AdS3S3T4}
ds^2 = \ell_{AdS}^2 { \Big[}\tilde{r}^2\big( -dT^2 + dY^2\big)+\frac{d\tilde{r}^2}{\tilde{r}^2}+d\Omega_3 ^2{ \Big]} + |d\vec{z}|^2  ~.
\end{equation}
This is asymptotically AdS$_3 \times \bS^3 \times \bT^4$ with the length scale set by the fivebrane charge $Q_5$
\begin{equation}
	\ell_{AdS} = \sqrt{n_5 \alpha'}~.
\end{equation}
We will calculate the expectation values of the harmonic functions and forms in the grand canonical ensemble and reproduce the known result \rcite{Alday:2006nd,Balasubramanian:2008da,Raju:2018xue}
\begin{align}
\label{HarmonicGC}
	\langle H_1 (\vec{x}) \rangle &=\frac{Q_1}{|\vec{x}|^2}\left(1-e^{-\frac{6\tau_2 |\vec{x}|^2}{\pi \mu^2}} \right)
\nn\\[.2cm]
	\langle H_5 (\vec{x})\rangle &=\frac{Q_5}{|\vec{x}|^2}\left(1-e^{-\frac{6\tau_2  |\vec{x}|^2}{\pi \mu^2}} \right)
\\[.5cm]
	\langle \vec{A}\rangle &= 0 ~.
\nn
\end{align}
The radial coordinate characterizing the fuzzy boundary of the thermal blob is
\begin{equation}
	r_b = \sqrt{\frac{\pi }{6\tau_2}}\mu~,
\end{equation}
where $\tau_2$ is a chemical potential for the total excitation level of the profile $\vec f(v)$, and
\begin{equation}
\label{mudef}
\mu = \frac{g_s (\alpha')^2}{R_y \sqrt{V_4}}~.
\end{equation}

\subsection{Averages of the harmonic functions}
\label{subsec:HarmFnAves}

We now present a detailed path integration derivation of equations (\ref{HarmonicGC}) which is applicable for any chemical potential for the charge $Q_1$. 
Ultimately we will focus on the case of small chemical potential relevant to the large $N=n_1n_5$ limit. 

The source profile $\vec f(v)$ parametrizes the 1/2-BPS phase space of purely NS backgrounds (an additional set of four ``internal'' profile functions parametrizes 1/2-BPS configurations with R-R excitations).  Geometric quantization of this phase space turns the Fourier coefficients of the profile functions into a set of decoupled harmonic oscillator creation/annihilation operators~\cite{Rychkov:2005ji}.
The path integral approach we utilize gives the same results as the operator approach used in previous literature on the average harmonic functions \rcite{Alday:2006nd,Balasubramanian:2008da,Raju:2018xue}. 

The approach of Lunin and Mathur was to begin with the F1-P 1/2-BPS solutions where the momentum wave on the fundamental string has the profile $\vec f$, and then perform a sequence of U-dualities to arrive at the NS5-F1 solutions~\eqref{A}.  One can instead start directly with the NS5-P duality frame, where $\vec f(v)$ is the profile of a momentum wave on the fivebranes, and then T-dualize to NS5-F1 where $\vec f(v)$ characterizes a string winding condensate carried by the fivebranes.  The fivebranes compactified on $\bT^4$ are an effective string, and we wish to quantize this set of excitations, which form a closed subspace of the set of all fivebrane excitations.%
\footnote{
Note that the profile function $\vec{f}(v)$ does not have a dynamical zero mode in the decoupling limit.  The fivebranes have a tension $\propto g_s^{-2}$, and the decoupling limit involves taking the asymptotic string coupling to zero, making the fivebranes infinitely heavy; the center-of-mass motion of the fivebranes thus decouples.
}

We wish to study the ensemble of 1/2-BPS states at some large total momentum on this effective string.  To that end, we introduce a chemical potential for the excitation level; in other words, we consider the conventional worldsheet path integral on a Euclidean torus with the modulus $\tau$, described by the Polyakov action with the transverse $\bR^4$ as the target space. This way we study the effects of four bosonic fields describing excursions of the fivebranes in transverse space (this interpretation is appropriate in the NS5-P duality frame). The integration is over closed curves of these excursions. We will then adapt the relevant expressions to describe chiral bosonic states preserving half of the spacetime supersymmetry generators that annihilate the worldsheet CFT ground states.

Starting with the expectation value of $H_5 (\vec{x})$, we want to compute
\begin{equation}\label{GreensFnAve}
\left \langle \frac{1}{|\vec{x}-\vec{f}(z,\bar{z})|^2} \right\rangle =\frac{1}{Z(\tau,\bar{\tau})} \int \cD \vec{f} \; e^{-S_P [\vec{f}]}\,\frac{1}{|\vec{x}-\vec{f}(z,\bar{z})|^2}~,
\end{equation}
where the worldsheet position $z$ is related to $v$ through
\begin{equation}
\label{zv}
z=e^{i\frac{2\pi v}{L}},
\end{equation}
the Polyakov action is given by
\begin{equation}
\label{effstringact}
S_P [\vec{f}] = \frac{1}{4\pi \mu ^2} \int d^2 \sigma~ \partial_a \vec{f} \cdot \partial^a \vec{f}~.
\end{equation}
Normally, the string tension is denoted $\alpha'$. Here, one is instructed to describe an effective ``dual string'' (the fivebrane compactified on $\bT^4$) whose inverse tension $\mu^2$ is determined by the moduli as in~\eqref{mudef}.%
\footnote{More precisely, in the NS5-P frame, $\tilde\mu^2=g_s^2(\alpha')^2/V_4$ sets the inverse tension of the fivebrane wrapped on $\bT^4$; the factor of $R_y$ in the NS5-F1 frame expression~\eqref{mudef} is due to the T-dual string coupling.}

The notation $Z(\tau,\bar{\tau})$ represents the factor of the partition function of the theory on a torus worldsheet of modulus $\tau$, excluding zero modes:
\begin{equation}
  Z(\tau,\bar{\tau})=\frac{1}{|\eta(\tau)|^{2D}}~.
\end{equation}
We work in a convention where the expectation value of the unit operator is equal to one.
It is convenient to use the identity
\be
\frac{1}{|\vec{x}-\vec{f}|^2} = \int_0 ^{\infty} \!ds \,
\frac{1}{(\pi s)^{{D}/{2}}}\int d^D b~ e^{-\frac{1}{s}|\vec{b}|^2 -s|\vec x|^2 + 2(s\vec x+i\vec{b}) \cdot \vec{f}}
\ee
(the substitution $D=4$ will be made later). Substituting into the path integral~\eqref{GreensFnAve} results in%
\footnote{We follow the conventions of \rcite{Polchinski:1998rq} (Volume I section 6.2).}
\begin{align}
\label{harmfn2exptl}
	\left\langle \frac{1}{|\vec{x}-\vec{f}(z,\bar{z})|^2} \right\rangle  &= \frac{1}{Z(\tau,\bar{\tau})}\int D\vec{f} \int_0 ^{\infty} ds \int \frac{d^D b}{\left(\pi s\right)^{\frac{D}{2}}}\, e^{-S_P[\vec{f}]-\frac{|\vec{b}|^2}{s}-s|\vec{x}|^2 + 2\left(s\vec{x}+i\vec{b}\right) \cdot \vec{f}(z,\bar{z})}~.
\end{align}
The profile $\vec{f}$ can be expanded in terms of eigenfunctions $\Phi_i(z,\bar z)$ of the Laplacian on the torus (with eigenvalues $-\omega_i ^2$)
\begin{equation}\label{expansion}
	\vec{f} (z,\bar{z}) = \sum_i \vec{X}_i\Phi_i (z,\bar{z})~;
\end{equation}
the path integral then becomes a set of decoupled Gaussian integrals over the coefficients $\vec{X}_i$. 
The result can be expressed in terms of the worldsheet Green's function 
\begin{equation}\label{FormalExpressionGreen}
	G'(z_1,z_2,\bar{z}_1,\bar{z}_2) = \sum_{i\neq0} \frac{\pi\mu^2}{\omega_i ^2}\Phi_i (z_1,\bar{z}_1) \Phi_i (z_2,\bar{z}_2)~.
\end{equation}
On the torus, this evaluates to \rcite{Polchinski:1998rq} 
\begin{equation}\label{GreensTorus}
	G'(w,w',\bar{w},\bar{w}') = -\frac{\mu^2}{2}\log\bigg| \theta_1 \Big( \frac{w-w'}{2\pi}\Big|\tau \Big) \bigg| ^2+\frac{\mu^2}{4\pi \tau_2}\text{Im}\left(w-w'\right)^2+k(\tau,\bar{\tau})~.
\end{equation}
Orthogonality with the zero mode (the constant function on the torus) determines
\begin{equation}
	k(\tau,\bar{\tau}) = \frac{\mu^2}{2}\log\left(|\eta(\tau)|^2\right)~.
\end{equation}
The renormalized version of $G'$ at coincident points is obtained by subtracting the leading logarithmic ultraviolet divergence from the limit of coincident points of $G'$. A finite function which is a sum of a holomorphic term and an anti-holomoprphic term emerges. The holomorphic term in the result reads
\begin{equation}
\label{Gr0}
	G^r (0|q) =-\frac{\mu^2}{2}\log\left( \frac{\theta_1 '(0|\tau)}{2\pi} \right)+k(\tau)=-\frac{\mu^2}{2}\log\left( q^{-\frac{1}{12}}\eta(q)^{2}\right)~.
\end{equation}
Taking a degeneration limit with $\tau_1=0$ results in
\begin{equation}\label{SizeSquare}
	G ^r (0|q) \approx \mu^2~ \frac{1}{1-q}\sum_{m=1} ^{\infty} \frac{1}{m^2} =\frac{\pi^2 \mu^2}{6 (1-q)}= \frac{\pi \mu^2}{12 \tau_2}~.
\end{equation}
The renormalization scheme of subtracting the lograrithmic divergence is not unique; instead one could have introduced an ultraviolet cutoff $\epsilon$ on the worldsheet and consider the Green's function for a pair of points separated by $\epsilon$. In the operator approach, one can show that the same $G^r (0|q)$ emerges as a consequence of calculating the expectation value of the normal ordered exponential operator $:\!e^{i\vec{k} \cdot \vec{\hat{f}}}\!:$ 
\begin{equation}
  \frac{1}{Z(\tau)} \text{tr} \Big(:\!e^{i \vec{k}\cdot \vec{\hat{f}} }\!\!: q^{\hat{L}_0} \Big) =e^{-\frac{1}{2}|\vec{k}|^2 G^r (0|q)}  
\end{equation}
where $\hat{L}_0$ is the zeroth Virasoro generator. Moreover, in the operator approach one can consider the non-normal ordered operator and introduce a cutoff on the mode number $m$ that labels states, $m\leq \Lambda$, with $\epsilon = \frac{1}{\Lambda}$.
The distinction between normal ordering or not normal ordering does not matter as long as one first takes the limit of $\tau_2 \to 0^+$ and then $\epsilon\to 0$, because then the $\frac{1}{\tau_2}$ term in the expansion (\ref{SizeSquare}) wins over any additive logarithmic correction. Throughout this paper we consider this order of limits, and not the reverse.

Going back to the Green's function and plugging in the equations above,
\begin{align}
\Bigg\langle \frac{1}{|\vec{x}-\vec{f}(z,\bar{z})|^2} \Bigg\rangle & =\frac{1}{Z(\tau,\bar{\tau})} \Big(\prod_i \int d^D\!X_i \Big) \int_0 ^{\infty} ds \int \frac{d^D b}{\left(\pi s\right)^{\frac{D}{2}}} 
\\
  &\hskip 1cm
\times \exp\left(-\sum_i \frac{\omega_i ^2}{2\pi \mu^2}\vec{X}_i \cdot \vec{X}_i-\frac{|\vec{b}|^2}{s} -s|\vec{x}|^2 + 2\left(s\vec{x}+i\vec{b}\right) \cdot \sum_i \vec{X}_i \Phi_i (z,\bar{z})\right)~.
\nn
\end{align}
Performing the $\vec{X}$-integration gives rise to
\begin{align}
	\left\langle \frac{1}{|\vec{x}-\vec{f}(z,\bar{z})|^2} \right\rangle  &=  \int_0^{\infty} ds \int \frac{d^D b}{\left(\pi s\right)^{\frac{D}{2}}} \,e^{-\frac{|\vec{b}|^2}{s}-s|\vec{x}|^2   +2 G^r (0|q)\left(s\vec{x}+i\vec{b}\right) \cdot \left(s\vec{x}+i\vec{b}\right)}~.
\end{align}
Note that the result is independent of the worldsheet position $z,\bar{z}$.
It follows that
\begin{align}
	\left\langle \frac{1}{|\vec{x}-\vec{f}(z,\bar{z})|^2} \right\rangle  &=  \int_0 ^{\infty} ds~ e^{-s|\vec{x}|^2   + 2s^2 |\vec{x}|^2 G^r (0|q)} \int \frac{d^D b}{\left(\pi s\right)^{\frac{D}{2}}} \, e^{-\left(\frac{1}{s}+2G^r (0|q)\right)|\vec{b}|^2+4is G^r (0|q) \vec{b}\cdot  \vec{x} }. 
\end{align}
Shifting the $\vec{b}$ variable to $\vec{\tilde{b}} \equiv \vec{b}-\frac{2is G^r (0|q)}{\frac{1}{s}+2G^r (0|q)} \vec{x} $ and doing the $\vec{\tilde{b}}$ integration yield
\begin{align}
	\left\langle \frac{1}{|\vec{x}-\vec{f}(z,\bar{z})|^2}\right \rangle &= \int_0 ^{\infty} ds~ e^{-s|\vec{x}|^2   + 2s^2 |\vec{x}|^2 G^r (0|q) - \frac{4s^2 G^r (0)^2 |\vec{x}|^2}{\frac{1}{s}+2G^r (0|q)}} \frac{1}{\left(1+ 2s G^r (0|q)\right)^{\frac{D}{2}}}
 \nonumber\\[.2cm]
	&  =\int_0 ^{\infty} ds~ \frac{1}{\left(1+ 2s G^r (0|q)\right)^{\frac{D}{2}}}e^{-\frac{|\vec{x}|^2}{2G^r (0|q)}\left(1- \frac{1}{1+2s G^r (0|q)}\right) }~.
\end{align}
The integrand is analytic if  $\text{Im}(G^r (0|q))=0$ and $\text{Re}\left(G^r (0|q) \right)>0$, which is the case in the degeneration limit $\tau_2 \to 0^+$ with $\tau_1=0$ that we will consider. As long as $\text{Im}(G^r (0|q))\neq0$, no essential singularity plagues the integrand.
Let us focus on $D=4$ and change variables: 
\begin{equation}
	y \equiv \frac{|\vec{x}|^2}{2G^r (0|q)\big(1+2s G^r (0|q)\big)}~.
\end{equation}
Then
\begin{align}\label{Result1}
	\left\langle \frac{1}{|\vec{x}-\vec{f}(z,\bar{z})|^2}\right \rangle  &=\frac{1-e^{-\frac{|\vec{x}|^2}{2G^r (0|q)}}}{|\vec{x}|^2}~.
\end{align}
Substituting the small chemical potential limit in eq.~(\ref{SizeSquare}) into eq.~(\ref{Result1}), averaging over the worldsheet position and multiplying by the NS5 charge $Q_5$ result in%
\begin{equation}\label{Result1prime}
	\big\langle H_5 (\vec{x})\big\rangle =\oint \frac{dz}{2\pi i z}\left\langle \frac{Q_5}{|\vec{x}-\vec{f}(z,\bar{z})|^2}\right \rangle =\frac{Q_5}{|\vec{x}|^2}\left( 1-e^{-\frac{6\tau_2|\vec{x}|^2}{\pi \mu^2}}\right)~.
\end{equation}
At large distances $|\vec{x}|\gg \sqrt{G^r(0|q)}$ the result approaches the correct asymptotic falloff condition ${Q_5}/{|\vec{x}|^2}$; at small distances $|\vec{x}|\ll \sqrt{G^r(0|q)}$ or small $\tau_2\ll 1$ , the averaged Green's function is finite and approximately constant at the value ${6 \tau_2}/({\pi \mu^2})$.

Next, consider the following quantity:
\begin{equation}
	\mathcal{A}_i(v,\vec{x}) \equiv -\frac{\partial_v f_i(v)}{|\vec{x}-\vec{f}(v)|^2}~,
\end{equation}
 We are interested in calculating
\begin{equation}
	\big\langle 	\mathcal{A}_i(v,\vec{x}) \big\rangle =  -\int_0 ^{\infty} ds\int \frac{d^D b}{(\pi s)^{\frac{D}{2}}} e^{-s|\vec{x}|^2 -\frac{|\vec{b}|^2}{s}} \left\langle \partial f_i  e^{i \vec{k}\cdot \vec{f}}\right\rangle~,
\end{equation}
where the operator on the right is taken at some point in the worldsheet and
\begin{equation}
\label{kdef}
	\vec{k} = -2i(s\vec{x} + i \vec{b})~.
\end{equation}
It is shown that the renormalized correlation function in the integrand above vanishes
\begin{equation}
\left \langle \partial f_i(z_1)  e^{i \vec{k}\cdot \vec{f}(z_2,\bar{z}_2)}\right \rangle_{ren}=0~.
\end{equation}
The renormalization is done by subtracting a pole singularity from the limit of the correlator as $z_1\to z_2$. 
\begin{equation}
	\left\langle \partial f_i(z_1)  e^{i \vec{k}\cdot \vec{f}(z_2,\bar{z}_2)}\right\rangle = ik_i \partial_{z_1} G'(z_1,\bar{z}_1,z_2,\bar{z}_2|\tau) \left\langle e^{i\vec{k}\cdot \vec{f}}\right \rangle~.   
\end{equation} 
For nearly coincident points, $v_1\to v_2$ 
\begin{equation}
	\partial_{v_1} G'\left( \frac{v_{12}}{2\pi}|\tau\right) = -\frac{\mu^2}{2}\frac{ \frac{\partial}{\partial \nu} \theta_1\left( \nu|\tau\right)_{\nu=0}}{2\pi\theta_1 \left( \frac{v_{12}}{2\pi}|\tau\right)}=
	-\frac{\mu^2}{2}\frac{\frac{1}{v_{12}}}{1+\frac{1}{2}\frac{1}{2\pi}\frac{\theta_1''(0|\tau)}{\theta_1 ' (0|\tau)}v_{12}+\frac{1}{6}\frac{1}{4\pi^2} \frac{\theta_1 '''(0|\tau)}{\theta_1'(0|\tau)}v_{12}^2+\dots}
\end{equation}
It follows that the renormalized derivative of the Green's function is 
\begin{equation}
	\partial G^r (0|q) = \frac{\mu^2}{8\pi\theta_1'(0|\tau)} \theta_1''(0|\tau)=0.
\end{equation}
To explain the last equality, the Jacobi theta function satisfies the diffusion equation 
\begin{equation}\label{Diffusion}
\frac{\partial ^2}{\partial w^2}\theta_1 \left(\frac{w}{2\pi}|\tau \right)=\frac{i}{\pi} \frac{\partial}{\partial \tau }\theta_1 \left(\frac{w}{2\pi}|\tau\right)~.
\end{equation}
For $\nu=0$, $\theta_1 (0|\tau)=0$ and so is $\frac{\partial}{\partial \tau }\theta_1 (0|\tau)=0$.
The conclusion is that
\begin{equation}
	\big\langle 	\mathcal{A}_i \big\rangle =0~.
\end{equation}
This is physically sensible by thinking about members of the ensemble of the bound state as coming in pairs that mirror each other around each axis $i$ with opposite momentum or winding. Then the Kalb-Ramond and graviphoton fields that they produce cancel on average; a parity symmetry emerges on average.

To evaluate the expectation value of the last harmonic function $H_1 (\vec{x})$, consider the operator
\begin{equation}
	K(v,\vec{x}) = \frac{|\partial_v \vec{f}|^2}{|\vec{x}-\vec{f}(v)|^2}~.
\end{equation}
Computing the expectation value of $K(v,\vec{x})$ amounts to writing
\begin{equation}
	\big\langle K(v,\vec{x}) \big\rangle = \int_{0} ^{\infty}ds \int \frac{d^D b}{(\pi s)^{\frac{D}{2}}} e^{-s |\vec{x}|^2 -\frac{|\vec{b}|^2}{s}} \left\langle \partial \vec{f} \tight\cdot \partial \vec{f} \, e^{i\vec{k}\cdot \vec{f}}\right\rangle~,
\end{equation}
with $\vec{k}$ given by~\eqref{kdef}.
We saw that the renormalized contraction between $\partial f_i$ to $e^{i\vec{k}\cdot\vec{f}}$ vanishes. It follows that the renormalized contraction we need to compute is between the two derivatives of $\vec{f}$, and multiply the result by the renormalized average of the exponential:
\begin{equation}\label{Contractions}
	 \left\langle \partial \vec{f} \tight\cdot \partial \vec{f} \, e^{i\vec{k}\cdot \vec{f}}\right\rangle =  \left\langle \partial \vec{f} \tight\cdot \partial \vec{f} \right\rangle_{ren} \left \langle e^{i\vec{k}\cdot \vec{f}}\right\rangle_{ren}~.
\end{equation}
The expectation value of the operator on the left factor on the R.H.S of (\ref{Contractions}) admits a $w^{-2}$ double pole singularity as $w\to 0$; its renormalized value is the expectation value of the normal-ordered stress energy tensor  (times $-\mu^2$) of the effective string. 
The relevant expression is available in \rcite{Polchinski:1998rq}, but rather than writing it here, one can take a shortcut.
A useful relation between the charges carried by the bound state is
\begin{equation}
   Q_1=\frac{Q_5}{L}\int_0 ^{L} |\partial_v  \vec{f}(v)|^2 dv~.
\end{equation}
Therefore,
\begin{equation}
\label{H1vev}
\big\langle H_1 (\vec{x}) \big\rangle = \frac{Q_5}{L}\int_0 ^L \!\big\langle K(v,\vec{x}) \big\rangle \,dv=  \frac{Q_1}{|\vec{x}|^2}\left( 1- e^{-\frac{6\tau_2|\vec{x}|^2}{\pi \mu^2}}\right)~.
\end{equation}
Alternatively, one can view the coefficient of ${|\vec{x}|^{-2}}$ falloff at large distances as the renormalized value of the fundamental string charge.

The advantage of our derivation is that it shows that for any finite chemical potential, the scale $\frac{\mu^2 \pi}{12\tau_2}$ is replaced by
\begin{equation}
    G^r (0|q) = -\mu^2 \sum_{n=1} ^{\infty} \log(1-q^n)=\mu^2 \sum_{n=1} ^{\infty} \frac{q^n}{n(1-q^n)} ~,~ q=e^{-2\pi \tau_2}.
\end{equation}
Modular invariance implies that for small $\tau_2$, the leading order correction to $\frac{\mu^2 \pi}{6\tau_2}$ is $\mu^2 \log(\tau_2)$, which is negative. This means that the size of the solution decreases when increasing slightly the chemical potential; this makes sense because states of high string charge are suppressed. 
Note however that this logarithmic correction comes in at the same order as the cutoff dependence that we renormalized away in evaluating $G^r(0,q)$ in eq.~\eqref{Gr0}, and so suffers from renormalization scheme dependence.

\subsection{Moving to the microcanonical ensemble}
\label{subsec:MovingToTheMicrocanonicalEnsemble}

We would like to calculate the averaged harmonic functions in the microcanonical ensemble when the fixed charges are large. The upshot is that a saddle point approximation relates $\tau_2$ with an order one number times $(n_1 n_5)^{-1/2}$, therefore one should make such a substitution in the results (\ref{HarmonicGC}). 

As a short introduction to this subsection, it is useful to recall a few basic facts. Generally, the partition function $Z(\beta)$ of any canonical ensemble is related to the density of states $\rho(E)$ through
\begin{equation}
	Z(\beta) = \int_0 ^{\infty} dE\, \rho(E) \,e^{-\beta E}~.
\end{equation}
(The ground state energy is set to zero.)
The inverse Laplace transform allows one to extract the density of states from the knowledge of the partition function
\begin{equation}
	\rho(E) = \int_C \frac{d\beta}{2\pi i }\, Z(\beta) \,e^{\beta E}~.
\end{equation}
In this equation, $C$ is a  vertical contour in the $\beta$-complex plane to the right of possible singularities of $Z(\beta)$.
Thermal expectation values transform to microcanonical expectation values (around the narrow energy window $\left[E\tight-\frac{1}{2}\Delta E,E\tight+\frac{1}{2}\Delta E\right]$) utilizing the same inverse Laplace transform:
\begin{equation}
	\big\langle A \big\rangle _E = \int_C \frac{d\beta}{2\pi i } \, \big\langle A \big\rangle^{~}_{\beta} \,e^{\beta E}~.
\end{equation}
In writing $\langle A\rangle_{\beta} $ one should not normalize it by dividing by the partition function.
For our purposes, the notations $\beta = 2\pi \tau_2$ is the chemical potential and $E = N$, where $N$ is the total level or charge of the state:
\begin{equation}\label{MicroCanonical}
	\big\langle A \big\rangle _{N} = \int_C \frac{d\tau_2}{i } \big\langle A \big\rangle (\tau_2) \,e^{2\pi \tau_2 N }~.
\end{equation}
Now, the following expression for the small chemical potential partition function on a torus worldsheet with $D$ physical target space dimensions, can be derived utilizing modular invariance,
\begin{equation}
	Z(\tau_2 \to 0^+) =\tau_2 ^{\frac{D}{2}}e^{\frac{\pi D}{12 \tau_2 }}~. 
\end{equation}
It is useful to recall the computation of the density of states that follows from it, via
eq.~(\ref{MicroCanonical}), 
\begin{equation}\label{DOS}
	 \rho(N) =-i\int_C d\tau_2 \,e^{2\pi \tau_2 N }\tau_2 ^{\frac{D}{2}}e^{\frac{\pi D}{12 \tau_2 }}\equiv-i\int_C d\tau_2 \,e^{g(\tau_2)}~,
\end{equation}
where
\begin{equation}
	g(\tau_2) \equiv 2\pi N\tau_2 +\frac{\pi D}{12\tau_2} + \frac{D}{2}\log(\tau_2)~.
\end{equation}
Conventionally, the saddle point approximation is used to evaluate the integral in eq.~(\ref{DOS}). At large $N$, the condition $g'(\tau_2^*)=0$ leads to
\begin{align}\label{tauSaddle}
	\tau_2 ^* &\approx \frac{\sqrt{D}}{\sqrt{24 N}} -\frac{D}{8\pi N}+ \dots 
\\
	g(\tau_2 ^*) &= 2\pi \sqrt{\frac{D}{6}N}-\frac{D}{4}\log(N)+...
\end{align}
We choose the integration contour $C$ to contain the point $\tau_2 ^*$ and parallel to the imaginary axis $\text{Im}(\tau_2)$. Then, one may rotate back to real $\tau_2$ by multiplying the ``original $\tau_2$'' by $-i$. This leads to
\begin{equation}\label{DOS2}
	\rho(N) \sim  N^{-\frac{D+3}{4}}e^{2\pi \sqrt{\frac{D}{6}N}}~.
\end{equation}
Moving on to the unnormalized expectation value of the 4D Green's function,
\begin{align}
\lim_{\tau_2\to 0^+}
    \bigg\langle \frac{1}{|\vec{x}-\vec{f}|^2}\bigg \rangle =\frac{1}{|\vec{x}|^2}\tau_2 ^{2}e^{\frac{\pi }{3 \tau_2 }}\left( 1-e^{-\frac{6\tau_2|\vec{x}|^2}{\pi \mu^2}}\right)~,
\end{align}
we want to apply an inverse Laplace transform on this function. The first term transforms trivially to the density of states computed above. One can normalize it away, so that ${|\vec{x}|^{-2}}$ is the first term in the expression we want to calculate. Regarding the second term, note that an ``effective level'' enters the computation 
\begin{equation}
	N' \equiv N -\frac{3|\vec{x}|^2}{\pi^2 \mu^2}~.
\end{equation}
As long as $N\gg {|\vec{x}|^2}/{\mu^2}$, the saddle point approximation is reliable. In this case,
\begin{align}\label{AveHarmMicroIntermediate}
	\bigg\langle \frac{1}{|\vec{x}-\vec{f}|^2}\bigg \rangle \Bigg|_{N\gg \frac{|\vec{x}|^2}{\mu^2}} &= \frac{1}{|\vec{x}|^2 N^{\frac{7}{4}}}e^{2\pi \sqrt{\frac{2}{3}N}}\left( 1-e^{2\pi \sqrt{\frac{2}{3}\left(N- \frac{3|\vec{x}|^2}{\pi^2 \mu^2}\right)}-2\pi \sqrt{\frac{2}{3}N}}\right)~.
\end{align}  
Dividing by $N^{-\frac{7}{4}}e^{2\pi \sqrt{\frac{2}{3}N}}$ and expanding the difference in the argument of the exponential on the R.H.S of the last equation yield
\begin{align}
\label{AveHarmMicro}
	\bigg\langle \frac{1}{|\vec{x}-\vec{f}|^2}\bigg \rangle\Bigg|_{N\gg \frac{|\vec{x}|^2}{\mu^2}} &= \frac{1}{|\vec{x}|^2}\left( 1-e^{-\frac{\sqrt{6} |\vec{x}|^2}{\pi \mu^2 \sqrt{N}}}\right)~.
\end{align}  
In the opposite regime $N\ll {|\vec{x}|^2}/{\mu^2}$, the exponential suppression $e^{-\frac{6 \tau_2 |\vec{x}|^2}{\pi \mu^2} }$ renders the second term very small compared to the first one. In fact, already from eq.~(\ref{AveHarmMicro}) one sees that the scale at which the two terms are comparable is $|\vec{x}|\sim N^{\frac{1}{4}}\mu$. The conclusion is that the average harmonic functions for the ensemble in question, where $N=n_1 n_5$, are
\begin{align}\label{HarmonicMC}
	\big\langle H_1 (\vec{x}) \big\rangle &=\frac{Q_1}{|\vec{x}|^2}\Big(1-e^{-\frac{\sqrt{6} |\vec{x}|^2}{\pi \mu^2 \sqrt{n_1 n_5}}}\Big)
\nn\\
	\big\langle H_5 (\vec{x}) \big\rangle &=\frac{Q_5}{|\vec{x}|^2}\Big(1-e^{-\frac{\sqrt{6} |\vec{x}|^2}{\pi \mu^2 \sqrt{n_1 n_5}}} \Big)
\\[.2cm]
	\big\langle \vec{A} \,\big\rangle &= 0 ~.
\nn
\end{align}
These equations describe a spherically symmetric blob of radius 
\begin{equation}
\label{rbdef}
    r_b = \mu \sqrt{\pi} \left( \frac{n_1 n_5}{6}\right)^{\frac{1}{4}}~.
\end{equation}
One can understand the scale $r_b$ as follows.  The ensemble under consideration is a thermal ensemble of left-moving excitations on the effective string governed by~\eqref{effstringact}.  The typical mode number in such 
a thermal state is of order $\sqrt{N}$, and one can crudely model the typical configuration as a random walk whose step size is the thermal wavelength.  There are $\sqrt{N}$ steps of this size, leading to an r.m.s. radius of the random walk of order $N^{\frac14}$, which is what we see in~\eqref{rbdef}. 

The resulting exponential of twice the dilaton is everywhere constant and small:
\begin{equation}
	e^{2\phi} = g_s ^2\frac{H_5}{H_1}=\frac{n_5 V_4}{n_1(\alpha')^2}~.
\end{equation}
The NS-NS flux components are given by
\begin{equation}
    H_{rty} =-\frac{\partial_r H_1}{H_1^2}~, 
\end{equation}
and defining angles through $x_1 +ix_2= r\sin(\theta)e^{i\phi}$ and $x_3+ix_4 = r\cos(\theta) e^{i\psi}$,
\begin{equation}
     H_{\theta \psi \phi}=r^3 \cos(\theta)\sin(\theta)\frac{\partial}{\partial r} H_5 (r).
\end{equation}
The 6D part of the line element of the microcanonical ensemble averaged geometry is thus given by:
\begin{equation}
	ds^2 _6 = \frac{r^2}{Q_1\big(1-e^{-{r^2}/{r_b ^2}} \big)}\left( -dt^2 + dy^2\right)+ \frac{Q_5\big(1-e^{-{r^2}/{r_b ^2}} \big)}{r^2}\left(dr^2 + r^2 d\Omega_{3} ^2 \right) ~.
\end{equation}
We use coordinates suitable for describing physics in the decoupling limit
\begin{equation}
	r= a \tilde{r}~~,~~~~ y= R_y Y ~~,~~~~ t= R_y T~,
\end{equation}
where
\be
\label{adef}
a = \frac{\sqrt{Q_1Q_5}}{R_y}  ~.
\ee
In particular, 
\begin{equation}\label{tilderb}
    \tilde{r}_b = \frac{\mu}{a} \sqrt{\pi} \left( \frac{n_1 n_5}{6}\right)^{\frac{1}{4}}=\frac{\sqrt{\pi}}{(6 n_1 n_5)^{\frac{1}{4}}}~.
\end{equation}
The line element becomes
\begin{equation}
\label{MetricAveragedNS5F1}
	ds_6 ^2 = \ell_{AdS} ^2\left[\frac{ \tilde{r}^2}{1-e^{-{ \tilde{r}^2}/{\tilde{r}_b ^2}}}\left( -dT^2 + dY^2\right)+ \frac{1-e^{-{\tilde{r}^2}/{\tilde{r}_b ^2}}}{\tilde{r}^2}\left(d\tilde{r}^2 + \tilde{r}^2 d\Omega_{3} ^2 \right)\right]~.
\end{equation}
The dilaton remains constant, and one should drop (via gauge transformation) the constant $-1$ in the equation for B-field in the decoupling limit. Then
\begin{equation}
\label{aveBTY}
  H_{\tilde{r}TY} =\partial_{\tilde{r}}   B^{(2)} _{TY} ~,~ B^{(2)} _{TY} = -\ell_{AdS} ^2 \frac{\tilde{r}^2}{1-e^{-{\tilde{r}^2}/{\tilde{r}_b ^2}}}~.
\end{equation}
The proper size of the Y-circle in the string frame is given by:
\begin{equation}
	\text{Size}(\bS^1 _Y)=	\sqrt{g_{YY}} = \sqrt{n_5 \alpha'} \frac{\tilde{r}}{\sqrt{1-e^{-{\tilde{r}^2}/{\tilde{r}_b ^2}}}}~.
\end{equation}
The rescaled radial coordinate where the circle is approximately string-sized is
\begin{equation}
\label{rd}
	\tilde{r}_{d} = \frac{1}{\sqrt{n_5}}~.
\end{equation}
Note that because $n_5\ll n_1$, the blob radius is much less than the duality radius $\tilde{r}_b \ll\tilde{r}_{d}$.
Furthermore, since the 10D Planck length scale is given by
	\begin{equation}
		\ell_{P}^{(10D)} = \left((2\pi)^7 \frac{n_5V_4 (\alpha')^2}{2n_1} \right)^{\frac{1}{8}}~,
	\end{equation}
the boundary of the thermal blob has the property that
\begin{equation}
		\frac{\text{Size}(\bS^1 _y)|_{\tilde{r}_b}}{\ell_{P}^{(10D)}} \propto \left(\frac{n_5(\alpha')^2}{n_1 V_4}\right)^{\frac{1}{8}} \ll 1	~.
\end{equation}
Therefore, NS5-F1 is the wrong duality frame far away from the thermal blob -- it breaks down already at $\tilde{r}_{d}\gg \tilde{r}_b$.
The $y$-circle shrinks below the 10D Planck scale in the NS5-F1 duality frame. The shrinking of the spatial circle below the string scale in the geometry of the black hole with the same charges and the need to find an appropriate duality frame was discussed in \rcite{Martinec:1999gw}. In the context of two-charge horizonless solutions, the same issue was emphasized in \rcite{Chen:2014loa}. 

The D1-D5 frame is also the wrong duality frame for describing the solution~\eqref{HarmonicGC} near the source.
The logic of~\rcite{Martinec:1999gw,Chen:2014loa} suggests that the appropriate duality frame for the description of the vicinity of the source is the NS5-P frame. While heuristic arguments in \rcite{Chen:2014loa} led to using several U-duality operations in a chase for a weakly coupled and weakly curved description, which resulted in a strongly-curved phase, here we find that a single T-duality operation suffices to attain this goal successfully.

\subsection{The solution in the NS5-P frame}
\label{subsc:TheSolutionInTheNS5-PFrame}

For the NS5-P duality frame to be the appropriate effective description, the string coupling should be small and the gradients of the supergravity fields should be small in string units.  Let us apply the Buscher rules of T-duality \cite{Buscher:1987sk} to the solution in the NS5-F1 frame with respect to the Y-circle, to arrive at a solution in the NS5-P frame.

For the general NS5-F1 Lunin-Mathur geometries~\eqref{LMgeom}, the T-dual NS5-P geometry is
\begin{align}
\label{NS5-Psol}
 ds^2 &= -d\uhat\,d\vhat + H_1\, d\vhat^2 +2\,\sfA_i\,dx^i\,d\vhat +H_5\, d\vec{x}^2 + |d\vec{z}|^2 
\nn\\[.2cm]
dB &= d\vhat\wedge *_\perp d\sfA + *_\perp dH_5  
\\[.2cm]
e^{2\phi} &= g_s ^2 H_5 ~.
\nn
\end{align}
For the ensemble geometry, the harmonic functions remain those given above in eq.~\eqref{HarmonicGC}.
Applying the Buscher rules directly to the geometry~\eqref{MetricAveragedNS5F1}, \eqref{aveBTY},
one finds that the T-dual line element can be written 
as
\begin{align}
\label{NS5-P2ndRep}
ds^2 _{\textrm{NS5-P}} &=-n_5 \alpha' \frac{\tilde{r}^2}{1-e^{-{\tilde{r}^2}/{\tilde{r}_b ^2}}}\,dT^2+ \frac{\alpha'}{n_5} \frac{1-e^{-{\tilde{r}^2}/{\tilde{r}_b ^2}}}{\tilde{r}^2}\left[d\tilde{Y}+n_5 \frac{\tilde{r}^2}{1-e^{-{\tilde{r}^2}/{\tilde{r}_b ^2}} } dT\right]^2 
\nonumber\\[.2cm]
 &\hskip 2cm
+ n_5 \alpha' \frac{1-e^{-{\tilde{r}^2}/{\tilde{r}_b ^2}}}{\tilde{r} ^2}\Big(d\tilde{r}^2 + \tilde{r}^2 d\Omega_3 ^2 \Big)+d|\vec{z}|^2
\end{align}
(with $\tilde Y$ the coordinate T-dual to $Y$).
This metric has the form~\eqref{NS5-Psol} with $\uhat=T$, $\vhat=\tilde Y$.
Curvature invariants of the geometry, like the Ricci curvature and the square root of the Kretschmann scalar, are suppressed by $Q_5 ^{-1}$. 

The geometry (\ref{NS5-P2ndRep}) and associated B-field and dilaton do not define an exactly conformal sigma model~-- there will be non-trivial $\alpha'$ corrections to the background.  The gradients of the background in string units are all of order $1/n_5$, however, so these corrections will be small in the limit of large~$n_5$.

\begin{figure}[ht]
\centering
  \begin{subfigure}[b]{0.4\textwidth}
  \hskip 0cm
\includegraphics[width=\textwidth]{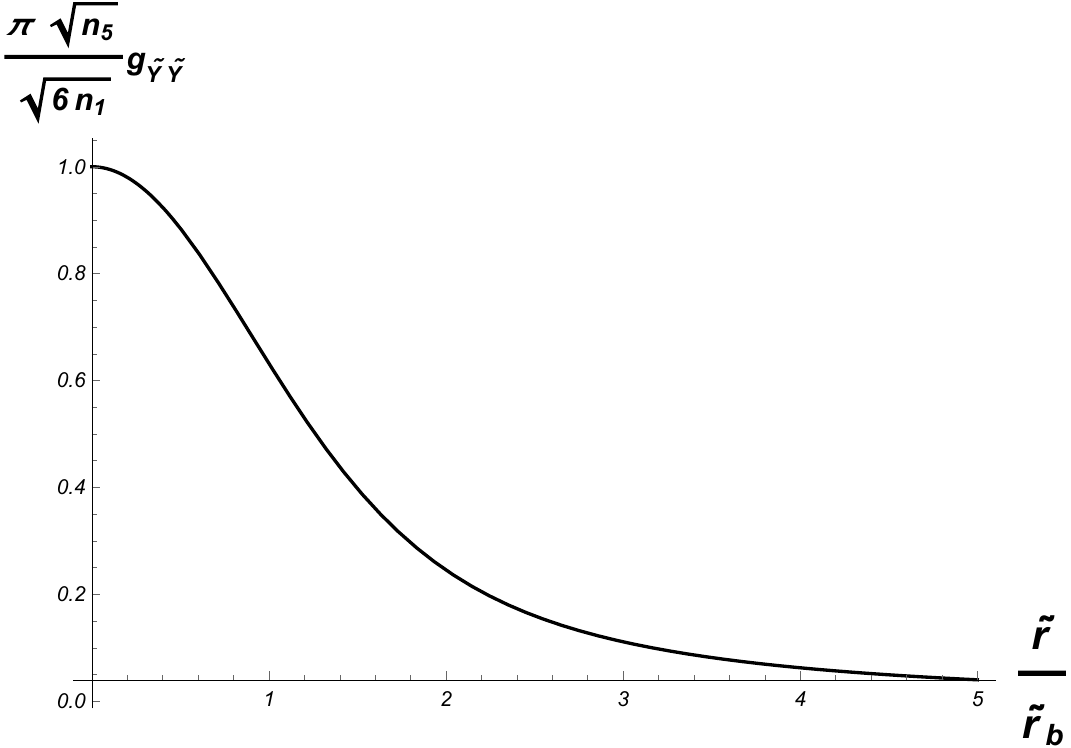}
    \caption{ }
    \label{fig:YtildeTube}
  \end{subfigure}
\hskip 2cm
  \begin{subfigure}[b]{0.4\textwidth}
  \vskip -2cm
    \includegraphics[width=\textwidth]{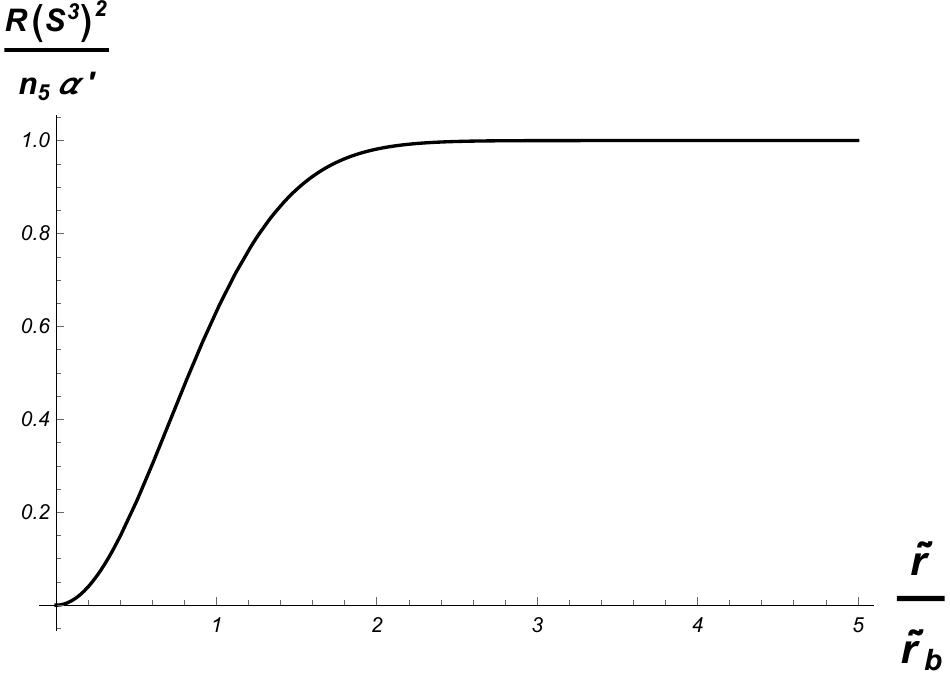}
    \caption{ }
    \label{fig:RS3}
  \end{subfigure}
\caption{ \it a) The $\tilde Y$-circle metric as a function of a dimensionless radial coordinate that saturates near the core of the solution. b) The radius squared of the $\bS^3$ factor of the geometry is fixed away from the source to be $\sqrt{n_5 \alpha'}$, and decreases to zero at the origin smoothly. 
}
\label{fig:YtildeTubeS3}
\end{figure}

The near-source geometry in this duality frame has the structure of a ``bag of gold''.  The $\tilde{Y}$-circle grows as the radius decreases, from the string scale at the radius $\tilde r_d=\frac{1}{\sqrt{n_5}}$, eq.~\eqref{rd}, saturating at the value set by~\eqref{NS5-P2ndRep} as $r\to 0$
\begin{equation}
\alpha'\sqrt{\vphantom{t}g_{\tilde{Y}\tilde{Y}}}=\alpha'\sqrt{g^{YY} } \propto \left( \frac{n_p}{n_5}\right)^{\frac{1}{4}}
\end{equation}
starting at the blob radius $\tilde r_b$.
This value is much larger than the string scale for $n_5\ll n_p$ (which is a condition for the NS5-P duality frame to be valid).
Thus, the Y-circle starts from the string scale at $\tilde{r}_d$ and saturates at $\sqrt{\alpha'} \big( \frac{n_p}{n_5}\big)^{\frac{1}{4}}$ at the origin. The invariant radial distance over which the circle grows in this frame is approximately given by 
\be
L_{\rm bag}\approx \sqrt{n_5 \alpha'} \Big[1+\frac{1}{4}\log\Big(\frac{n_p}{n_5} \Big)\Big].
\ee
A plot of $g_{\tilde{Y}\tilde{Y}}$ and the radius squared of the $\bS^3$ factor of the geometry are shown in figure~\ref{fig:YtildeTubeS3}. 

On the other hand, the $\bS^3$ factor of the geometry is large outside the blob, with radius $\sqrt{n_5 \alpha'}$.
As one dives inside the blob, the radius of the sphere starts to shrink, vanishing smoothly at the origin. This behavior is explained by the NS-NS magnetic three-form flux threading the sphere, which is supporting the sphere from shrinking. This flux is sourced by the fivebranes, which are wandering about inside the blob radius, such that in the ensemble average the source of flux smoothly fills the spherical blob.  As we go to smaller radius inside the blob, a Gaussian surface encloses less and less of the fivebrane charge, and so the three-sphere starts to shrink in size since there is less magnetic $H$-flux to support it.%
\footnote{In the NS5-P frame, the fivebranes are seen as explicit sources for the magnetic three-form flux via the harmonic function $H_5$, and the momentum is carried by the fivebrane.  Upon ensemble averaging, the fivebrane singularity is smoothed out, but the magnetic $H_3$ flux threading the angular sphere still measures the amount of fivebrane charge inside a given radius; the geometry still has a distributed, explicit source.
After T-duality to the NS5-F1 frame, there are no explicit sources for the $H_3$ flux~-- the wandering fivebranes have been transformed into a distribution of KK monopoles, whose core is a coordinate singularity~\rcite{Mathur:2005zp}.  There are no explicit charged sources in any individual geometry, as the fivebranes have in some sense been dissolved into flux, nevertheless, upon ensemble averaging the dipolar KKM structure averages away, and one ends up with the same feature that in the averaged geometry, $H_3$ flux is decreasing as one penetrates further into the source blob.
}

In performing T-dualities, one runs the risk of encountering strong string coupling. However, this is not the case for the problem under consideration: Plugging the value of the dilaton in the deep interior of the geometry predicted by the attractor mechanism $	e^{2\phi(r\to 0)} = \frac{n_5 V_4}{n_p(\alpha')^2}$ and the size of the Y-circle,
\begin{equation}
	e^{\tilde{\phi}_b} =  \left(\frac{n_5 V_4 }{n_p (\alpha')^2}\right)^{\frac{1}{4}} \left(\frac{V_4}{(\alpha')^2}\right)^{\frac{1}{4}}~.
\end{equation}
The factor on the right hand side of the last equation can be made small: $V_4 \sim (\alpha')^2$ and $n_5 \ll n_p$.  The conclusion is that string theory admits a weakly-coupled and weakly-curved description near the source.

The redshift factor for a ``co-rotating observer'', for which the combination in the square brackets of eq.~(\ref{NS5-P2ndRep}) vanishes, is predicted to be
\begin{equation}
	\frac{1}{\sqrt{|g_{TT}(r<r_b)|}} \;\Biggl|_{d\tilde{Y} + \sqrt{\frac{n_5}{n_p} }dT = 0} ~\sim~ \left(\frac{n_p}{n_5}\right)^{\frac{1}{4}} \,\gg~ 1 ~.
\end{equation}
While the solution has no horizon, there is a large redshift in its core.

Reference \cite{Alday:2006nd} put a stretched horizon at the radius of the blob and pointed out that its area in Planck units is proportional to $N^{\frac{3}{4}}$, much bigger than the statistical mechanical entropy of the system $\propto \sqrt{N}$. As this reference mentioned, the geometry is not that of a black hole, and so we see no reason why one should have put a stretched horizon there in the first place.  

One may introduce a stretched horizon for the classical, massless BTZ black hole at $\tilde{r}_{stretch}=\frac{1}{\sqrt{n_1 n_5}}$ to reproduce the scaling of the statistical mechanical entropy. However, the typical microstates contributing to the ensemble are not the vacuum all the way down to this radius, but rather have explicit fivebrane sources already out at the blob radius.  The objects being described are fivebrane stars rather than black holes.

Note that the light excitations in this ``bag of gold'' are momentum modes, which are string winding modes in the NS5-F1 frame -- we are rearranging the string condensate carried by the fivebranes.  At very low energy, this amounts to motion along the 1/2-BPS configuration space parametrized by the $f^I(v)$.


\section{Separation between strands of the fivebranes}
\label{sec:SeparationStrands}
A well-known phenomenon in string theory is that a collection of coincident NS5 branes are characterized by a dilaton that grows linearly in the proper distance towards them (in string units).  This phenomenon persists even in the stringy regime of two fivebranes, which can be described by the worldsheet NSR formalism.  In the NS5-P frame, D2-branes stretching between the intersecting fivebranes become tensionless ``W-strings''.  In the NS5-F1 frame, two-sphere factors of the geometry shrink and the tension of D3 branes wrapping them (as well as a direction in $\bT^4$) goes to zero. Consequently, string perturbation theory breaks down and a different description has to be found. Reference \rcite{Martinec:2022okx} speculated that this different description predicts the formation of a small black hole at the intersection point, or a splitting of the original fivebrane into two.

On the other hand, when fivebranes are separated along a circle in transverse space, the string coupling remains weak throughout \rcite{Giveon:1999px,Giveon:1999tq}. What happens when putting these solitonic objects in an ensemble of fixed chemical potential in the limit of small chemical potential?  We will see that the strands of the multiply wound fivebrane are typically separated by an invariant distance scale comparable to an order one number times $\sqrt{n_5 \alpha'}$, allowing one to use string perturbation theory to reliably describe them. This is consistent with the result of the previous subsection, that the string coupling is weak even when approaching the ensemble averaged source.

\begin{figure}[ht]
\centering
\includegraphics[scale=0.4]{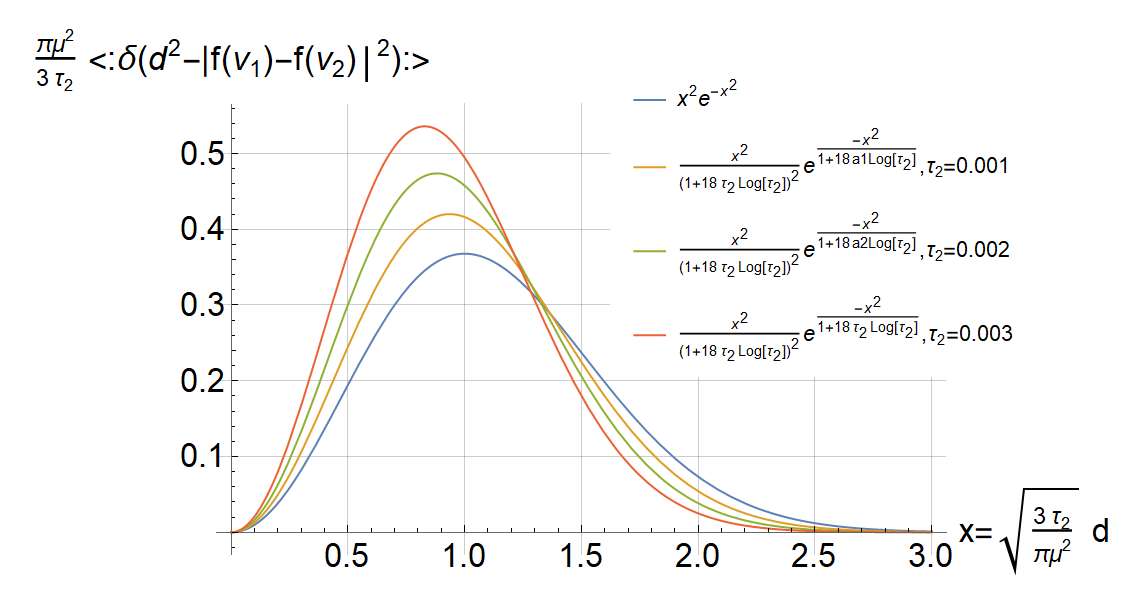}
\caption{ 
 When $d$ is small, the likelihood that points along the profile are separated by $d$ is small. As $d\to\infty$, it is exponentially unlikely to find pairs of points separated by $d$. The likelihood peaks at $d\approx \frac{\sqrt{\pi}\,\mu}{\sqrt{3\tau_2}}$.  
 }
 \label{Separation}
\end{figure}

We would like to calculate the typical scale of separation between points on the fluctuating fivebrane profile among typical members of the ensemble:
\begin{equation}
	\left\langle :\delta\left(d^2 - \big|\vec{f}(v_1)\tight-\vec{f}(v_2)\big|^2\right):\right\rangle~.
\end{equation}
Fourier transforming the delta function
\begin{equation}\label{Separation1}
	\left\langle :\delta\left(d^2 - \big|\vec{f}(v_1)-\vec{f}(v_2)\big|^2\right):\right\rangle=\frac{1}{2\pi} \int_{-\infty} ^{\infty} ds\; \Big\langle :e^{is\left(d^2 - |\vec{f}(v_1)-\vec{f}(v_2)|^2\right)}:\Big\rangle~
\end{equation}
and then transforming the Gaussian in $\vec{f}$ using
\begin{equation}
	e^{-is |\vec{f}(v_1)-\vec{f}(v_2)|^2} = \int\frac{d^D k}{(i \pi s)^{\frac{D}{2}}} \, e^{i\frac{|\vec{k}|^2}{s}+2i\vec{k}\cdot (\vec{f}(v_1)-\vec{f}(v_2))}~,
\end{equation}
eq.~(\ref{Separation1}) becomes
\begin{equation}\label{Separation2}
	\left	\langle :\delta\left(d^2 - \big|\vec{f}(v_1)-\vec{f}(v_2)\big|^2\right):\right\rangle=\frac{1}{2\pi} \int_{-\infty} ^{\infty} ds~ e^{is d^2}\int\frac{d^D k}{(i \pi s)^{\frac{D}{2}}}\, e^{i \frac{|\vec{k}|^2}{s}} \left\langle :e^{ 2i\vec{k}\cdot\left( \vec{f}(v_1)-\vec{f}(v_2)\right)}:\right\rangle~.
\end{equation}
The expectation value of the exponential operator on the RHS is given by
Appendix \ref{Appendix:DetailsTorusGreensFunction}, which demonstrates that in the small chemical potential limit $\tau_2\ll 1$,
\begin{equation}\label{TwoPointResult}
	\left\langle :e^{ 2i\vec{k}\cdot\left( \vec{f}(v_1)-\vec{f}(v_2)\right)}:\right\rangle \to e^{-E |\vec{k}|^2} ~,~ E = 2\mu^2 \left(\frac{\pi}{6\tau_2}+\log(\tau_2)\right).
\end{equation}
Note that the same considerations regarding normal-ordering, renormalization scheme dependence, \etc. discussed in section~\ref{subsec:HarmFnAves}, apply here as well~-- these only come in at the subleading level of the logarithm.
The $\vec{k}$-integral takes the form
\begin{equation}\label{GaussianIntegral}
	\int d^D k ~ e^{-E|\vec{k}|^2 +\frac{i}{s}|\vec{k}|^2} = \frac{\pi^{\frac{D}{2}}}{\left( E-\frac{i}{s}\right)^{\frac{D}{2}}}~.
\end{equation} 
Substituting eq. (\ref{GaussianIntegral}) into eq. (\ref{Separation2}) yields 
\begin{equation}\label{Separation4}
	\left\langle :\delta\left(d^2 - \big|\vec{f}(v_1)-\vec{f}(v_2)\big|^2\right):\right\rangle=\frac{1}{2\pi} \int_{-\infty} ^{\infty} ds~ e^{is d^2} \frac{1}{\big(1+iEs\big)^{\frac{D}{2}}}~.
\end{equation}
For $D=4$ this can be written as
\begin{equation}\label{Separation5}
	\left\langle :\delta\left(d^2 - \big|\vec{f}(v_1)-\vec{f}(v_2)\big|^2\right):\right\rangle	= -\frac{1}{2\pi E^2} \int _{-\infty} ^{\infty} ds ~ e^{id^2 s} \frac{1}{(s-s_*)^2}~,
\end{equation}
where
\begin{equation}
	s_* \equiv \frac{i}{E}.
\end{equation}
The Taylor series of the exponential around $s=s_*$ contains the terms
\begin{equation}\label{ExponentTaylor}
	e^{id^2 s} = 1+id^2 e^{-\frac{d^2}{E}}(s-s_*)+\dots
\end{equation}
Since $e^{isd^2}=e^{id^2 \text{Re}(s)-d^2 \text{Im}(s)}$, the contour of the integral (namely the real line) can be supplemented by a semicircle of infinitely large radius. The semicircle is located in the positive $\text{Im}(s)>0$ region in the complex plane associated with $s$. Since $\text{Re}(E)>0$, a pole is present inside that region. Plugging eq.~(\ref{ExponentTaylor}) into eq.~(\ref{Separation5}) and using the residue theorem, one obtains 
\begin{equation}\label{Separation6}
	\left\langle :\delta\left(d^2 - \big|\vec{f}(v_1)-\vec{f}(v_2)\big|^2\right):\right\rangle=\frac{d^2}{E^2} \,e^{- \frac{d^2}{E}}\;~.
\end{equation}
The result in eq.~(\ref{Separation6}) is consistent with the fact that the delta-function has a unit integral:  Integrating $d^2$ from zero to infinity and changing the variable $z\equiv \frac{d^2}{E}$, $\int_0 ^{\infty} \!ze^{-z}dz =1$ as required. 
Using the result (\ref{TwoPointResult}) of appendix \ref{Appendix:DetailsTorusGreensFunction}, it follows that
\begin{align}\label{Separation7}
	&\frac{\pi \mu^2}{3\tau_2}\left\langle :\delta\left(d^2 - \big|\vec{f}(v_1)-\vec{f}(v_2)\big|^2\right):\right\rangle \approx
\frac{1}{\left(1+\frac{6}{\pi}\tau_2 \log(\tau_2)\right)^2} \,x^2 \,\exp\bigg[{-\frac{x^2}{1+\frac{6}{\pi}\tau_2 \log(\tau_2)}}\bigg]~,
\nn
\end{align}
where
\begin{equation}
	x\equiv \frac{\sqrt{3\tau_2}\, d}{\sqrt{\pi}\,\mu}~.
\end{equation}
To leading order, 
\begin{align}\label{Separation8}
	&\frac{\pi \mu^2}{3\tau_2}\left\langle :\delta\left(d^2 - \big|\vec{f}(v_1)-\vec{f}(v_2)\big|^2\right)\!:\right\rangle\approx x^2 e^{-x^2}~.
\end{align}
This result is plotted in figure \ref{Separation}. One can see that the points on the profile are typically of distance $x\approx 1$. Including the effect of the $\log(\tau_2)$ correction reduces the value of $d$ where the maximum is located and increases the value of that maximum: Larger chemical potentials contract the extent of the bound state in target space. 

The observable we have computed treats the strands of the fivebranes as if they were living in flat space, whereas they propagate in a curved spacetime. Nonetheless, the approximation of flat space becomes valid near the source where the metric components saturate at constant values. 

One can understand the length scale for the typical distance between strands
\begin{equation}
 d_{typical} = \frac{\sqrt{\pi} \mu}{\sqrt{3\tau_2}} \propto \mu (n_1 n_5 )^{\frac{1}{4}}\propto r_b~,
\end{equation}
in the following way. If the strands are separated radially at $\{r_1=0,r_2=r_b\}$, $\{r_1=0.5r_b,r_2=1.5r_b\}$ and $\{r_1=r_b,r_2=2r_b\}$, then the invariant distances can be computed using the radial-radial component of the metric which we found in the previous subsection. These distances are determined by
\begin{equation}
    L = \int_{\tilde{r}_1} ^{\tilde{r}_2} \sqrt{n_5 \alpha'}\frac{\sqrt{1-e^{-{\tilde{r}^2}/{\tilde{r}_b ^2}}}}{\tilde{r}} \,d\tilde{r} =\{0.92,0.79,0.63\}\,\sqrt{n_5 \alpha'}~.
\end{equation}

The conclusion is that the strands are separated over a characteristic invariant length scale comparable with $\sqrt{n_5 \alpha'}$. It is therefore permitted to use the same weakly-coupled string theory description for them, without resorting to another description (\eg\ low-energy gauge theory or 11D M-theory) of a strongly-coupled phase of nearly-coincident fivebranes.  The phenomenon of brane self-intersection, which is nearby in the configuration space of circular supertubes, does not generically occur in more general microstates. There is evidence that black hole physics is tightly related to the self-intersection of fivebranes, when they are described by non-abelian gauge theories. The result of the previous section that the grand canonical ensemble average of the two-charge states is a horizonless configuration is consistent with the lack of self-intersections we see in this ensemble average.

\section{New rotating solution}
\label{sec:RotatingSolution}

\begin{figure}[ht]
    \centering
    \includegraphics{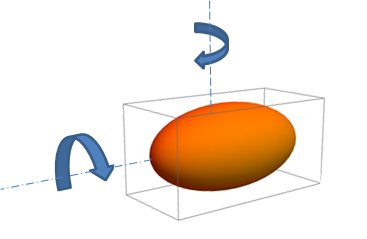}
    \caption{Structure of the ensemble average of NS5-F1 with fixed but different angular potentials.}
    \label{RotatingNS5F1}
\end{figure}

In this section we derive the average harmonic functions in the half-BPS ensemble of fixed chemical potential and angular potentials in two orthogonal planes. 
The excitations of the fivebranes are again restricted to the four transverse directions so that the background is purely NS.  The main result is that we obtain a smooth, horizonless, supersymmetric solution that exhibits ellipsoidal structure normal to the compactification manifold. The spread of the bound state with equal but nonzero angular potentials is a sphere of radius larger than the sphere obtained at zero angular potentials, in a way we predict below. Breaking the symmetry by distinct angular potentials (\eg\ zero rotation in one plane and nonzero rotation in the other) gives rise to the ellipsoid. The topology of this solution is generated by zero angular momentum states participating in the ensemble of fixed angular potentials - a phenomenon we explain in the next section. Figure~\ref{RotatingNS5F1} depicts the main result schematically. The physical properties of the solution are summarized in subsection \ref{subs:Properties}.

\subsection{Preliminaries: Action and partition function}
\label{Preliminaries:subsec}
To introduce the ensemble we wish to study, it is useful to review the action description and the partition function describing rotating strings. In the present context, strands of NS5 branes transverse to $\bS^1 _y \times \bT^4$ are described by an effective string on a worldsheet torus, with four target space position fields.

The torus worldsheet is spanned by the vectors $2\pi \hat{\bf e}_{1}$ and $2\pi(\tau_1 \hat{\bf e}_{1} + \tau_2 \hat{\bf e}_{2})$. A parametrization of a cell in the torus has $0\leq \sigma_1 \leq 2\pi$ and $0\leq \sigma_2 \leq 2\pi \tau_2$.
The $\sigma_1$ coordinate is treated as a ``spacelike'' coordinate while $\sigma_2$ is considered to be a ``timelike'' coordinate.  
We would like to describe fivebrane waves that depend on a torus coordinate, such that the fivebrane rotates about a plane, subjected to a given chemical potential conjugate to the angular momentum. 
We start with the Polyakov action on a torus:
\begin{equation}
	I = \frac{1}{4\pi \mu^2} \sum_i \int d^2 \sigma ~ \partial_{a} X^i \partial^{a} X_i  .
\end{equation}
We parametrize the plane of rotation in the target space by $X^1$ and $X^2$. The angular momentum is defined in terms of the wedge product between the position and its conjugate momentum. The momentum conjugate to $X_i(\sigma_1,\sigma_2)$ is 
\begin{equation}
	P_i(\sigma_1,\sigma_2) = \frac{\delta I}{\delta (\partial_{\sigma_2} X_i(\sigma_1,\sigma_2))} = \frac{1}{2\pi \mu^2} \partial_{\sigma_2} X_i(\sigma_1,\sigma_2)~.
\end{equation}
The angular momentum is the spatial integral at some constant $\sigma_2$ (``time slice'') of the product:
\begin{equation}\label{AngularMomentumij}
	L_{12} = -\frac{i}{2\pi \mu^2}\int_0 ^{2\pi} d\sigma_1 \left( X_1 \partial_{\sigma_2} X_2 - X_2 \partial_{\sigma_2} X_1 \right)~. 
\end{equation}
We integrate over the timelike direction the product $i\omega_{12} L_{12}$, where $\omega_{12}$ stands for the angular potential:
\begin{equation}\label{OmegaL}
	i\int_0 ^{2\pi \tau_2} d \sigma_2 \,\omega_{12}\, L_{12}~.
\end{equation}
While the inclusion of the term in eq.~(\ref{OmegaL}) is supposed to subject an extended object to a given angular potential, naively it breaks conformal invariance on the worldsheet.  In order to preserve conformal invariance, we gauge the rotational symmetry, and interpret $\omega_{12}$ as a constant background gauge field. Writing
\begin{equation}\label{PolarCoordinates}
	X_1 \pm i X_2 = r\,e^{\pm i \phi}~,
\end{equation} 
the Polyakov action reads
\begin{equation}
	I = \frac{1}{4\pi \mu^2} \sum_a \int d^2 \sigma \left( \partial_a r \partial^a r+r^2 \partial_a \phi \partial^a \phi\right)~. 
\end{equation}
The angular chemical potential term eq.~(\ref{OmegaL}) becomes
\begin{equation}
i\int_0 ^{2\pi \tau_2} d \sigma_2\, \omega_{12}\,L_{12}=	\frac{1}{2\pi \mu^2} \int d^2 \sigma ~ \omega_{12} \,r^2\, \partial_2 \phi~.
\end{equation}
The gauging of rotations involves adding to the theory a background gauge field $A$ with the Lagrangian description
\begin{equation}\label{TotalAction}
	I_{tot} = \frac{1}{4\pi \mu^2}  \int d^2 \sigma \left( \sum_a\partial_a r \partial^a r+r^2 \gamma^{11} \partial_1 \phi \partial_1 \phi + r^2 \gamma^{22}\left( \partial_2 \phi + A\right)\left(\partial_2 \phi + A\right)\right)~. 
\end{equation}
(The notation $\gamma^{ab}$ here stands for the component of the inverse worldsheet metric.) Incorporating a chemical potential for the angular momentum amounts to setting a background gauge field
\begin{equation}
	A = \omega_{12}~,
\end{equation}
in other words a Wilson line on the ``timelike'' cycle of the torus.

To perform the path integral, one again expands in eigenfunctions of the wave operator
\begin{align}\label{XiSum}
	X_i (\sigma_1,\sigma_2) &= \sum_{n,m} X^i _{n,m} \Phi_{n,m}(\sigma_1,\sigma_2)
\\[.1cm] 
\label{EigenfunctionsDef}
	\nabla^2 \Phi_{n,m}(\sigma_1,\sigma_2) &= -\omega_{n,m} ^2 \Phi_{n,m}(\sigma_1,\sigma_2) ~,~ \omega_{n,m}^2 = k_1^2 + k_2^2,
\\[.2cm]
\label{EigenfunctionsExp}
	\Phi_{n,m}(\sigma_1,\sigma_2) &= \frac{1}{\sqrt{(2\pi)^2 \tau_2}} e^{i k_1 \sigma_1 + ik_2 \sigma_2}~~,~~~~
	k_1 = n ~,~~ k_2 = \frac{m-n\tau_1}{\tau_2}~.
\end{align}
The eigenfunctions satisfy an orthonormality condition
\begin{equation}\label{Orthonormality}
	\int_0 ^{2\pi} d\sigma_1\int_0 ^{2\pi \tau_2} d\sigma_2  \, \Phi_{n,m} (\sigma_1,\sigma_2)\, \Phi_{n',m'}(\sigma_1,\sigma_2) = \delta_{n,-n'} \delta_{m,-m'}~.
\end{equation}
The reality of the $X_i$ fields implies that $X^i _{-n,-m}=X^{i~*} _{n,m}$. 
A substitution of eq.~(\ref{XiSum}) into the Polyakov action results in
\begin{equation}
	I = \frac{1}{4\pi \mu^2} \sum_{n,m} \frac{|n \tau-m|^2 }{\tau_2 ^2} \left(X^1 _{n,m} X^1 _{-n,-m}+X^2 _{n,m} X^2 _{-n,-m}\right)~.
\end{equation}
We plug eq.~(\ref{XiSum}) into eq.~(\ref{AngularMomentumij}) to find
\begin{equation}
	\int_0 ^{2\pi \tau_2} d\sigma_2\, \omega_{12}\, L_{12} =
 \frac{\omega_{12}}{\pi \mu^2} \sum_{n,m} \frac{m-n \tau_1}{\tau_2} X^1 _{n,m} X^2 _{-n,-m}~.
\end{equation}
The last $A^2$-term is
\begin{equation}
	\frac{\omega_{12}^2}{4\pi \mu^2} \sum_{n,m} \left( X^1 _{n,m} X^1 _{-n,-m}+X^2 _{n,m} X^2 _{-n,-m}\right)~.
\end{equation}
It is useful to change variables to diagonalize the quadratic sum with respect to the target space dimensions:
\begin{equation}
	X_{n,m} ^+ = \frac{1}{\sqrt{2}}\left( X^1 _{n,m} + iX^{2} _{n,m}\right) ~~,~~~~  	X_{n,m} ^- = \frac{1}{\sqrt{2}}\left( X^1 _{n,m} - iX^{2} _{n,m}\right)~.
\end{equation}
The inverse transformation is given by:
\begin{equation}\label{InverseTransformation}
	X^1 _{n,m} = \frac{1}{\sqrt{2}}\left( X^+ _{n,m}+X^- _{n,m}\right)~~,~~~~ X^2 _{n,m} = \frac{1}{\sqrt{2}i}\left( X^+ _{n,m}-X^- _{n,m}\right)~.
\end{equation}
The sum of the actions above, namely eq.~(\ref{TotalAction}), now reads
\begin{align}
I_{tot}&=\frac{1}{4\pi \mu^2}\sum_{n,m} \left[ \left(\frac{|n\tau-m|^2}{\tau_2 ^2}+\omega_{12}^2\right)\left( X^+ _{n,m} X^- _{-n,-m}+X^- _{n,m}X^+ _{-n,-m} \right) \right.
\nonumber\\
&\hskip 3cm
 \left. -2\omega_{12}\frac{m-n\tau_1}{\tau_2}\left(X^+ _{n,m} X^- _{-n,-m} -X^- _{n,m} X^+ _{-n,-m}\right)\right]~.
\end{align}
Next, the partition function of the theory with field content $X_1,X_2$ can be computed. First one can treat the variables $X^1 _{n,m},X^2 _{n,m}, X^1 _{-n,-m},X^2 _{-n,-m}$ as independent, and then take a square root of the result because they are actually dependent due to the reality conditions $X^{\pm} _{n,m} = X^{\mp ~*} _{-n,-m}$. 
\begin{align}\label{PartitionFunctionPIAngularMomentum}
	Z = {\prod_{n,m}\,}' \int dX^+ _{n,m} dX^- _{n,m} e^{-I_{tot}[X^+ _{n,m},X^- _{n,m}]}={\prod_{n,m}\,}' \frac{4\pi^2 \mu^2}{ { \big|\frac{n\tau-m}{\tau_2}-\omega_{12}\big| \, \big|\frac{n\tau-m}{\tau_2}+\omega_{12}\big|^{\vphantom{|}} }}~.
\end{align}

Some comments about eq.~(\ref{PartitionFunctionPIAngularMomentum}): Above we assumed that $\omega_{12}\in \mathbb{R}$ in assembling the quadratic form to an absolute square. 
Second, note that for real nonzero $\omega_{12}$, zero modes exist if there is $m\in \mathbb{Z}$ such that
$	n=0 ~,~ \omega_{12} = \frac{m}{\tau_2},$, and then the prime notation appearing in eq.~(\ref{PartitionFunctionPIAngularMomentum}) indicates the exclusion of such modes from the product, otherwise there are no zero modes. Proceeding with the calculation, the inverse square root of the determinant in eq.~(\ref{PartitionFunctionPIAngularMomentum}) can be expressed as an exponential of a trace: 
\begin{equation}
	 Z= \exp\left[{\sum_{n,m}\,}'\log\Bigg( \frac{4\pi^2 \mu^2}{{ \big|\frac{n\tau -m}{\tau_2}-\omega_{12}\big| \, \big|\frac{n\tau-m}{\tau_2}+\omega_{12}\big|^{\vphantom{|}} }} \Bigg)\right]~.
\end{equation}
Below, the log of the product above is expressed as a sum of two logs and the following identity is used 
\begin{equation}
	\log(x) = \frac{\partial}{\partial p} x^p |_{p=0}
\end{equation}
implies that
\begin{equation}\label{NaivePartitionFunction}
	Z = \exp\left[\frac{1}{2}\frac{\partial}{\partial p} \sum_{n,m}~ 
	\frac{(4\pi^2 \mu^2)^p}{\left[\big|\frac{n\tau -m}{\tau_2}+\omega_{12}\big|^2\right] ^{p} }~\Bigg|_{p=0}+(\omega_{12}\to -\omega_{12})\right]~.
\end{equation}
We now use the second version of the classical Kronecker limit formula to calculate (\ref{NaivePartitionFunction}). Reference \rcite{ModularFunctions} contains useful formulae for the following function:
\begin{equation}
	G(s) \equiv \sum_{n,m} \frac{1}{|m \mu + n \nu + u\mu + v \nu|^{2s}}~,
\end{equation}
with $\mu,\nu$ being complex variables, while $u,v\in\mathbb{R}$ and not both integer. The classical Kronecker limit formula states that
\begin{equation}
	G'(0) = -2 \log\left|e^{\pi i u\frac{u\mu + v \nu}{\nu}}\frac{\theta_1 (\frac{u\mu + v\nu}{\nu}|\frac{\mu}{\nu})}{\eta(\frac{\mu}{\nu})}\right|~.
\end{equation}
Set $\mu = \tau, \nu=1, u=0, v= \omega_{12} \tau_2 $ so that 
\begin{equation}
	G'(0) = 2  \log\left|\frac{\eta(\tau)}{\theta_1 (\omega_{12}\tau_2|\tau)}\right| ~;
\end{equation}
also, $G(0)=0$. Both of these formulas will be proven in Appendix \ref{Appendix:DetailsRotatingGreenI}. It follows that 
\begin{equation}
\label{PartitionFunctionRotatingStringPI}
Z =e^{\log\left(\frac{|\eta(\tau)|}{ \left|\theta_1 \left( \tau_2\omega_{12}|\tau\right)\right|}\right)+(\omega_{12}\to -\omega_{12})} =\left| \frac{\eta(\tau)}{\theta_1 \left(\omega_{12}\tau_2|\tau\right)} \right| ^2~.
\end{equation}	

\subsection{One-point function of an exponential}
\label{subsec:OnePointFunctionOfAnExponential}

For the purpose of computing expectation values of harmonic functions in the presence of rotation, let us insert an exponential in the path integral as in the analysis of section~\ref{subsec:HarmFnAves} (specifically, eq.~\eqref{harmfn2exptl}), located at $(\sigma_1,\sigma_2)$.  The case under consideration is where an angular potential is turned on in a two-dimensional plane in the target space; the general formula for two independent angular potentials in two planes is deduced by taking a suitable product.
\begin{equation}\label{exponential1PointFunction}
	\left\langle e^{i\vec{k}\cdot \vec{X}}\right\rangle = \int DX^1 DX^2 e^{i(k_1 X^1 + k_2 X^2)} e^{-\frac{1}{4\pi \mu^2}\int d^2 \sigma \sum_{i=1,2}\left( \partial_a X^i \partial^a X^i  +\omega_{12}^2 (X^i)^2 \right)-\frac{i}{2\pi \mu^2}\int d\sigma_2 \omega_{12} L_{12}}~.
\end{equation}
Plugging the expansion (\ref{XiSum})
into (\ref{exponential1PointFunction}) and using eq.~(\ref{InverseTransformation}) yield
\begin{align}
	\left \langle e^{i \vec{k}\cdot \vec{X} (\sigma_1,\sigma_2)}\right \rangle &= \prod_{n,m} \int dX^1 _{n,m} \int dX^2 _{n,m} e^{i\sum_{n,m}(k_1 X^1 _{n,m} + k_2 X^2 _{n,m} ) \Phi_{n,m} (\sigma_1,\sigma_2)}\times
 \\
	&\hskip 1cm \times e^{-\frac{1}{4\pi \mu^2} \sum_{n,m} \left[ \left(\left|\frac{n\tau-m}{\tau_2}\right|^2+\omega_{12}^2\right) \left(X^1 _{n,m}X^1 _{-n,-m}+X^2 _{n,m}X^2 _{-n,-m} \right)-4i \omega_{12}\frac{m-n\tau_1}{\tau_2} X^1 _{n,m} X^2 _{-n,-m}\right]}~.
\nonumber 
\end{align}
Defining
\begin{equation}
	k_{12} ^* \equiv \frac{k_1 - ik_2}{\sqrt{2}} ~,~ k_{12} \equiv \frac{k_1 + ik_2}{\sqrt{2}}~,
\end{equation}
one obtains
\begin{align}\label{OnePointFunctionExponential0}
	\left \langle e^{i \vec{k}\cdot \vec{X} (\sigma_1,\sigma_2)}\right \rangle &= \prod_{n,m} \int dX^+ _{n,m} \int dX^- _{n,m} e^{i\sum_{n,m}(k_{12}^* X^+ _{n,m} + k_{12} X^- _{n,m} ) \Phi_{n,m} (\sigma_1,\sigma_2)}\times\nonumber\\
		&\hskip 1cm\times  e^{-\frac{1}{4\pi \mu^2} \sum_{n,m}  \left(\left|\frac{n\tau-m}{\tau_2}-\omega_{12}\right|^2 X^+ _{n,m}X^- _{-n,-m}+\left|\frac{n\tau-m}{\tau_2}+\omega_{12}\right|^2 X^- _{n,m}X^+ _{-n,-m} \right)}~.
\end{align}
In these integrals, $X^+ _{n,m} = (X^- _{-n-,m})^*$ and $X^- _{n,m} = (X^+)^*_{-n,-m}$ are complex variables. We perform a change of variables in which the phase of $k_{12} ^* \Phi_{n,m}$ is absorbed in $X^+ _{n,m}$ and the phase of $k_{12} \Phi_{n,m}$ is absorbed in $X^- _{n,m}$. The last line of eq.~(\ref{OnePointFunctionExponential0}) and the measure remain intact. Then
\begin{align}
	\left \langle e^{i \vec{k}\cdot \vec{X} (\sigma_1,\sigma_2)}\right \rangle &= \prod_{n,m} \int d\tilde{X}^+ _{n,m} \int d\tilde{X}^- _{n,m} e^{i\sum_{n,m}\big(|k_{12}| \tilde{X}^+ _{n,m} + |k_{12}| \tilde{X}^- _{n,m} \big) |\Phi_{n,m} (\sigma_1,\sigma_2)|}\nonumber\\
	&\hskip 2cm\times  e^{-\frac{1}{4\pi \mu^2} \sum_{n,m}  \left(\left|\frac{n\tau-m}{\tau_2}-\omega_{12}\right|^2 \tilde{X}^+ _{n,m}\tilde{X}^- _{-n,-m}+\left|\frac{n\tau-m}{\tau_2}+\omega_{12}\right|^2 \tilde{X}^- _{n,m}\tilde{X}^+ _{-n,-m} \right)}
 \nn\\[.3cm]
\label{OnePointFunctionExponentialOmega} 
&=  e^{-\frac{1}{2}(k_1^2 + k_2 ^2)\sum_{n,m} \left(\frac{\pi \mu^2  \Phi_{n,m} ^*(\sigma_1,\sigma_2) \Phi_{n,m} (\sigma_1,\sigma_2)}{\left|\frac{n\tau-m}{\tau_2}+\omega_{12} \right|^2}+ \frac{\pi \mu^2  \Phi_{n,m}^*(\sigma_1,\sigma_2) \Phi_{n,m} (\sigma_1,\sigma_2)}{\left|\frac{n\tau-m}{\tau_2}-\omega_{12} \right|^2}\right)}~.
\end{align} 
The fact that $|\Phi_{n,m}|^2 = \frac{1}{(2\pi)^2 \tau_2}$ implies
\begin{equation}
		\left \langle e^{i\vec{k}\cdot \vec{X}} \right \rangle =  e^{-\frac{1}{2} (k_1^2 + k_2 ^2) G^r (0,\tau,\omega_{12})}~,
\end{equation}
where
\begin{equation}
\label{GreensFunction}
G^r (0,\tau,\omega_{12})=	\frac{\mu^2}{4\pi}\sum_{n,m} \left(\frac{\tau_2}{\left|n\tau-m +\tau_2\omega_{12} \right|^2}+ \frac{\tau_2}{\left|n\tau-m - \tau_2\omega_{12} \right|^2}\right)~.
\end{equation}
This function is elliptic: It is periodic under $\tau_2\omega_{12} \to \tau_2\omega_{12}+1~,~ \tau_2\omega_{12}\to \tau_2\omega_{12}+\tau $.  It is also modular invariant: It is unchanged when $\tau\to \tau+1,\omega_{12}\to \omega_{12}$; and when $\tau \to -\frac{1}{\tau} ~,~ \omega_{12} \tau_2 \to \frac{\omega_{12}\tau_2}{\tau}$.

In appendix \ref{Appendix:DetailsRotatingGreenI} we evaluate 
\begin{equation}\label{GswDefinition}
	G(s,w,\tau) = \sum_{n,m} \frac{1}{|m\tau - n + w|^{2s}}~,
\end{equation}
in the vicinity of $s=1$, with $w\in \mathbb{R}$. Looking back at eq.~(\ref{GreensFunction}), we are interested in:
\begin{align}\label{Propagator}
	G^{r}(0,q,\omega)&=
	\frac{ \mu^2 \tau_2}{4\pi} \left[G (s=1,w=\omega_{12} \tau_2,\tau)+G (s=1,w=-\omega_{12} \tau_2,\tau)\right]~.
\end{align}
In that appendix we elaborate on a calculation that yields the following holomorphic part of the Green's function at coincident points 
\begin{equation}
\label{Gr02}
	G^{r}(0,q,\omega) = \frac{1}{2}\mu^2 \sum_{m=1} ^{\infty} \left( \frac{q^m e^{2\pi  i \omega}}{1-q^m e^{2\pi i \omega}}+\frac{q^m e^{-2\pi  i \omega}}{1-q^m e^{-2\pi i \omega}}\right) \equiv \mu^2\zeta(q,\omega)~.
\end{equation}
One can interpret this as a measure of the spatial extent (squared) of the rotating bound state in the rotation plane.

The one-point function of an exponential reads
\begin{equation}
\label{RotatingExptl}
	\left\langle e^{i \vec{k}\cdot\vec{X} (\sigma_1,\sigma_2)} \right\rangle = e^{-\frac{1}{2}|\vec{k}|^2 G^r (0,q,\omega)}=e^{-\frac{1}{2}|\vec{k}|^2 \mu^2 \zeta(q,\omega)}~.
\end{equation}
Let us take a small chemical potential scaling limit  
\begin{equation}
\label{ScalingLimit}
\omega\to0^+ 
~~,~~~~
\tau_2 \to 0^+
~~,~~~~
\gamma = \frac{2\pi i\omega}{\log(q) } = \text{fixed} ~. 
\end{equation}
For $\tau_1=0$, the scaling limit involves fixing~$\frac{\omega}{\tau_2}$. Then $e^{2\pi i\omega} = q^{\gamma}$ and
\begin{equation}
	\zeta\left(q,\frac{\gamma \log(q)}{2\pi}\right) = \frac{1}{2}\sum_{m=1} ^{\infty} \frac{1}{m}\left( \frac{q^{m+\gamma}}{1-q^{m+\gamma}}+\frac{q^{m-\gamma}}{1-q^{m-\gamma}}\right)~.
\end{equation}
The limit in question is performed in
\begin{equation}\label{HighTlimit}
	\zeta\left(q,\frac{\gamma \log(q)}{2\pi}\right) \to \frac{1}{2} \sum_{m=1} ^{\infty} \frac{1}{m(1-q)}\left( \frac{1}{m+\gamma}+\frac{1}{m-\gamma}\right) = \frac{1}{1-q} \sum_{m=1}^{\infty} \frac{1}{m^2 - \gamma^2}~.
\end{equation}
A well-defined limit requires $\gamma \notin \mathbb{Z}/\{0\}$. To evaluate the series, it is useful to recall the product representation of the $\sin(x)/x$ function:
\begin{equation}\label{sincProduct}
	\frac{\sin(\pi \gamma)}{\pi \gamma} = \prod_{m=1}^{\infty} \left( 1- \frac{\gamma^2}{m^2}\right)~.
\end{equation}
Taking the logarithm of eq.~(\ref{sincProduct}) and the derivative of it with respect to $\gamma$ yield
\begin{equation}\label{SumEquation}
	\pi \cot(\pi \gamma) - \frac{1}{\gamma} = -2\sum_{m=1} ^{\infty}\frac{\gamma}{m^2 - \gamma^2}~.
\end{equation}
Consequently, eq.~(\ref{HighTlimit}) becomes
\begin{equation}\label{HighTlimit2}
	\zeta\left(q,\frac{\gamma \log(q)}{2\pi}\right) \to \frac{1}{1-q} \frac{1-\pi \gamma \cot(\pi \gamma)}{2\gamma^2}~.
\end{equation}
For $\gamma\approx 0$, the leading term in the Taylor series is $\frac{1}{1-q} \frac{\pi^2}{6}$, which one can also obtain from the preceding calculation at $\gamma=0$. In the vicinity of integers, the function $\frac{1-\pi \gamma \cot(\pi \gamma)}{2\gamma^2}$ goes to $\pm \infty$. In addition, this is an even and surjective function that admits infinitely many zeros.  We consider the range $0\leq \gamma <1$ where the function increases monotonically.

\subsection{Averaged harmonic function \texorpdfstring{$H_5 (\vec{x})$}{TEXT}}
\label{subsec:AveragedHarmonicFunctionH5}

The above result~\eqref{RotatingExptl} allows us to calculate the averaged harmonic function $\langle H_5 (\vec{x}) \rangle $ associated with the transverse metric. This subsection is devoted for the calculation of this quantity.
\begin{align}
	\bigg \langle \frac{1}{|\vec{x} - \vec{f}|^2} \bigg \rangle = \int_0 ^{\infty} ds~ \left \langle e^{-s|\vec{x}-\vec{f}|^2}\right \rangle &=
	\int_0 ^{\infty} ds \int \frac{d^4 b}{(\pi s)^2} e^{-s|\vec{x}|^2 - \frac{|\vec{b}|^2}{s}}\left \langle e^{i \vec{k}\cdot \vec{f}}\right \rangle~,
\end{align}
with
\begin{equation}
	\vec{k} = -2i (s\vec{x} + i\vec{b})~.
\end{equation}
Eq.~(\ref{RotatingExptl}) implies
\begin{equation}
\bigg \langle \frac{1}{|\vec{x} - \vec{f}|^2} \bigg \rangle =	\int_0 ^{\infty} ds~ e^{-s |\vec{x}|^2}\int \frac{d^4 b}{(\pi s)^2} e^{-\frac{|\vec{b}|^2}{s}}e^{2 \mu^2 \left[(s \vec{x}+i\vec{b} )_{12} ^2 \zeta(q,\omega_{12})+(s \vec{x}+i\vec{b} )_{34} ^2 \zeta(q,\omega_{34}) \right]}~.
\end{equation}
The $b$-integrals yield
\begin{align}
	\bigg \langle \frac{1}{|\vec{x} - \vec{f}|^2} \bigg \rangle &=	\int_0 ^{\infty} ds~ \frac{1}{1+2 s\mu^2 \zeta(q,\omega_{12})}\frac{1}{1+2 s\mu^2 \zeta(q,\omega_{34})}e^{-s |\vec{x}|^2} 
 \\
&\hskip 1cm
 \times e^{2\mu^2 s^2 \left[(x_1 ^2 + x_2 ^2)\zeta(q,\omega_{12})+(x_3 ^2 + x_4 ^2)\zeta(q,\omega_{34})\right]} e^{-4\mu^4 s^3 \left(\frac{(x_1 ^2+ x_2 ^2)\zeta(q,\omega_{12})^2}{1+2\mu^2 s \zeta(q,\omega_{12})}+ \frac{(x_3 ^2+ x_4 ^2)\zeta(q,\omega_{34})^2}{1+2\mu^2 s \zeta(q,\omega_{34})}\right)}~.
\nn
\end{align}
Combining the arguments in the exponentials through common denominators brings about
\begin{align}\label{RotatingGreen}
	\bigg \langle \frac{1}{|\vec{x} - \vec{f}|^2} \bigg \rangle &=	\int_0 ^{\infty} ds~ \frac{1}{\left(1+2 s\mu^2 \zeta(q,\omega_{12})\right)\left(1+2 s\mu^2 \zeta(q,\omega_{34})\right)}  e^{-s\left[\frac{x_1 ^2 + x_2^2}{1+2\mu^2 s \zeta(q,\omega_{12})}+\frac{x_3^2 +x_4^2}{1+2\mu^2 s \zeta(q,\omega_{34})} \right]}~.
\end{align}
This function admits a maximum at the center because the eigenvalues of the Hessian matrix of this function are all negative. For instance,
\begin{align}\label{RotatingGreenHessian}
	\frac{\partial ^2}{\partial x_1 ^2}\bigg \langle \frac{1}{|\vec{x} - \vec{f}|^2} \bigg \rangle \bigg|_{\vec{x}=0} &=	-2\int_0 ^{\infty} ds~ \frac{s^2}{\left(1+2 s\mu^2 \zeta(q,\omega_{12})\right)^3\left(1+2 s\mu^2 \zeta(q,\omega_{34})\right)} <0 ~.
\end{align}
Note that we consider values of the angular potential such that $\zeta(q,\omega)\geq 0$. Negative values of $\zeta(q,\omega)$ can be arranged for, however, in this case one is beset by an essential singularity at $s=-\frac{1}{2\mu^2 \zeta(q,\omega)}$ making the integral divergent and so we deem these cases unphysical.  
Additionally, the partial derivatives with respect to $x_i$ for all $i=1,2,3,4$ are negative and thus the function decreases in all of the direction away from the center. Namely, the function in question attains a global maximum there. 
One can rewrite eq.~(\ref{RotatingGreen}) as
\begin{align}
	\bigg \langle \frac{1}{|\vec{x}-\vec{f}|^2}\bigg\rangle &= e^{-\frac{x_1^2 + x_2^2}{2\mu^2 \zeta(q,\omega_{12})}} e^{-\frac{x_3^2 + x_4^2}{2\mu^2 \zeta(q,\omega_{34})}}\int_0 ^{\infty} ds \frac{1}{1+2s\mu^2 \zeta(q,\omega_{12}) } \nonumber\\
	&\hskip 1cm\times \frac{1}{1+2s\mu^2 \zeta(q,\omega_{34}) } e^{\frac{x_1 ^2 + x_2 ^2}{2\mu^2 \zeta(q,\omega_{12})}\frac{1}{1+2\mu^2 s \zeta(q,\omega_{12})}}e^{\frac{x_3 ^2 + x_4 ^2}{2\mu^2 \zeta(q,\omega_{34})}\frac{1}{1+2\mu^2 s \zeta(q,\omega_{34})}}~.
\end{align}
The result is simple when the two chemical potentials are equal: $\omega_{12}=\omega_{34}=\omega$. In this case, one performs the change of variables:
\begin{equation}
	y\equiv \frac{1}{1+2\mu^2 \zeta(q,\omega)s}~,
\end{equation}
which implies
\begin{equation}\label{AveragedGreen'sFunctionRotating}
	\bigg \langle \frac{1}{|\vec{x}-\vec{f}|^2}\bigg\rangle =\frac{1-e^{-\frac{|\vec{x}|^2}{2\mu^2 \zeta(q,\omega)}}}{|\vec{x}|^2}~.
\end{equation}
This result displays a spherical structure for any angular potential, in contrast to other solutions sourced by rotating NS5-F1 bound states that have a ring shape \cite{Lunin:2001fv}. Those solutions have a fixed angular momentum which repels matter from the center, whereas in our case the angular potential is fixed, and as the next section shows, the ensemble of fixed angular potential admits large fluctuations of the angular momentum.
The size of this averaged bound state is roughly determined by the place where the exponential function decreases by a factor of $e$ from $1$. This is given by
\begin{equation}
	r =\mu \sqrt{2\zeta(q,\omega)}~,
\end{equation}
and is somewhat larger than the size of the non-rotating solution (the properties of the solution will be discussed further in section~\ref{subs:Properties}).

For $\omega_{12}\neq \omega_{34}$, one needs to evaluate
\begin{equation}
	e^{-\mathcal{A}-\mathcal{B}}\int_0 ^{\infty} dx \frac{e^{\frac{\mathcal{A}}{1+ax}+\frac{\mathcal{B}}{1+bx}}}{(1+ax)(1+bx)}~.
\end{equation}
If either $a=0,b\neq0$ or $a\neq0 ,b=0$ then the integral diverges logarithmically at large $x$. If $a<0$ or $b<0$ then the integrand suffers from an essential singularity. It is assumed below that $a\neq b,~ a,b>0$.
The change of variables $z\equiv \frac{1-\frac{b}{a}}{1+ax}+\frac{b}{a}$ leads to 
\begin{equation}\label{Original-Integral}
	e^{-\mathcal{A}-\mathcal{B}}\int_0 ^{\infty} dx \frac{e^{\frac{\mathcal{A}}{1+ax}+\frac{\mathcal{B}}{1+bx}}}{(1+ax)(1+bx)}= \frac{e^{\frac{b\mathcal{B}-\mathcal{A}a}{a-b}}}{a-b}\int_{\frac{b}{a}} ^1 \frac{1}{z} e^{\frac{a\mathcal{A}z}{a-b} - \frac{\mathcal{B}b}{(a-b)z}}dz~.
\end{equation}
Then the relevant integral is
\begin{equation}
	\bigg \langle \frac{1}{|\vec{x}-\vec{f}|^2}\bigg\rangle=	\frac{1}{2\mu^2 (\zeta_{12}-\zeta_{34})}\int_{\frac{\zeta_{34}}{\zeta_{12}}} ^1 \frac{1}{z}e^{\frac{|\vec{x}_{12}|^2}{2\mu^2 (\zeta_{12}-\zeta_{34})} (z-1)+\frac{|\vec{x}_{34}|^2}{2\mu^2 (\zeta_{12} - \zeta_{34})}\left(1-\frac{1}{z}\right)}dz~,
\end{equation}
where
\begin{equation}
    \zeta_{ij} = \zeta(q,\omega_{ij}) ~,~ ij=\{12,34\}. 
\end{equation}
We can evaluate the integral analytically for  $|\vec{x}_{34}|=0$ and any $|\vec{x}_{12}|\geq 0$. 
\begin{align}\label{AveragedGreensx34=0}
	&\frac{1}{2\mu^2 (\zeta_{12}-\zeta_{34})}\int_{\frac{\zeta_{34}}{\zeta_{12}}} ^1 \frac{1}{z}e^{\frac{|\vec{x}_{12}|^2}{2\mu^2 (\zeta_{12}-\zeta_{34})} (z-1)}dz =\left(w\equiv \frac{|\vec{x}_{12}|^2}{2\mu^2 (\zeta_{12}-\zeta_{34})} z\right)
\\
&\hskip1cm
 =\frac{1}{2\mu^2 (\zeta_{12}-\zeta_{34})}\,e^{-\frac{|\vec{x}_{12}|^2}{2\mu^2 (\zeta_{12} - \zeta_{34})}}\left[Ei\left( \frac{|\vec{x}_{12}|^2}{2\mu^2 (\zeta_{12}-\zeta_{34})}\right)-Ei\left(\frac{|\vec{x}_{12}|^2}{2\mu^2 (\zeta_{12}-\zeta_{34})}\frac{\zeta_{34}}{\zeta_{12}}\right)\right]
\nn
\end{align}
where $Ei$ is the exponential integral function.

We now consider this expression in the regimes where the argument of the exponential integral function is either large or small. To this end, the following asymptotics are useful  \cite{AbraSteg72}
\begin{align}
		Ei (y\gg1) = \frac{e^y}{y}\left( 1+\frac{1}{y}+O\left(\frac{1}{y^2}\right)\right)~,~
	Ei (y\ll1) = \gamma^{~}_E + \log(y)+O(y)~,
\end{align}
where $\gamma^{~}_E$ is the Euler constant. 
For large arguments, which are realized for large distances:
\begin{align}
	&\frac{1}{2\mu^2 (\zeta_{12}-\zeta_{34})}\,e^{-\frac{|\vec{x}_{12}|^2}{2\mu^2 (\zeta_{12} - \zeta_{34})}}\left[Ei\left( \frac{|\vec{x}_{12}|^2}{2\mu^2 (\zeta_{12}-\zeta_{34})}\right)-Ei\left(\frac{|\vec{x}_{12}|^2}{2\mu^2 (\zeta_{12}-\zeta_{34})}\frac{\zeta_{34}}{\zeta_{12}}\right)\right]\nonumber\\[.2cm]
	&\hskip2cm
 \approx\frac{1}{|\vec{x}_{12}|^2}\left(1-\frac{\zeta_{12}}{\zeta_{34}}\,e^{-\frac{|\vec{x}_{12}|^2}{2\mu^2 \zeta_{12}}}+O\bigg(\frac{\mu^2(\zeta_{12}-\zeta_{34})}{|\vec{x_{12}}|^2}\bigg)\right)~.
\end{align}
This behavior is familiar from previous calculations where one had $\zeta_{12}=\zeta_{34}$. 	
For short distances, $|\vec{x}_{12}|\ll \mu \sqrt{\zeta_{12}}$, 
\begin{align}
	&\frac{1}{2\mu^2 (\zeta_{12}-\zeta_{34})}e^{-\frac{|\vec{x}_{12}|^2}{2\mu^2 (\zeta_{12} - \zeta_{34})}}\left[Ei\left( \frac{|\vec{x}_{12}|^2}{2\mu^2 (\zeta_{12}-\zeta_{34})}\right)-Ei\left(\frac{|\vec{x}_{12}|^2}{2\mu^2 (\zeta_{12}-\zeta_{34})}\frac{\zeta_{34}}{\zeta_{12}}\right)\right]\nonumber\\
	&\hskip2cm
 \approx\frac{1}{2\mu^2 (\zeta_{12}-\zeta_{34})}\log\left( \frac{\zeta_{12}}{\zeta_{34}}\right)+O\left(\frac{|\vec{x}_{12}|^2}{\mu^2\zeta_{12}}\right)~.
\end{align}
Taking further the limit $\zeta_{34}\to \zeta_{12}$, one obtains $\frac{1}{2\mu^2 \zeta_{12}}$ which again agrees with the results above. When $\zeta_{12}\gg \zeta_{34}$, we see that a logarithmic enhancement occurs with respect to the case of equal angular potentials. This suggests that the length scale ``seen'' by the above averaged harmonic function is approximately
\be
 \frac{\mu \sqrt{\zeta_{12}}}{\sqrt{\log\big( \frac{\zeta_{12}}{\zeta_{34}}\big)}} ~.
\ee

The result for the harmonic function does not exhibit a divergence anywhere, which in particular implies that there is neither an event horizon nor a singularity in transverse space.  


\subsection*{Small chemical potential limit}
\label{subsec:SmallChemicalPotentialLimit}

Previously we showed in eq.~(\ref{HighTlimit2}) that as $q\to 1$, $\omega\to 0$ with a fixed $\gamma = \frac{2\pi i\omega}{\log(q)}$,
\begin{equation}
	\zeta_{12}\left(q,\frac{\gamma \log(q)}{2\pi}\right) \to \frac{1}{1-q} \frac{1-\pi \gamma \cot(\pi \gamma)}{2\gamma^2}\to \frac{1-\pi \gamma \cot(\pi \gamma)}{4\pi \tau_2\gamma^2}~.
\end{equation}
We consider zero chemical potential for the 3-4 plane. Using the small $x$ approximation $\cot(x)\approx \frac{1}{x}-\frac{x}{3}$,
\begin{equation}
	\zeta_{34} (q,\omega_{34}=0) = \frac{\pi }{12\tau_2 }~. 
\end{equation}
Consequently,
\begin{align}
	&	\left\langle \frac{1}{|\vec{x}_{12}-\vec{f}|^2}\right\rangle\to \frac{1}{2\mu^2 (\zeta_{12}-\zeta_{34})}e^{-\frac{|\vec{x}_{12}|^2}{2\mu^2 (\zeta_{12} - \zeta_{34})}}\left[Ei\left( \frac{|\vec{x}_{12}|^2}{2\mu^2 (\zeta_{12}-\zeta_{34})}\right)-Ei\left(\frac{|\vec{x}_{12}|^2}{2\mu^2 (\zeta_{12}-\zeta_{34})}\frac{\zeta_{34}}{\zeta_{12}}\right)\right]
 \nonumber\\[.3cm]
	& =\frac{6\tau_2}{\pi \mu^2 \left(\frac{3}{\pi^2 \gamma^2}\left(1-\pi \gamma \cot(\pi \gamma)\right)-1\right)}
\exp\bigg[{-\frac{6\tau_2 |\vec{x}_{12}|^2}{\pi\mu^2 \left(\frac{3}{\pi^2 \gamma^2}\left(1-\pi \gamma \cot(\pi \gamma)\right)-1\right) }}\bigg]
\nn\\
 &\hskip2cm
\times\left[Ei\Bigg( \frac{6\tau_2| \vec{x}_{12}|^2}{\pi\mu^2 \left(\frac{3}{\pi^2 \gamma^2}\left(1-\pi \gamma \cot(\pi \gamma)\right)-1\right)}\Bigg)\right. \nonumber\\
&\hskip4cm 
 \left.-\;Ei\Bigg(\frac{6\tau_2 |\vec{x}_{12}|^2}{\pi\mu^2 \left(\frac{3}{\pi^2 \gamma^2}\left(1-\pi \gamma \cot(\pi \gamma)\right)-1\right)}\frac{\pi^2\gamma^2}{3\left(1-\pi \gamma \cot(\pi \gamma)\right)}\Bigg)\right]~.\nonumber
\end{align}
We have thus found an analytical expression for the average harmonic function in the ensemble of fixed chemical potential and angular potentials in the scaling limit (\ref{ScalingLimit}) at the plane $x_3=x_4=0$. It displays a maximum at a central point and its support in transverse space increases as the angular potential parameter $\gamma$ increases. At the same time, the value of the maximum decreases, meaning that the bound state spreads when it rotates faster, consistent with a general expectation that highly-spinning objects are less localized than low-spinning ones. We will explain in the next section that the maximal value at the central point can be understood as a consequence of zero angular momentum states that participate in the ensemble.
\begin{figure}[hbt!]
\centering
\includegraphics[scale=0.3]{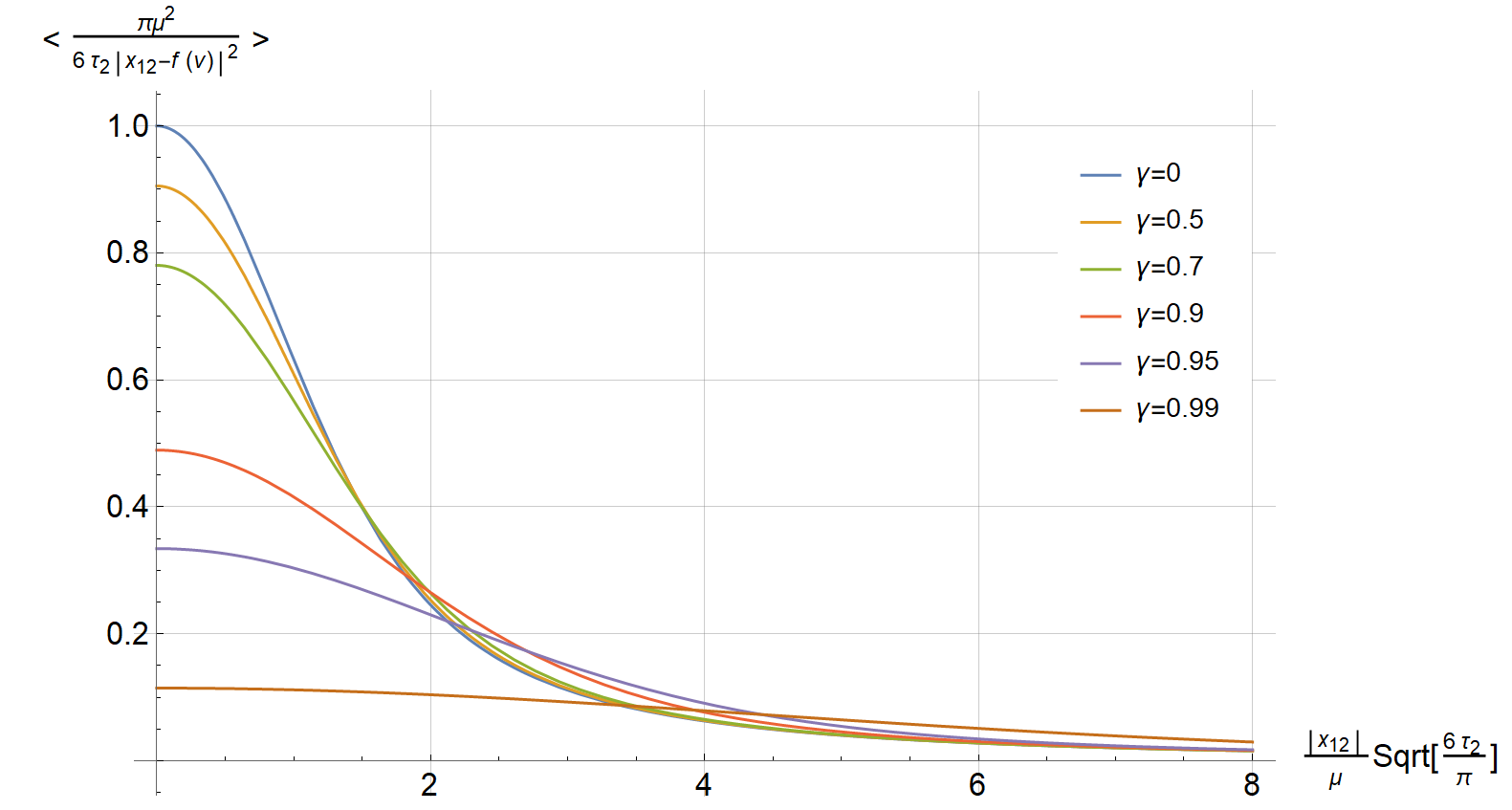}
	\caption{Averaged harmonic function for real values of $\gamma$ in the $x_1-x_2$ plane. Increasing the chemical potential for the angular momentum makes the maximal value of the averaged harmonic function decreasing and its support on the plane increasing.}
\end{figure}


\subsection{Averaged harmonic form \texorpdfstring{$\bf{A} (\vec{x})$}{TEXT}}
\label{subsec:AveHarmFormA}
The goal of this subsection is to derive the average harmonic form $A$ appearing in the metric~\eqref{LMgeom}, in the ensemble of fixed chemical potential for the string charge and fixed angular potentials. This expectation value is first related to $\big\langle \partial_v f_1 e^{i\vec{k}\cdot \vec{f}}\big\rangle$ and then we use an expression for the two-point function of exponential operators $e^{i\vec{k}_1 \cdot \vec{f}} e^{i \vec{k}_2 \cdot \vec{f}}$ to extract the result for $\big\langle \partial_v f_1 e^{i\vec{k}\cdot \vec{f}}\big\rangle$. We then evaluate the integral expressions for $A$, with the final result given in eq.~\eqref{Aresult}.

Let us now calculate the $x_1$ component of a quantity related to the $x_1$ component of the one-form $\A$, but without the $v$-integral
\begin{equation}\label{A1vev}
	\langle A_1 \rangle = \int_0 ^{\infty} ds \int \frac{d^4b}{(\pi s)^2} e^{-s|\vec{x}|^2 -\frac{|\vec{b}|^2}{s}}\left\langle \partial_v f_1 e^{i\vec{k}\cdot \vec{f}}\right\rangle~,
\end{equation} 
with
\begin{equation}
	\vec{k} =-2i(s\vec{x}+i\vec{b})~.
\end{equation}
This requires calculating $\left\langle \partial_v f_1 e^{i\vec{k}\cdot \vec{f}}\right\rangle$ first.
If one knows the two-point function of tachyon vertex operators, then it possible to extract this expectation value:
\begin{equation}\label{PartialfExpf}
	\left\langle \partial_v f_x (v_1) e^{i\vec{k}_2\cdot \vec{f}(v_2)} \right\rangle = -i \lim_{v_1\to v_2} \frac{\partial}{\partial v_1}\left( \frac{\partial}{\partial k_{1x}} \left\langle e^{i\vec{k}_1 \cdot \vec{f}(v_1)} e^{i\vec{k}_2 \cdot \vec{f}(v_2)} \right\rangle \bigg|_{k_{1x}=0} \right)~. 
\end{equation}
The purpose of Appendix \ref{app:rot details} is to show that the part of the two-point function of exponentials in the ensemble of fixed chemical potential and angular potential, coming purely from holomorphic oscillators, is given by
 \begin{align}\label{Rotating2PointFunction}
 	&\left\langle e^{ik_1 \cdot X(z_1) + ik_2 \cdot X(z_2)}\right\rangle \propto e^{-\frac{1}{2}(|\vec{k}_1|^2+|\vec{k}_2|^2)G^r(0,\omega,\tau)-\vec{k}_1 \cdot \vec{k}_2 G(z_1,z_2,\omega,\tau)-i\vec{k}_1 \wedge \vec{k}_2 \mathcal{G}(z_1,z_2,\omega,\tau)}~,
 \end{align}
where
\begin{equation}
	\vec{k}_1 \wedge \vec{k}_2 \equiv k_{1x}k_{2y} - k_{1y} k_{2x}~.
\end{equation}
When $\vec{k}_1 || \vec{k}_2$ this vanishes, in particular for $\vec{k}_1 = \pm\vec{k}_2$. Also:
\begin{align}
	G^r (0,\omega,\tau)&\equiv \frac{\mu^2}{2}\sum_{n=1} ^{\infty} \frac{1}{n} \left( \frac{q^n e^{2\pi i\omega }}{1-q^n e^{2\pi i \omega}}+\frac{q^n e^{-2\pi i\omega }}{1-q^n e^{-2\pi i \omega}}\right)
\\[.2cm]
\label{straightG}
	G(z_1,z_2,\omega,\tau) &= \frac{\mu^2}{4}\sum_{n=1} ^{\infty} \frac{1}{n} \left( \frac{q^n e^{2\pi i\omega }}{1-q^n e^{2\pi i \omega}}+\frac{q^n e^{-2\pi i\omega }}{1-q^n e^{-2\pi i \omega}}\right)\left( \left( \frac{z_1}{z_2}\right)^n + \left(\frac{z_2}{z_1}\right)^n\right)~.
\end{align}
This equation reduces to the previous one for equal points on the torus $z_1=z_2$. 
\begin{equation}
\label{curlyG}
	\mathcal{G}(z_1,z_2,\omega,\tau)=\frac{\mu^2}{4}\sum_{n=1} ^{\infty} \frac{1}{n}\left( \frac{q^n e^{2\pi i \omega}}{1-q^n e^{2\pi i \omega}}-\frac{q^n e^{-2\pi i \omega}}{1-q^n e^{-2\pi i \omega}}\right)\left[ \left( \frac{z_2}{z_1}\right)^n - \left( \frac{z_1}{z_2}\right)^n\right]~.
\end{equation}
Substituting eq.~(\ref{Rotating2PointFunction}) into the eq.~(\ref{PartialfExpf}) and using eqs.~(\ref{straightG})-(\ref{curlyG}), one obtains
\begin{align}\label{IntermediateA}
	\left\langle \partial_v f_x (v_2) e^{i\vec{k}_2\cdot \vec{f}(v_2)} \right\rangle &= - \lim_{v_1\to v_2} \left( -k_{2x} \partial_{v_1} G(v_1,v_2,\omega,\tau) - i k_{2y} \partial_{v_1} \mathcal{G}(v_1,v_2,\omega,\tau) \right) e^{-\frac{1}{2}G^r (0,\omega,\tau) |k_2|^2}
 \nonumber\\[.2cm] 
&	=-k_{2y} \frac{\mu^2}{2} \sum_{n=1} ^{\infty} \left( \frac{q^n e^{2\pi i \omega}}{1-q^n e^{2\pi i \omega}} - \frac{q^n e^{-2\pi i \omega}}{1-q^n e^{-2\pi i \omega}}\right)e^{-\frac{1}{2}G^r (0,\omega,\tau) |k_2|^2}~. 
\end{align}
Similarly,
\begin{align}
	&\left\langle \partial_v f_y (v_2) e^{i\vec{k}_2\cdot \vec{f}(v_2)} \right\rangle =
	k_{2x} \frac{\mu^2}{2} \sum_{n=1} ^{\infty} \left( \frac{q^n e^{2\pi i \omega}}{1-q^n e^{2\pi i \omega}} - \frac{q^n e^{-2\pi i \omega}}{1-q^n e^{-2\pi i \omega}}\right)e^{-\frac{1}{2}G^r (0,\omega,\tau) |k_2|^2}~. 
\end{align}
As we will show in section \ref{subsec:FixedJvsOmega}, the expectation value of the angular momentum operator admits the following series representation:
\begin{align}\label{SumforA}
	\langle \hat{J}_{12}\rangle (q,\omega_{12}) &= \sum_{m=1} ^{\infty}\left[ \frac{q^m e^{2\pi i\omega_{12}}}{1-q^m e^{2\pi i \omega_{12}}}-\frac{q^m e^{-2\pi i\omega_{12}}}{1-q^m e^{-2\pi i\omega_{12}}}\right] ~.
\end{align}
Returning to $\langle A_1\rangle $ of eq.~(\ref{A1vev}) with $\vec{k}=-2i (s\vec{x}+i\vec{b})$ and also using eq.~(\ref{IntermediateA}),
\begin{align}
	\langle A_1\rangle &= -\mu^2\langle \hat{J}_{12}\rangle (q,\omega_{12})\int_0^{\infty} ds \, e^{-s |\vec{x}|^2} \int \frac{d^4 b}{(\pi s)^2}\, e^{-\frac{|\vec{b}|^2}{s}} (sx_2 + ib_2) 
 \nonumber\\
	&\hskip1cm\times e^{-\sum_{j=1,2}\left( \frac{1}{s}+ 2\mu^2 \zeta(q,\omega_{12})\right)\left[ b_j - \frac{2i \mu^2 sx_j \zeta (q,\omega_{12})}{\frac{1}{s}+2\mu^2 \zeta(q,\omega_{12})}\right]^2 - \frac{4\mu^4 s^2 (x_1^2+x_2^2) \zeta(q,\omega_{12})^2}{\frac{1}{s}+2\mu^2 \zeta(q,\omega_{12})}}
	e^{2\mu^2 \zeta(q,\omega_{12}) s^2 (x_1^2+ x_2^2)}
 \nonumber\\
	&\hskip1cm \times e^{-\sum_{j=3,4}\left( \frac{1}{s}+ 2\mu^2 \zeta(q,\omega_{34})\right)\left[ b_j - \frac{2i \mu^2 sx_j \zeta (q,\omega_{34})}{\frac{1}{s}+2\mu^2 \zeta(q,\omega_{34})}\right]^2 - \frac{4\mu^4 s^2 (x_3^2+x_4^2) \zeta(q,\omega_{34})^2}{\frac{1}{s}+2\mu^2 \zeta(q,\omega_{34})}}
	e^{2\mu^2 \zeta(q,\omega_{34}) s^2 (x_3^2+ x_4^2)}~.
\end{align}
A few manipulations lead to
 \begin{align}
\langle A_1 \rangle &= -\mu^2\langle \hat{J}_{12}\rangle (q,\omega_{12})	\,e^{-\frac{x_1^2 + x_2^2}{2\mu^2 \zeta(q,\omega_{12})}} e^{-\frac{x_3^2 + x_4^2}{2\mu^2 \zeta(q,\omega_{34})}}
\\
&\hskip.7cm
\int_0 ^{\infty} ds \frac{sx_2}{(1+2s\mu^2 \zeta(q,\omega_{12}) )^2}\frac{1}{1+2s\mu^2 \zeta(q,\omega_{34}) }   
  \,e^{\frac{x_1 ^2 + x_2 ^2}{2\mu^2 \zeta(q,\omega_{12})}\frac{1}{1+2\mu^2 s \zeta(q,\omega_{12})}}\,e^{\frac{x_3 ^2 + x_4 ^2}{2\mu^2 \zeta(q,\omega_{34})}\frac{1}{1+2\mu^2 s \zeta(q,\omega_{34})}}~.
  \nn
\end{align}
For equal chemical potentials, the subscripts $12$ and $34$ can be omitted.  Changing variables from $s$ to 
\begin{equation}
 	y = \frac{|\vec{x}|^2}{2\mu^2 \zeta(q,\omega)} \frac{1}{1+2\mu^2 \zeta(q,\omega)s} 
 \end{equation}
simplifies the integral, with the result for the components $A_i$
\begin{align}
\label{Ai vevs}
\langle A_1 \rangle = -x_2 \mathcal{A}
~~&,~~~~
\langle A_2 \rangle = +x_1 \mathcal{A}
~~,~~~~
\langle A_3 \rangle = -x_4 \mathcal{A}
~~,~~~~
\langle A_4 \rangle = +x_3 \mathcal{A}
\nn\\[.3cm]
\mathcal{A} &=\frac{\mu^2}{|\vec{x}|^2}\langle \hat{J}\rangle (q,\omega) \,\frac{1-e^{-\frac{|\vec{x}|^2}{2\mu^2 \zeta(q,\omega)}}\left(1+\frac{|\vec{x}|^2}{2\mu^2 \zeta(q,\omega)}\right)}{|\vec{x}|^2}~.
\end{align}
When integrating along a variable $\tilde{v}= \frac{2\pi v}{L}$ and multiplying by $\frac{n_5}{L}$, the one-form $\A$ that appears in eq.~(\ref{LMgeom}) is given by $R_y$ times eq.~(\ref{Ai vevs}).
For distinct angular potentials, the following integral should be solved:
\begin{equation}
	I=e^{-\mathcal{A}-\mathcal{B}}\int_0 ^{\infty} \frac{x}{(1+ax)^2} \frac{1}{1+bx} e^{\frac{\mathcal{A}}{1+ax}+\frac{\mathcal{B}}{1+bx}}dx~.
\end{equation}
For $a>b$, the change of variables
\begin{equation}
	z = \frac{1-\frac{b}{a}}{1+ax}+ \frac{b}{a}~,
\end{equation}
gives rise to
\begin{equation}
	I = \frac{e^{\frac{\mathcal{B}b-\mathcal{A}a}{a-b}}}{(a-b)^2} \int_{\frac{b}{a}}^1 dz\left( \frac{1}{z}-1\right) e^{\frac{a\mathcal{A}}{a-b}z- \frac{b \mathcal{B}}{(a-b)z}}~.
\end{equation}
For $\vec{x}_{34}=0$ and general $\vec{x}_{12}$, calculations similar to the ones performed in subsection \ref{subsec:AveHarmFormA} give
\begin{align}\label{Aresult}
     &\langle A_{\phi}(\vec{x}_{12},\vec{x}_{34}=0) \rangle = \frac{|\vec{x}_{12}|^2\langle J_{1 2} \rangle}{2(\zeta_{12}-\zeta_{34})} 	  \times \nonumber\\
	&\hskip 1cm \left[ \frac{e^{-\frac{|\vec{x}_{12}|^2}{2\mu^2 (\zeta_{12}-\zeta_{34})}}}{2\mu^2(\zeta_{12}-\zeta_{34})}\left(Ei \left( \frac{|\vec{x}_{12}|^2}{2\mu^2 (\zeta_{12}-\zeta_{34})}\right)-Ei \left( \frac{|\vec{x}_{12}|^2}{2\mu^2(\zeta_{12}-\zeta_{34})} \frac{\zeta_{34}}{\zeta_{12}}\right)\right)-\frac{1}{|\vec{x}_{12}|^2 } \left( 1- e^{-\frac{|\vec{x}_{12}|^2}{2\mu^2 \zeta_{12}}} \right)\right]
\nn\\[.3cm]
&\hskip3cm
    A_r(\vec{x}_{12},\vec{x}_{34}=0)= A_{\psi} (\vec{x}_{12},\vec{x}_{34}=0)=A_{\theta}(\vec{x}_{12},\vec{x}_{34}=0)=0~,
\end{align}
 where the angles $\phi$, $\psi$ and $\theta$ are defined through
 \begin{equation}
    x_1 + ix_2 = r \cos(\theta) e^{i \phi}  ~,~ x_3 + ix_4 = r \sin(\theta) e^{i\psi}.
 \end{equation}

One can take the scaling limits $\tau_2\to 0^+,\omega_{12},\omega_{34}\to 0$ with fixed ratios   $\gamma_{ij} = \frac{\omega_{ij}}{\tau}$. In the absence of rotation in the $x_3-x_4$ plane, 
\begin{equation}
	\zeta(q,\omega_{12})\to \frac{1-\pi \gamma \cot(\pi \gamma)}{4\pi \tau_2\gamma^2} 
 ~~,~~~~
	\zeta (q,\omega_{34}) \to \frac{\pi }{12\tau_2 }~. 
\end{equation}
As expected, we have obtained results proportional to the expectation value of the angular momentum and with spatial support along $\mu \sqrt{\zeta(q,\omega_{12})}$. Using this, the angular momentum of the solution is computed in subsection \ref{subs:Properties}.

\subsection{Averaged harmonic function  \texorpdfstring{$H_1 (\vec{x})$}{TEXT}}
\label{subsec:AveHarmFnH1}

This subsection completes the calculations of the list of harmonic functions and forms, focusing on the expectation value of the harmonic function $H_1 (\vec{x})$. This is related to an integral of $\big\langle \left(\partial_v f_j\right)^2 e^{i\vec{k}\cdot \vec{f}} \big\rangle$, and the three-point function of exponentials in the ensemble in question allow one to evaluate it; the final result is given in eq.~\eqref{H1result}.

We would like to calculate
\begin{equation}\label{Kvev}
	\big\langle K(v,\vec{x}) \big\rangle = \sum_{j=1} ^4 \bigg \langle  \frac{\left(\partial_v f_j\right)^2}{|\vec{x}-\vec{f}|^2}\bigg \rangle=	\sum_{j=1} ^4 \int_0 ^{\infty} ds e^{-s|\vec{x}|^2} \int \frac{d^4b}{(\pi s)^2} e^{-\frac{|\vec{b}|^2}{s}}\left \langle \left(\partial_v f_j\right)^2 e^{i\vec{k}\cdot \vec{f}} \right\rangle~,
\end{equation}
with $\vec{k}= -2i (s\vec{x}+i\vec{b})$.
The expectation value in \eqref{Kvev} can be evaluated using standard manipulations along the lines of the previous subsections, again using the Green's functions~\eqref{straightG}, \eqref{curlyG}, resulting in
\be
\big\langle K(v,\vec x) \big\rangle = K_1 + K_2,
\ee
where
 \begin{align}\label{K1}
	 &K_1=\int_0 ^{\infty} ds~ e^{-\frac{x_1^2+x_2^2}{2\mu^2 \zeta(q,\omega_{12})}} e^{\frac{x_1 ^2+x_2^2}{2\mu^2 \zeta(q,\omega_{12})}\frac{1}{1+2\mu^2 \zeta(q,\omega_{12})s}} e^{-\frac{x_3^2+x_4^2}{2\mu^2 \zeta(q,\omega_{34})}} e^{\frac{x_3 ^2+x_4^2}{2\mu^2 \zeta(q,\omega_{34})}\frac{1}{1+2\mu^2 \zeta(q,\omega_{34})s}} 
  \nonumber\\
	 &\hskip2cm\times\frac{\mu^2 q\frac{\partial}{\partial q}\left( \log(Z(q,\omega_{12}))+\log(Z(q,\omega_{34}))\right)}{\left(1+2\mu^2 \zeta(q,\omega_{34})s\right) \left(1+2\mu^2 \zeta(q,\omega_{12})s\right)}=	\mu^2 \langle L_0 \rangle \left \langle \frac{1}{|\vec{x}-\vec{f}|^2}\right \rangle ~.
\end{align}
and
\begin{align}\label{K2}
	&K_2=\mu^4\int_0 ^{\infty}ds~e^{-\frac{x_1^2+x_2^2}{2\mu^2 \zeta(q,\omega_{12})}} e^{\frac{x_1 ^2+x_2^2}{2\mu^2 \zeta(q,\omega_{12})}\frac{1}{1+2\mu^2 \zeta(q,\omega_{12})s}}e^{-\frac{x_3^2+x_4^2}{2\mu^2 \zeta(q,\omega_{34})}} \frac{se^{\frac{x_3 ^2+x_4^2}{2\mu^2 \zeta(q,\omega_{34})}\frac{1}{1+2\mu^2 \zeta(q,\omega_{34})s}}}{(1+2\mu^2 \zeta(q,\omega_{12})s)(1+2\mu^2 \zeta(q,\omega_{34})s)}
 \nonumber\\
	&\hskip1cm
 \times\left\{\left[ \frac{\partial}{\partial \omega_{12}}\log(Z(q,\omega_{12}))\right]^2 \left[\frac{s (x_1^2+x_2^2)}{(1+2\mu^2 \zeta(q,\omega_{12})s)^2} - \frac{1}{1+2\mu^2 \zeta(q,\omega_{12})s}\right]+\right. \nonumber\\
	&\hskip2cm
 +\left.\left[ \frac{\partial}{\partial \omega_{34}}\log(Z(q,\omega_{34}))\right]^2 \left( \frac{s (x_3^2+x_4^2)}{(1+2\mu^2 \zeta(q,\omega_{34})s)^2}-\frac{1}{1+2\mu^2 \zeta(q,\omega_{34})s}\right) \right\} ~.
\end{align}

It is possible to evaluate the integrals in the case of $\omega_{12}= \omega_{34}$; afterward, we will exhibit a rather more involved expression for the answer in the case $\omega_{12} \neq \omega_{34}$ and $\vec{x}_{34}=0$. 
Using eq.~(\ref{AveragedGreen'sFunctionRotating}), equation (\ref{K1}) becomes
\begin{equation}\label{K1equalOmega}
    	K_1(\omega_{12}=\omega_{34})=2\mu^2 q \frac{\partial}{\partial q}\left( \log(Z(q,\omega))\right) \frac{1-e^{-\frac{|\vec{x}|^2}{2\mu^2 \zeta(q,\omega)}}}{|\vec{x}|^2}~.
\end{equation}
To evaluate eq.~(\ref{K2}), let us define
\begin{equation}
	I=e^{-\mathcal{A}}\int_0 ^{\infty} dx \frac{x^2 e^{\frac{\mathcal{A}}{1+ax}}}{(1+ax)^4}~~,~~~~	
 \mathcal{A} \equiv \frac{|\vec{x}|^2}{2\mu^2 \zeta(q,\omega)} ~~,~~~~ a\equiv 2\mu^2 \zeta(q,\omega)~.
\end{equation}
Then
\begin{align}
&	I=  
\frac{1}{a^2}\frac{\partial^2}{\partial\mathcal{A}^2}\int_0 ^{\infty}  \frac{1}{(1+ax)^2}e^{-\frac{\mathcal{A} a x}{1+ax}}dx=
\frac{2}{a^3 \mathcal{A}^3}\left( 1-e^{-\mathcal{A}}\left( 1 + \mathcal{A} +\frac{1}{2}\mathcal{A}^2\right)\right)~,
\end{align}
and so  
\begin{equation}\label{K2equalOmega}
	K_2(\omega_{12}=\omega_{34})=\frac{1}{4\zeta(q,\omega)^2} \left(\frac{\partial}{\partial \omega}\log(Z(q,\omega))\right)^2e^{-\frac{|\vec{x}|^2}{2\mu^2 \zeta(q,\omega)}}~.
\end{equation}
Assembling eqs.~(\ref{K1equalOmega}) and (\ref{K2equalOmega}), multiplying by the charge of the one-branes and averaging over the worldsheet locations lead to the expectation value of the harmonic function $H_1$: 
\begin{align}\label{ResultRotatingK}
	&\big\langle H_1 (\vec{x}) \big\rangle (q,\omega) = 	 Q_1 \Bigg( \frac{1-e^{-\frac{|\vec{x}|^2}{2\mu^2 \zeta(q,\omega)}}}{|\vec{x}|^2}-\frac{1}{4 \zeta(q,\omega)^2}\frac{\langle J\rangle^2}{a^2} e^{-\frac{|\vec{x}|^2}{2\mu^2 \zeta(q,\omega)}}\Bigg)  ~,
\end{align}
where $a$ was defined in~\eqref{adef}.
The factor of $\frac{1 }{a^2}$ arises from a change of variables $ \tilde{v}\equiv \frac{2\pi}{L}v$.
Next, consider $\omega_{12}\neq \omega_{34}$. 
Eq.~(\ref{K1}) involves an integral we encountered previously:
\begin{equation}\label{Original-IntegralAgain}
	I(a,b,\mathcal{A},\mathcal{B})\equiv e^{-\mathcal{A}-\mathcal{B}}\int_0 ^{\infty} dx \frac{e^{\frac{\mathcal{A}}{1+ax}+\frac{\mathcal{B}}{1+bx}}}{(1+ax)(1+bx)}= \frac{e^{\frac{b\mathcal{B}-\mathcal{A}a}{a-b}}}{a-b}\int_{\frac{b}{a}} ^1 \frac{1}{z} e^{\frac{a\mathcal{A}z}{a-b} - \frac{\mathcal{B}b}{(a-b)z}}dz~
\end{equation}
with
\begin{align}
\mathcal{A} = \frac{|\vec{x}_{12}|^2}{2\mu^2 \zeta (q,\omega_{12})} ~,~~~ a= 2\mu ^2 \zeta(q,\omega_{12}) ~,~~~ \mathcal{B} = \frac{|\vec{x}_{34}|^2}{2\mu^2 \zeta(q,\omega_{34})} ~,~~~ b=2\mu^2 \zeta(q,\omega_{34}).
\end{align}
Eq.~(\ref{K2}) can then be computed 
\begin{align}
   & K_2=\mu^4 \left[\langle J_{x_1,x_2}\rangle  ^2\left(\frac{\partial}{\partial |\vec{x}_{12}|^2}+|\vec{x}_{12}|^2 \frac{\partial^2}{\partial (|\vec{x}_{12}|^2 )^2}\right) 
   \right.  
   \nonumber\\  
   &\hskip2cm \left.  
   +\;\langle J_{x_3 x_4}\rangle ^2 \left(\frac{\partial}{\partial |\vec{x}_{34}|^2}+|\vec{x}_{34}|^2 \frac{\partial^2}{\partial (|\vec{x}_{34}|^2 )^2}\right)\right]I(\vec{x}_{12},\vec{x}_{34},q,\omega_{12},\omega_{34}) .
\end{align}
For $\vec{x}_{34}=0$ and $|\vec{x}_{12}|\geq 0$, the full result for the average harmonic function $H_1 (\vec{x})$ is 
\begin{align}
\label{H1result}
\frac{1}{Q_1}\big\langle H_1 (\vec{x}) \big\rangle &=  \frac{1}{2\mu^2\zeta_{12}-\zeta_{34})}e^{-\frac{|\vec{x}_{12}|^2}{2\mu^2 (\zeta_{12} - \zeta_{34})}}\left[Ei\left( \frac{|\vec{x}_{12}|^2}{2\mu^2 (\zeta_{12}-\zeta_{34})}\right)-Ei\left(\frac{|\vec{x}_{12}|^2}{2\mu^2 (\zeta_{12}-\zeta_{34})}\frac{\zeta_{34}}{\zeta_{12}}\right)\right] 
\nonumber\\[.3cm]
&\hskip.5cm
 + \frac{1}{a^2} \frac{\langle J_{x_1 x_2} \rangle ^2-\langle J_{x_3 x_4} \rangle ^2}{2 (\zeta_{12}-\zeta_{34})} \left\{ \frac{|\vec{x}_{12}|^2 -2\mu^2 (\zeta_{12}-\zeta_{34})}{ 4\mu^2(\zeta_{12}-\zeta_{34})^2}  e^{-\, \frac{|\vec{x}_{12}|^2}{2\mu^2 (\zeta _{12} - \zeta_{34})}} \right.
\\
&\hskip1.5cm
\left.\left[Ei\left( \frac{|\vec{x}_{12}|^2}{2\mu^2 (\zeta_{12}-\zeta_{34})}\right)-Ei\left(\frac{|\vec{x}_{12}|^2}{2\mu^2 (\zeta_{12}-\zeta_{34})}\frac{\zeta_{34}}{\zeta_{12}}\right)\right]   -\;\frac{1}{2 (\zeta_{12}-\zeta_{34})} \left(1-e^{-\frac{|\vec{x}_{12}|^2}{2\mu^2 \zeta_{12}}}\right)  \right\} 
\nn\\
&\hskip.5cm
+\frac{1}{4(\zeta_{12}-\zeta_{34})a^2} e^{-\frac{|\vec{x}_{12}|^2}{2\mu^2 \zeta_{12}}} \left(\frac{\langle J_{x_1 x_2}\rangle ^2}{\zeta_{12} } -\frac{\langle J_{x_3 x_4}\rangle^2}{\zeta_{34}}\right).
\nn
\end{align}
When the two angular potentials are equal, one recovers the result of eq.~(\ref{ResultRotatingK}).

To recapitulate, we have found the ensemble averaged harmonic functions that give rise to a new smooth and horizonless solution describing spinning NS5-P bound states. It is specified by the charges, average angular momenta and the spatial extents where the ellipsoidal source is.

\subsection{Properties of the solution}
\label{subs:Properties}

For simplicity, we consider the case of equal angular potentials. Then the harmonic functions and forms are given by:
\begin{equation}
    H_5 (\vec{x}) = Q_5 \frac{1-e^{-\frac{|\vec{x}|^2}{2\mu^2 \zeta (q,\omega)}}}{|\vec{x}|^2},
\end{equation}
\begin{equation}
    A_1 = -\mathcal{A}x_2 ~,~  A_2 = \mathcal{A} x_1 ~,~ A_3 = -\mathcal{A} x_4 ~,~ A_4 = \mathcal{A}x_3 
\end{equation}
where
\begin{equation}\label{1form}
 \mathcal{A} = R_y \, \mu^2 \langle J \rangle \frac{1-e^{-\frac{|\vec{x}|^2}{2\mu^2 \zeta(q,\omega)}}\left(1+\frac{|\vec{x}|^2}{2\mu^2 \zeta(q,\omega)}\right)}{|\vec{x}|^4}.
\end{equation}
Finally,
\begin{equation}
H_1 (\vec{x}) = Q_1 \Bigg[\frac{1-e^{-\frac{|\vec{x}|^2}{2\mu^2 \zeta (q,\omega)}}}{|\vec{x}|^2} - \frac{ \langle J\rangle^2}{4\zeta(q,\omega)^2 a^2} e^{-\frac{|\vec{x}|^2}{2\mu^2 \zeta(q,\omega)}}\Bigg].
\end{equation}
Here are the limiting values of these expressions when approaching the origin $|\vec{x}|\ll \mu \sqrt{\zeta(q,\omega)}$:
\begin{equation}\label{LimitOrigin}
    H_5 \to \frac{Q_5}{2\mu^2 \zeta(q,\omega)}     
~~,~~~~
\mathcal{A} \to \frac{R_y \langle J\rangle}{4\mu^2 \zeta(q,\omega)^2}
~~,~~~~
    H_1 \to \frac{Q_1}{2\mu^2 \zeta(q,\omega)} - \frac{ Q_1\langle J \rangle^2}{4\zeta(q,\omega)^2a^2}.
\end{equation}
We have shown that in the small chemical potential limit,
\begin{equation}
	\zeta(q,\omega)\to \frac{1-\pi \gamma \cot(\pi \gamma)}{4\pi \tau_2\gamma^2}~, 
 \end{equation}
 and will show in the next section that
\begin{eqnarray}\label{AngularMomentumPre}
	\langle \hat{J} \rangle \to\frac{1}{2\pi \tau_2} \frac{-1+\pi \gamma \cot(\pi \gamma)}{\gamma}.
\end{eqnarray}
When taking the decoupling limit and then the scaling limit of small chemical potential, the first term in $H_1$ in eq. (\ref{LimitOrigin}) proportional to $\frac{Q_1}{\mu^2}$ dominates the second term, as long as $N=n_5 n_1\to \infty$. The full functions $H_1,\mathcal{A},H_5$ interpolate between the limiting values in eq. (\ref{LimitOrigin}) and the power-law falloff associated with asymptotically AdS$_3$. 

There are several features of note: First, no singularities plague the solution corresponding to the ensemble average, as $H_1,H_5,\mathcal{A}$ all saturate near the source at finite values. Second, the geometry has no horizon because $H_5 (\vec{x})$ which is the radial-radial component of the metric never diverges. Third, the solution is weakly-curved due to the fact that the characteristic length of variation $\mu \sqrt{\zeta}\gtrsim \frac{\mu}{\sqrt{\tau_2}}$ corresponds to invariant distances greater than $\sqrt{n_5 \alpha'}$. Fourth, there does not exist an ergo-region since there is no divergence of $H_1$ at any location in transverse space (see eq. (\ref{LMgeom})). Fifth, it can be checked that the Killing vectors that generate rotations in the 1-2 and 3-4 planes, are always spacelike - implying that closed timelike curves are absent.

The exponential of twice the dilaton is given by eq. (\ref{LMdil})
\begin{equation}
    e^{2\phi} = g_s ^2\frac{Q_5}{Q_1} \,\frac{1}{1- \langle J\rangle ^2\frac{|\vec{x}|^2 \exp\big[-\frac{|\vec{x}|^2}{2\mu^2 \zeta(q,\omega)}\big]}{4\zeta(q,\omega)^2 a^2 \big(1-\exp\big[-\frac{|\vec{x}|^2}{2\mu^2 \zeta(q,\omega)}\big]\big)}}\approx \frac{n_5 V_4}{n_1 (\alpha')^2}.
\end{equation}
The approximation made in the last relation follows from taking $n_1 n_5 \gg 1$.
Therefore, the solution is weakly-coupled throughout space. Just as in the non-spinning ensemble average, the size of the Y-circle becomes sub-Planckian near the core and one has to make a transition to the NS5-P duality frame as in section \ref{subsc:TheSolutionInTheNS5-PFrame} - and the dilaton stabilizes at a value corresponding to weak coupling near the source. 

The size of the bound state, $\sqrt{2}\mu\sqrt{\zeta(q,\omega)} $ can be evaluated in the small chemical potential scaling limit as in eq.~(\ref{HighTlimit2}). The ratio of the size of the spinning bound state and the non-spinning is plotted as a function of $\gamma = \frac{\omega}{\tau}$.

\begin{figure}[hbt!]
	\begin{center}
 \vskip.5cm
	\includegraphics[scale=0.6]{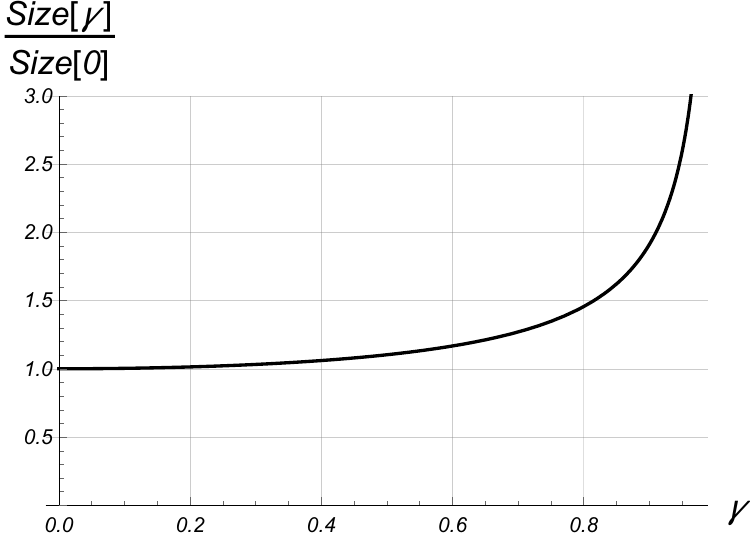}
\end{center}
\caption{The size of the rotating source in units of the non-rotating source vs the angular velocity parameter $\gamma$. As expected, rotation increases the spatial extent of the extended object. }
\end{figure}

We see that at any but the most extreme values of the angular potential, the rotating blob struggles to exceed the size of the non-rotating blob, and so any potential ring structure would be hidden underneath this existing natural size of the ensemble.  At the extreme right where it is possible for such a structure to emerge, the lowest rotating mode approaches a critical point, and rather than having a sharply-defined condensate it undergoes large fluctuations in its occupation number; the state can be thought of as a superposition of rings of a wide variety of sizes, and so the ensemble geometry is more of a pancake than a ring.

One can read off the angular momentum of the solution $J_{sol}$ by using its standard relation with the falloff condition of the metric spatial-time component of a D-dimensional spacetime:
\begin{equation} 
 g_{t\phi} \approx \frac{\nu}{{r}^{D-3}} ~,~ J_{sol}=\frac{\omega_{D-2}~ \nu \sin^2 (\theta)}{ 8\pi G_N}.
\end{equation}
Utilizing eq.~(\ref{1form}) we obtain
\begin{equation}
J_{sol} = \frac{\pi R_y \mu^2 \langle J\rangle}{4G_N ^{5D}}=  \langle J \rangle.
\end{equation}


\section{Fixing the angular momentum}
\label{sec:FixedAngularMomenta}

Previously we observed that the solution corresponding to an ensemble average with fixed angular potentials has an ellipsoidal shape.  How do we reconcile this result with the fact that generally one would expect the angular momentum to generate a barrier forcing the fivebranes away from the origin?  Typically one would expect a ring-shaped ensemble geometry, as seen in~\rcite{Lunin:2002qf}.

In this section we show how to reproduce the effect of repulsion from the rotation axis in an ensemble with fixed angular momentum about one plane of rotation.
We perform an approximate calculation of the quantity 
$ \big\langle \delta^4\big(\vec{x}-\vec{f}\big)\big \rangle$, 
which conveys where the bound state is, and show that it lies away from the rotation axis, with the scale of distance set by the the angular momentum. 
Further, we point out that the difference in structure in the two conjugate ensembles can be explained due to order one (or more) relative fluctuations of the angular momentum in the ensemble of fixed angular potentials. 
Finally, we calculate the first two terms in the $\frac{1}{N}$-expansion of the expectation value of the harmonic function $H_5 (\vec{x})$ in a particular state of fixed angular momentum. 

\subsection{Switching to an ensemble with fixed angular momentum}
\label{subsec:SwitchingEnesembles}

We begin by calculating the average of this delta-function observable in the grand canonical ensemble of fixed chemical potential and angular potentials. 
We require the small chemical potential scaling limit of the partition function, which can be computed from its behavior under modular transformations \rcite{Russo:1994ev} 
\begin{eqnarray}\label{HighTZ}
	Z_D(\tau_2,\omega) \to \tau_2^{\frac{D}{2}} e^{\frac{\pi  D}{12\tau_2}}\frac{ \pi \gamma }{\sin\left(\pi \gamma \right)}.
\end{eqnarray}
Additionally, the expectation value of the exponential operator in the grand canonical ensemble is useful for the purpose of computing the average of the delta-function. This was computed in subsection \ref{subsec:OnePointFunctionOfAnExponential} and including the factor $\frac{\pi \gamma}{\sin(\pi \gamma)}$ is necessary when passing between ensembles; thus one has%
\footnote{Again, any issues with operator normalization, scheme-dependence, \etc, enter the discussion at subleading order at large $N$.}
\begin{equation}\label{RotatingExptl3}
	\Big\langle\, \!:e^{i\vec{k}\cdot \vec{f}}:\! ~\Big\rangle_{\omega_{12},\omega_{34}} = \frac{\pi \gamma}{\sin(\pi \gamma)} e^{-\frac{1}{2} \mu^2 \left[ (k_1 ^2 + k_2 ^2) \zeta(q,\omega_{12})+(k_3 ^2 + k_4 ^2) \zeta(q,\omega_{34})\right]},
\end{equation}
where
\begin{equation}\label{HighTlimit2c}
	\zeta_{12}(q,\omega) \to \frac{1}{4\pi \tau_2} \frac{1-\pi \gamma \cot(\pi \gamma)}{\gamma^2}. 
\end{equation}
These expressions allow one to calculate the expectation value of the delta-function
\begin{align}
\big \langle \delta^4 (\vec{f} - \vec{x})\big\rangle &= \frac{1}{(2\pi)^4} \int d^4 k ~e^{-i \vec{k} \cdot \vec{x}} \frac{\pi \gamma}{\sin(\pi \gamma)}e^{-\frac{1}{2} \mu^2 \left[ (k_1 ^2 + k_2 ^2) \zeta(q,\Omega_{12})+(k_3 ^2 + k_4 ^2) \zeta(q,\Omega_{34})\right]}
\nonumber\\[.2cm]
&=\frac{\pi \gamma}{\sin(\pi \gamma)}\frac{1}{(2\pi)^2 \mu^4 \zeta_{12} \zeta_{34}} \,e^{-\frac{|\vec{x}_{12}|^2}{2\mu^2 \zeta_{12}} - \frac{|\vec{x}_{34}|^2}{2\mu^2 \zeta_{34}}} .
\end{align}
For simplicity we take $\zeta_{34}\to 0$ (alternatively, one can consider an ensemble where $\omega_{34}$ and $J_{12}$ are fixed)
 \begin{align}
\left \langle \delta^4 (\vec{f} - \vec{x})\right\rangle_{\omega} (\zeta_{34}\to0) &= \frac{1}{2\pi \mu^2 \zeta_{12}} \,\frac{\pi \gamma}{\sin(\pi \gamma)} \,e^{-\frac{|\vec{x}_{12}|^2}{2\mu^2 \zeta_{12}}} \;\delta(x_3)\,\delta(x_4) ~.
\nn\\[.2cm]
&= \frac{ 2\pi  \tau_2}{\mu^2}\,\frac{\gamma^3}{ \sin(\pi \gamma)-\pi \gamma \cos(\pi \gamma)} \,e^{-\frac{2\pi \tau_2\gamma ^2  \sin(\pi \gamma)|\vec{x}_{12}|^2}{\mu^2 \left(\sin(\pi \gamma)-\pi \gamma \cos(\pi \gamma)\right)}} \;\delta(x_3)\,\delta(x_4) ~.
\end{align}
Let us prepare to make a transition to an ensemble in which the angular momentum is fixed.  The expectation value of any operator $\O$ in the ensemble of fixed real angular potential is dressed with an exponential as shown in the following equation
\begin{equation}\label{Oomega}
	\langle \O \rangle_{\omega}=\text{tr} \left( \hat{\O} e^{i \omega \hat{J}}\right).
\end{equation}
Eq.~(\ref{Oomega}) implies that in the conjugate ensemble of fixed angular momentum, the expectation value is given by an inverse Fourier transform
\begin{equation}
	\langle \O \rangle_J = \frac{1}{2\pi} \int_{-\infty} ^{\infty}  e^{-i\omega J} 	\langle \O \rangle_{\omega} \,d\omega~. 
\end{equation}  
Two comments are in order. First, the trace operation in eq. (\ref{Oomega}) implies that states of different angular momenta $J$ contribute to  $\langle \O \rangle_{\omega}$; we later specify the variance of this distribution. Second, we have been working with a purely imaginary $\omega = i\tau_2 \gamma$ which permits the interpretation of $\mu \sqrt{\zeta}$ as the size of the rotating source. This means that an inverse Laplace transform is required to isolate the coefficient of $e^{-\tau_2 \gamma J}=e^{i\omega J}$ in the expectation value $\langle \O \rangle_{\omega}$. One can rotate $\gamma\to i\gamma$ and phrase the calculation in terms of an inverse Fourier transform, which we choose to do below, similarly to ~\cite{Russo:1994ev}.
For $\O = \delta ^4 (\vec{f}-\vec{x})$, and changing variables through $x=\pi \gamma$
\begin{align}\label{IntegralForDeltFunction}
	&\left \langle \delta^4 (\vec{f} - \vec{x})\right\rangle_J =  \delta(x_3)\,\delta(x_4)\,\frac{ 2 \tau_2 ^2}{\pi^2 \mu^2}\int _{-\infty} ^{\infty} dx \,e^{ i \frac{1}{\pi}x\tau_2 J} \frac{x^3}{x\cosh(x)-\sinh(x)} \,e^{-\frac{2 \tau_2 x^2  \sinh(x)|\vec{x}_{12}|^2}{\pi\mu^2 \left(x \cosh(x)-\sinh(x)\right)}}  ~.
\end{align}
It is useful to approximate the integrand by expanding the argument in the exponential in Taylor series around $x=0$. 
The following formulas are useful for to this end:
\begin{equation}
\frac{x^2 \sinh(x)}{x\cosh(x) - \sinh(x)}= 3+\frac{x^2}{5}+\dots
\quad,\qquad
\frac{x^3}{x\cosh(x) - \sinh(x)}=3 - \frac{3x^2}{10}+\dots
\end{equation}
We denote 
\begin{equation}
\label{pdef}
p\equiv \frac{2\tau_2 |\vec{x}_{12}|^2}{\pi \mu^2}.
\end{equation}
When $p\ll 1$, a part of the integrand multiplying $e^{ i \frac{1}{\pi}x\tau_2 J}$ admits a Taylor expansion and the leading order result vanishes, because
	\begin{equation}
		\int_{-\infty} ^{\infty} e^{ i \frac{1}{\pi}x\tau_2 J} \,dx=2\pi\, \delta \Big(\frac{1}{\pi}J \tau_2\Big)=0.
	\end{equation}
Perturbative corrections that scale like positive powers of $x^2 \times |\vec{x}|_{12} ^2$ would be derivatives of the above delta-function with respect to $J \tau_2$, which vanish too.  As one dials $p$ to be greater than roughly one, the exponential can be approximated by a Gaussian. We plot below the exact integrand and its Gaussian approximation, removing the  oscillating exponential factor.
\begin{figure}[hbt!]
	\begin{center}
 \vskip.5cm
	\includegraphics[scale=0.3]{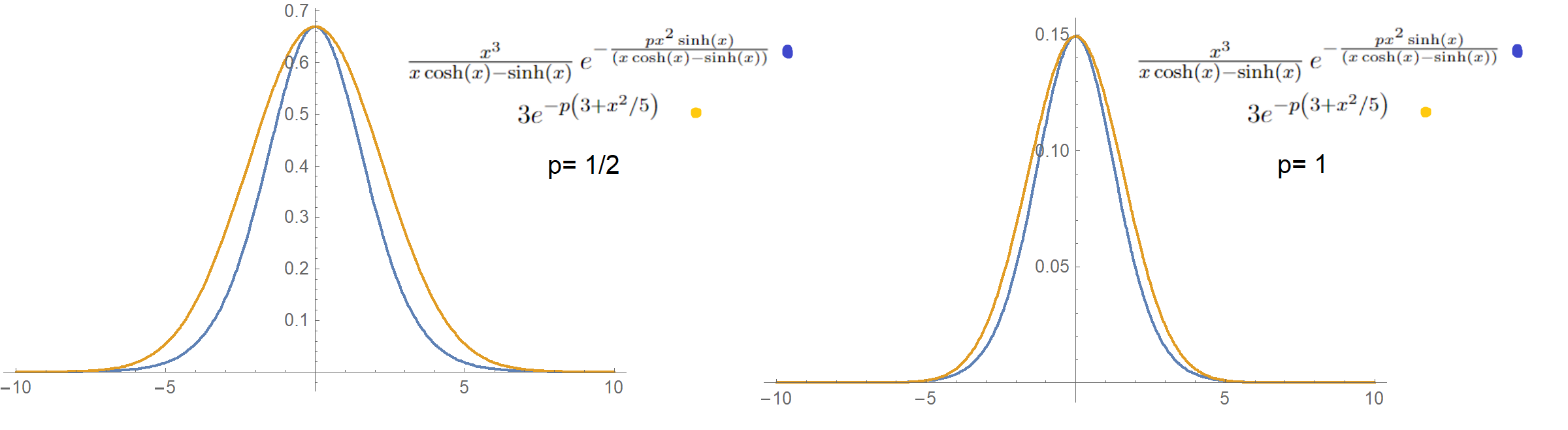}
\end{center}
\caption{These comparisons show that the  integrand (blue) of eq. (\ref{IntegralForDeltFunction}) can be approximated by a Gaussian (orange) for order one values of $p=\frac{2\tau_2 |\vec{x}_{12}|^2}{\pi \mu^2}$.}
\end{figure}

\noindent
Now,
\begin{align}
 \int_{-\infty} ^{\infty} dx\, e^{-\frac{2\tau_2 |\vec{x}_{12}|^2 x^2}{5\pi \mu^2}}e^{i\frac{1}{\pi}\tau_2 J x}=\frac{\pi\mu\sqrt{5}}{|\vec{x}_{12}|\sqrt{2\tau_2}} e^{-\frac{5\tau_2 J^2 \mu^2}{8\pi |\vec{x}_{12}|}}.
\end{align}
Consequently, the (normalized) expectation value of the delta-function is approximated by
\begin{align}
    \left\langle \delta^4 (\vec{x}-\vec{f}) \right\rangle_J \approx \frac{\sqrt{6\tau_2}}{\pi^{2}|\vec{x}_{12}|\mu} e^{-\frac{6 \tau_2}{\pi \mu^2 |\vec{x}_{12}|^2} \left( |\vec{x}_{12}|^2-\sqrt{\frac{5}{48}}J \mu^2\right)^2} \delta(x_3)\delta(x_4).
\end{align}
This displays the effect of the potential barrier associated with the presence of an angular momentum: As $|\vec{x}_{12}|\to0^+$, the expression decays exponentially to zero.  A characteristic radius set by $\mu \sqrt{J}$ emerges.

\subsection{Fixed angular potential vs fixed angular momentum}
\label{subsec:FixedJvsOmega}

In this subsection we compute the relative fluctuations of the angular momentum in the scaling limit of the grand canonical ensemble with fixed chemical potential and angular potentials. 
In a diagonal basis, the angular momentum operator can be expressed in terms of the difference between occupation numbers associated with two orthogonal polarizations: 
\begin{equation}
\hat{J} = \sum_{m=1} ^{\infty} \left( \hat{n}_{xm} -\hat{n}_{ym}\right). 
\end{equation}

The system we study is analogous to a thermal ensemble of infinitely many oscillators labeled by integer numbers, where each number is analogous to a winding number in the NS5-F1 duality frame.
The partition function of a system with inverse temperature $\beta = 2\pi \tau_2$ (which we interpret as a chemical potential for the string charge)  associated with a single oscillator of index $m$ in the $x$-direction where the angular momentum operator is diagonal reads
\begin{equation}
	Z_{mx} = \sum_{n=0} ^{\infty} e^{-\beta m n+2\pi i \omega n}=\frac{1}{1-e^{2\pi i\omega- \beta m}}.
\end{equation}
The averaged occupation number is
\begin{equation}
	\langle \hat{n}_{xm}\rangle = \frac{1}{2\pi i}\frac{\partial}{\partial \omega} \log(Z_1) = \frac{1}{e^{m\beta-2\pi i\omega}-1}=\frac{q^m e^{2\pi i \omega}}{1-q^m e^{2\pi i\omega}}.
\end{equation}
Similarly,
\begin{equation}
	\langle \hat{n}_{ym}\rangle  = \frac{q^m e^{-2\pi i\omega}}{1-q^m e^{-2\pi i \omega}}.
\end{equation}
The effect of the angular potential is to bias the relative number of modes occupying each oscillator with respect to the $x$ and $y$ directions.
The bosonic modes are governed by the conventional Bose-Einstein distribution when interpreting the oscillator index $m$ as the energy of the mode and writing $2\pi i\omega=\mu \beta$.  The lowest-lying mode $|n_{1x}\rangle$ starts to condense when the angular potential approaches the inverse temperature - this is a phenomenon that plays an important role as we explain below. Next, consider quantum fluctuations of the occupation numbers.
\begin{equation}
	\Delta n_x ^2\equiv \langle n_x ^2 \rangle - \langle n_x \rangle^2 = \frac{1}{2\pi i}\frac{\partial}{\partial \omega} \langle n_x \rangle=\frac{q^m e^{2\pi i \omega}}{\left(1-q^m e^{2\pi i \omega}\right)^2}.
\end{equation}
Next, the knowledge of the averages of occupation numbers allows one to calculate the expectation values of the angular momentum:
\begin{equation}
	\langle \hat{J}_{xy} \rangle = \left\langle \sum_{m=1} ^{\infty}  \left(\hat{n}_{xm}-\hat{n}_{ym}\right) \right\rangle = \sum_{m=1} ^{\infty} \left( \frac{q^m e^{2\pi i \omega}}{1-q^m e^{2\pi i \omega}}-\frac{q^m e^{-2\pi i\omega}}{1-q^m e^{-2\pi i \omega}}\right).
\end{equation}  
As a check, this vanishes when turning-off the angular potential. In the limit $\tau_2 \to 0^+, \omega \to 0^+$ and $\gamma = \frac{\omega}{i \tau_2}=\text{fixed}$, 
\begin{eqnarray}\label{AngularMomentum}
	\langle \hat{J}_{xy} \rangle = -\sum_{m=1} ^{\infty} \frac{2\gamma}{2\pi \tau_2} \frac{1}{m^2 - \gamma^2}=\frac{1}{2\pi \tau_2} \frac{-1+\pi \gamma \cot(\pi \gamma)}{\gamma}.
\end{eqnarray}
The last transition follows from eq. (\ref{SumEquation}). 
The absolute value of the expectation value of the angular momentum operator is much bigger than one whenever $0.1<\gamma$.
Recalling the relation $\tau_2 \sim \frac{1}{\sqrt{n_1 n_5}}$, the expectation value $\langle J \rangle$ is proportional to $\sqrt{n_1 n_5}$ for $\gamma$ not too close to $1$. 
Next, consider the quantum fluctuations of the angular momentum operator:
\begin{equation}
	\Delta J_{xy} ^2 = \sum_{m=1} ^{\infty}\left( \Delta n_{xm}^2 +\Delta n_{ym}^2 \right)=\sum_{m=1} ^{\infty} \left[\frac{q^m e^{2\pi i \omega}}{(1-q^{m} e^{2\pi i\omega})^2}+\frac{q^m e^{-2\pi i \omega}}{(1-q^{m} e^{-2\pi i \omega})^2}\right].
\end{equation}
In the scaling limit,
\begin{equation}
	\Delta J_{xy} ^2 \to \frac{1}{4\pi^2 \tau_2 ^2}\sum_{m=1} ^{\infty} \left( \frac{1}{(\gamma+m)^2}+ \frac{1}{(-\gamma+m)^2}\right)=\frac{1}{4\pi^2 \tau_2 ^2} \left(\zeta(2,1+\gamma)+\zeta(2,1-\gamma)\right).
\end{equation}
Here $\zeta(s,a)$ is the Hurwitz zeta function defined in eq.~(\ref{Hurwitz}).
We plot $\frac{\Delta J_{xy}}{\langle J_{xy}\rangle}(\gamma)$ as a function of $\gamma$. In all the range $0<\gamma<1$, the ensemble of fixed angular potential would predict results different from the ensemble of fixed angular momentum because the minimal value of the relative fluctuations is around $0.928$ for $\gamma\approx 0.84$. For $\gamma\approx 0$ the relative fluctuations are much greater than the minimal value. In particular, this means that states of low angular momentum are taken into account - which are more localized near the center than higher angular momentum states.

 \begin{figure}
\centering
\includegraphics[scale=0.5]{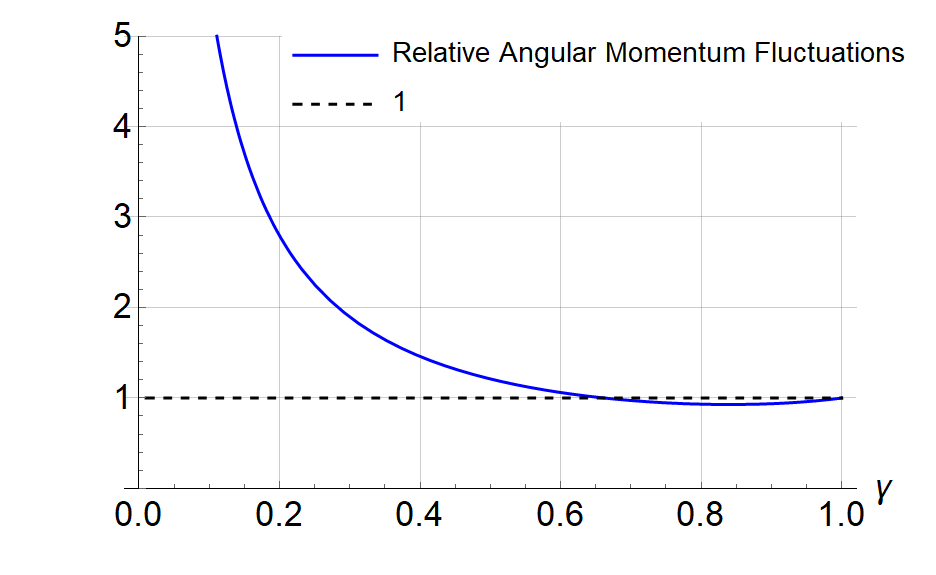}
\caption{The horizontal axes in this graph is $\gamma=\frac{\omega}{i \tau_2}$ and the vertical axis is the relative fluctuations of the angular momentum in the ensemble of fixed angular potentials, in the scaling limit.}
\label{FluctuationsAngularMomentumFIG}
\end{figure}

\subsection{Maximally-spinning supertube}
\label{subsec:MaximallySpinningSupertube}
The purposes of this subsection are to show that a particular state of fixed angular momentum displays a ring structure, and to find the leading correction in the $\frac{1}{N} $ expansion to metric components in transverse space. For simplicity, we choose the state of maximal angular momentum.

The one-point function of the harmonic function $H_5 (\vec{x})$, which determines the metric components in transverse space in the NS5-F1 duality frame, is related to the one-point function of an exponential as we have seen. 
We thus want to calculate first the one-point function of a exponential operator in the maximally spinning state, but we can start by writing several general expressions for the more general microcanonical ensemble of fixed angular momentum $J$ and level $N$ in a $D=2$ target-space:
\begin{equation}
	\sum_{j=1} ^{D}\sum_{m=1} ^N m n_{jm} = N ~,~ \sum_{m=1} ^{N} (n_{xm}-n_{ym}) = J.
\end{equation}
We denote states satisfying the constraints above by $|\{ n_{jm}\}\rangle$. 
The scalar product of the vector $\vec{k}$ with the profile function in 2D can be expressed in terms of the creation and annihilation operators associated with the diagonal angular momentum.  In equation,
\begin{equation}
		k_x \hat{f}_x + k_y \hat{f}_y = \frac{i \mu}{\sqrt{2}}\sum_{n=1} ^{\infty} \frac{1}{\sqrt{n}} \left( k_{xy} ^* \hat{b}_{xn} z^{-n} - k_{xy} \hat{b}_{xn} ^{\dagger} z^n+k_{xy} \hat{b}_{yn} z^{-n} - k_{xy} ^*\hat{b}_{yn} ^{\dagger }z^n\right).
\end{equation}
Here,
\begin{equation}
    k_{xy} = \frac{k_x - ik_y}{\sqrt{2}}.
\end{equation}
Then a factor in the matrix element we compute is
\begin{align}
	&\Big\langle \{ n_{xm}\} \Big| \Big(:e^{i \vec{k} \cdot \vec{f}}: \Big)_x \Big| \{ n_{xm}\} \Big\rangle = 	\Big\langle \{ n_{xm}\} \Big| e^{\frac{\mu}{\sqrt{2}} \sum_{xm} \frac{1}{\sqrt{m}} \hat{b}_{xm} ^{\dagger} z^m k_{xy} } e^{-\frac{\mu}{\sqrt{2}} \sum_{xm} \frac{1}{\sqrt{m}  z^m} \hat{b}_{xm} k_{xy} ^*  }  \Big| \{ n_{xm}\} \Big\rangle    
 \\[.2cm]
	&\hskip2cm = e^{\frac{1}{2}\mu^2 |k_{xy} |^2 \sum_{m\in \{ n_{xm}\}} \frac{1}{m} }	\Big\langle \{ n_{xm}\} \Big| e^{- \frac{\mu}{\sqrt{2}} \sum_{xm} \frac{1}{\sqrt{m}  z^m} \hat{b}_{jm} k_{xy} ^*  } e^{\frac{\mu}{\sqrt{2}} \sum_{xm} \frac{1}{\sqrt{m}} \hat{b}_{jm} ^{\dagger} z^m k_{xy}  } \Big| \{ n_{xm}\} \Big\rangle  ~. 
 \nn
\end{align}
In the last transition we used an identity that follows from the Baker-Campbell-Hausdorff formula.

We now insert the identity operator between the two exponentials. The completeness relation expressed using coherent states implies that
\begin{align}
		\Big\langle \{ n_{xm}\} \Big| \big(\!:e^{i \vec{k}\cdot \vec{f} }:\!\big)_x \Big| \{ n_{xm}\} \Big\rangle = \prod_{m}e^{\frac{1}{2}\mu^2 |k_{xy}|^2  \frac{1}{m} }    \int \frac{d^2 \beta_{xm}}{\pi}\frac{|\beta_{xm}|^{2n_{xm}}}{n_{xm}!} e^{-|\beta_{xm}|^2} e^{-\frac{\mu}{\sqrt{2m}} \frac{k_{xy}^* \beta_{xm}}{z^m} +\sqrt{\frac{\alpha'}{2m}}z^m k_{xy} \beta_{xm}^*}.
\end{align}
A useful integral representation of the Laguerre polynomial is
\begin{equation}
L_n (x_0^2+y_0 ^2)=\frac{1}{\pi n!}    \int_{-\infty} ^{\infty} dx \int_{\infty} ^{\infty} dy~ (x^2+y^2)^{n} e^{-(x-ix_0)^2-(y-iy_0)^2}.
\end{equation}
Therefore the integrals over the coherent state parameters give rise to Laguerre polynomials
\begin{align}\label{ExponentialMicro}
	\big\langle \{ n_{jm}\} \big| :e^{i \vec{k} \cdot \vec{f}}: \big| \{ n_{jm}\} \big\rangle =\prod_{m} L_{n_{xm}} \left( \frac{\mu^2}{4m}(k_x ^2 + k_y ^2)\right)L_{n_{ym}} \left( \frac{\mu^2}{4m}(k_x ^2 + k_y ^2) \right).
\end{align}
Additional factors associated with other target-space dimensions can be incorporated in a simple way. A microcanonical average involves a sum over all partitions of $N$ that also satisfy the constraint of a fixed angular momentum.

Consider the state $n_{1x}=N$ and $n_{jm}=0$ for $j\neq x$ or $m>1$. Denote this state by $|N,N\rangle$, where the labels indicate the level and angular momentum. Then
\begin{align}\label{MaximalAngularMomentum}
		\big\langle N,N \big| :e^{i \vec{k} \cdot \vec{f}}: \big| N,N \big\rangle = L_{N} \left( \frac{\mu^2}{4} (k_x ^2 + k_y^2)\right).
\end{align}
Now,
\begin{equation}\label{HarmonicExponentialRelation}
	\bigg \langle :\frac{1}{|\vec{x} - \vec{f}|^2}: \bigg\rangle = \frac{1}{4\pi^2}\int d^4 k \frac{e^{i \vec{k}\cdot \vec{x}}}{|\vec{k}|^2}  \left\langle :e^{-i \vec{k} \cdot  \vec{f}}: \right\rangle. 
\end{equation}
Eqs. (\ref{MaximalAngularMomentum}) and (\ref{HarmonicExponentialRelation}) imply that
\begin{align}
	\Big \langle N,N \Big| :\frac{1}{|\vec{x} - \vec{f}|^2} : \Big| N,N\Big\rangle = \frac{1}{4\pi^2}\int d^4 k \frac{e^{i \vec{k}\cdot \vec{x}}}{|\vec{k}|^2}  L_{N} \left( \frac{\mu^2}{4} (k_x ^2 + k_y^2)\right).
\end{align}
Parametrizing the momenta as 
\begin{align}
	k_x = k \cos(\phi)~~&,~~~~ k_y = k \sin(\phi) ~~,~~~~ k_3 = q \cos(\psi) ~~,~~~~ k_4 = q \sin(\psi).
\nn\\
	\vec{k}\cdot \vec{x} &= k |\vec{x}|_{12} \cos(\phi-\phi') + q |\vec{x}|_{34} \cos(\psi-\psi')  ~~,
\end{align}
one has
\begin{align}
	\Big \langle N,N \Big| :\frac{1}{|\vec{x} - \vec{f}|^2}: \Big| N,N\Big\rangle = \frac{1}{4\pi^2}\int_0 ^{2\pi}\!d\phi \int_0 ^{2\pi} \!d\psi \int_0 ^{\infty} \!dk \int_0 ^{\infty}\!dq \,q\,k \frac{e^{i \vec{k}\cdot \vec{x}}}{k^2 +q^2} \, L_{N} \Big( \frac{\mu^2}{4} k^2\Big)  ~.
\end{align}
We perform the angular integrations using the formula 
\begin{equation}
	J_0 (z) = \frac{1}{2\pi} \int_0 ^{2\pi} e^{i z \cos(\theta)} d\theta ~,
\end{equation}
and find
\begin{align}
	\Big \langle N,N \Big| :\frac{1}{|\vec{x} - \vec{f}|^2} : \Big| N,N\Big\rangle =  \int_0 ^{\infty} \!dk \int_0 ^{\infty}\!dq \,q\,k \frac{J_0 (k |\vec{x}|_{12})J_0 (q |\vec{x}|_{34})}{k^2 +q^2} \, L_{N} \Big( \frac{\mu^2}{4} k^2\Big).
\end{align}
Next, the $q$-integration is performed thanks to \cite{Prudnikov}

\begin{equation}
	\int_0 ^{\infty}  \frac{x}{x^2 +z^2} J_0 (cx)dx = K_0 (cz).
\end{equation}
Therefore,
\begin{align}
	\Big \langle N,N \Big| :\frac{1}{|\vec{x} - \vec{f}|^2}: \Big| N,N\Big\rangle =  \int_0 ^{\infty} \!dk \,k \,J_0 (k |\vec{x}|_{12})\,K_0 (k |\vec{x}|_{34}) \,L_{N} \Big( \frac{\mu^2}{4} k^2\Big).
\end{align}
The definition of the Laguerre polynomial and a few known formulae imply
\begin{equation}\label{LaguerreVSBessel}
	L_N \left(\frac{x}{N}\right) \approx J_0 (2\sqrt{x})-\frac{1}{2 N} x J_2 (2\sqrt{x}) + O\left( \frac{1}{N^2}\right)~.
\end{equation}
Using eq. (\ref{LaguerreVSBessel}) and writing
\begin{equation}
    \mu^2 = \frac{a^2}{N},
\end{equation}
lead to
\begin{align}\label{ApproximationHarmonicFunction}
\nn\\
&	\Big \langle N,N \Big| :\frac{1}{|\vec{x} - \vec{f}|^2}: \Big| N,N\Big\rangle \approx  \int_0 ^{\infty} \!dk \,k \,J_0 (k |\vec{x}|_{12})\,K_0 (k |\vec{x}|_{34}) \,\left(J_0 (k a) -\frac{k^2 a^2}{8N} J_2 (k a) +...\right).
\end{align}
An integral identity that helps computing the leading order term in eq. (\ref{ApproximationHarmonicFunction}) is \cite{Gradshteyn} 
\begin{equation}\label{GradshteynRhyzik}
	\int_0 ^{\infty} \!dx \,x \,K_0 (ax) \,J_{\nu} (bx) \,J_{\nu} (cx) =\frac{l_1 ^{\nu}}{l_2 ^{\nu}\left(l_2 ^2 - l_1^2\right)},
\end{equation}
where
\begin{equation}
	l_1 \equiv \frac{1}{2}\left(\sqrt{(b+c)^2+a^2}-\sqrt{(b-c)^2+a^2}\right)~~,~~~~
		l_2 \equiv \frac{1}{2}\left(\sqrt{(b+c)^2+a^2}+\sqrt{(b-c)^2+a^2}\right) ~.
\end{equation}
It follows that this contribution reads
\begin{align}
	\Big \langle N,N \Big| :\frac{1}{|\vec{x} - \vec{f}|^2}: \Big| N,N\Big\rangle &\approx \frac{1}{\sqrt{(|x_{12}|+a)^2+|x_{34}|^2}\sqrt{(|x_{12}|-a)^2+|x_{34}|^2}}\nonumber\\
	&=\frac{1}{\sqrt{\left( |\vec{x}|^2 +a^2\right)^2-4a^2 |\vec{x}_{12}|^2}}.
\end{align}
The leading order $1/N$ correction is written in terms of
\begin{align}
-\frac{a^2}{8N} \int_0 ^{\infty} \!dk \,k^3  \,J_0 (k |\vec{x}|_{12})\,K_0 (k |\vec{x}|_{34}) \,J_2 (ka).
\end{align}
To evaluate this, one can apply two parametric derivatives with respect to $a$ on eq. (\ref{GradshteynRhyzik}) and use
\begin{equation}\label{BesselRecursion}
	J_0 '(z)=-J_1 (z) ~,~ J_1' (z)= \frac{1}{z}J_1 (z)-J_2 (z).
\end{equation}
One obtains that the leading order correction in the $\frac{1}{N}$ expansion to $H_5 (\vec{x})$ reads
\begin{align}
& -\frac{Q_5}{8 N}\left( a^2\frac{\partial ^2}{\partial a^2} - a \frac{\partial }{\partial a}\right)\frac{1}{\sqrt{\left( |\vec{x}|^2 +a^2\right)^2-4a^2 |\vec{x}_{12}|^2}}=-\frac{Q_5 a^4}{N} \frac{(-|\vec{x}_{12}|^2+|\vec{x}_{34}|^2 +a^2)^2 - 2|\vec{x}_{12}|^2 |\vec{x}_{34}|^2}{\left((|\vec{x}|^2 +a^2)^2-4a^2 |\vec{x}_{12}|^2\right)^{\frac{5}{2}}}.
\end{align}
At the plane of the ring $\vec{x}_{34}=0$, this simplifies to $-\frac{a^4}{N} \frac{1}{|a^2 - |\vec{x}_{12}|^2|^3} $, possessing a cubic singularity where the fivebranes are. 
This is not an indication that $1/N$ corrections are making the geometry more singular, but rather are modifying the dependence of the harmonic functions on parameters such as $a$ at order $1/N$, and the series expansion in $a$ breaks down near the source.  A resummation of that expansion is expected to have no worse singularity as a function of $\vec x$ than the classical solution.


\section{Conclusions}
\label{sec:Conclusions}

We draw several main conclusions from the calculations we have done. First, unlike the D5-D1 or NS5-F1 frames, the NS5-P duality frame is appropriate for describing the near-source regime of ensemble averages of two-charge bound states. We found that the 1/2-BPS NS5-P ensemble involves a spherical bound state with radius comparable to $\sqrt{Q_5}$ (which is the $AdS_3$ radius in the NS5-F1 frame).  
But at a scale parametrically larger by a factor of order $(n_p/n_5)^{\frac14}\gg1$, the $y$-circle of the geometry has a proper size of order the string scale; at larger radii, the NS5-F1 frame is appropriate, while at smaller radii the NS5-P frame provides the effective description.

At this outer limit of the validity of the NS5-P description, the T-dual spatial $\tilde{y}$-circle has string scale size; it then expands as we move towards smaller radius, while the large three-sphere factor of the geometry is approximately constant in size.  Starting at the bound state radius, the size of the $\tilde y$-circle saturates, while the three-sphere smoothly shrinks to zero at the center. From the point of view of a far away observer, the frequency of any wave emitted by the source would seem redshifted by a large factor $\big({n_p}/{n_5}\big)^{\frac{1}{4}}$.   

Our second key result is that strands of the fivebranes are separated on average by invariant distances proportional to $\sqrt{Q_5}$ and thereby admit a weakly coupled string theory description. In particular, the probability that strands are of distance $d$ much smaller than the bound state radius $r_b$ goes to zero like $d^2$. 

A third finding is that, when the strands rotate with fixed angular velocities, their bulk structure is an ellipsoid, in contrast to rings that were previously encountered. And fourth, members of the ensemble with a fixed angular potential admit at least $90\%$ relative fluctuations in the angular momentum, making them qualitatively different from members of ensembles of fixed angular momenta.

This two-charge system has been proposed as a model for black hole microstates~\rcite{Lunin:2001jy,Lunin:2002qf,Mathur:2005zp}.  The original idea is that the black hole ``ensemble geometry'' would be replaced by a collection of individual horizonless microstate geometries~-- horizonless purportedly because the individual pure microstate has zero entropy.  

This ``fuzzball'' proposal has been the subject of much debate in the literature, even for the BPS two-charge ensemble.  It was suggested in~\rcite{Sen:1995in,Dabholkar:2004yr,Dabholkar:2004dq,Sen:2004dp} that $\alpha'$ corrections to the effective action yield a modification of the extremal black hole solution that has a finite area horizon whose area exactly matched the two-charge microstate entropy, raising the question of why this solution should be discarded (on the other hand,~\rcite{Cano:2018hut} argued that certain $\alpha'$ corrections do not result in a modified horizon).

The issue of whether these two-charge microstates should be thought of as black holes was revisited in~\rcite{Sen:2009bm}, with the conclusion that in the D5-D1 frame there is no black hole solution and so the microstates should not be thought of as black hole microstates; more recently,~\rcite{Mathur:2018tib} argued that the microstates are ``fuzzballs'' in all duality frames, while it was pointed out in~\rcite{Raju:2018xue} that observables in generic two-charge microstates differ from those of the corresponding black hole and suggested that this made the black hole interpretation of two-charge BPS states problematic.

In a sense, the authors of~\rcite{Sen:2009bm,Mathur:2018tib,Raju:2018xue} are all coming to the same basic conclusion, while emphasizing different aspects of it~-- that the black hole geometry is {\it not} an appropriate characterization of the ensemble of 1/2-BPS microstates.
The results of our analysis provide more evidence for this conclusion.%
\footnote{It is perhaps too much to hope that the issue is now settled.}
The generic two-charge NS5-F1 microstate consists of a fivebrane wandering a region which has a sensible supergravity description in the T-dual NS5-P frame, where its proper size is of order~$\sqrt{\nfive\alpha'}$.  Thus while supergravity in the NS5-F1 frame breaks down near the source, perturbative string theory is just fine.  The area surrounding the region occupied by such a state is parametrically larger than the ``stretched horizon'' scale of the extremal black hole whose area would account for the microstate degeneracy~\rcite{Alday:2006nd}.  The fivebrane strands are typically well-separated, and as a result there are typically no loci of strong-coupling dynamics.  Basically, the source is spread over a region that lies outside the scale of any would-be horizon, and thus should not be thought of as a black hole any more than one should describe an electron or an 't Hooft-Polyakov monopole as a black hole~-- their wavefunctions are spread over a scale much larger than the would-be horizon of a black hole of that mass, and so just as we wouldn't describe these particles/solitons as small black holes, neither should we think of the two-charge states that way.  Instead, these states are better thought of as BPS ``fivebrane stars''.

Nevertheless, we have in hand a system of macroscopic (though non-gravitational) entropy and extraordinarily large redshift at the source; they are in some sense rather close to being black holes, and it would be interesting to explore further to what extent they approximate the properties of black holes when excited away from extremality.  For instance, it was emphasized in~\rcite{Martinec:2023plo,Martinec:2023iaf} (reinterpreting~\rcite{Massar:1999wg}) that the probability of emission of a Hawking quantum from a black hole is largely governed by considerations of phase space; in order to model many black hole properties, it might be sufficient to simply have a localized object with a large internal phase space and chaotic mixing in the dynamics.

\section*{Acknowledgements}
We thank 
Nadav Drukker, Harvey Reall and Arkady Tseytlin
for discussions.
YZ thanks the Perimeter Institute and the organizers of Strings 2023 for their hospitality and the opportunity to present results written in this work. The research of YZ is supported by the Blavatnik fellowship, and was supported by the Adams fellowship and the German Research Foundation through a
German-Israeli Project Cooperation (DIP) grant ``Holography and the Swampland''.
The work of EJM is supported in part by DOE grant DE-SC0009924.


\appendix
\section{Details on the torus Green's function}
\label{Appendix:DetailsTorusGreensFunction}

We would like to compute the correlation function of the exponential that appears in section  \ref{sec:SeparationStrands} eq. (\ref{TwoPointResult}) in small chemical potential limit. To this end, we utilize an operator approach where the torus boundary conditions are implemented by tracing states:
\begin{equation}
	\left \langle : e^{i \vec{k}_1 \cdot \vec{f}(z_1)+i\vec{k}_2 \cdot \vec{f}(z_2) }: \right \rangle_{T^2} = \text{Tr}\left(: e^{i \vec{k}_1 \cdot \vec{f}(z_1)+i\vec{k}_2 \cdot \vec{f}(z_2) }: q^{L_0} \right).
\end{equation}
Choosing a coherent state basis $|\lambda_{jm}\rangle$ where $m$ labels oscillators and $j$ labels polarization, the trace can be written as
\begin{align}
\label{IntegralCoherentStates}
	\left \langle : e^{i \vec{k}_1 \cdot \vec{f}(z_1)+i\vec{k}_2 \cdot \vec{f}(z_2) }: \right \rangle_{T^2} =\prod_{jm}\int \frac{d^2 \lambda_{jm}}{\pi}e^{-|\lambda_{jm}|^2}\langle \lambda_{jm}|: e^{i \vec{k}_1 \cdot \vec{f}(z_1)+i\vec{k}_2 \cdot \vec{f}(z_2) }: |q^m\lambda_{jm}\rangle.
\end{align}
The $j^{\rm th}$ component of the $\vec{f}$ operator admits a mode expansion
\begin{equation}
	\hat{f}_j(z) = i\frac{\mu}{\sqrt{2}}\sum_{m=1} ^{\infty} \frac{1}{\sqrt{m}}\left( \frac{\hat{a}_{jm}}{z^m} - \hat{a}_{j m} ^{\dagger} z^m \right).
\end{equation}
The normal-ordered exponential operator is given by
\begin{equation}
	: e^{i \vec{k}_1 \cdot \vec{f}(z_1)+i\vec{k}_2 \cdot \vec{f}(z_2) }\!:~ = \prod_{jm}e^{\mu\sqrt{\frac{1}{2m}}  \hat{a}_{jm} ^{\dagger} (k_{1j}z_1 ^m + k_{2j}z_2 ^m) }e^{-\mu\sqrt{\frac{1}{2m}}  \hat{a}_{jm} (k_{1j}z_1 ^{-m} + k_{2j}z_2 ^{-m}) }.
\end{equation}
Now this is substituted into eq. (\ref{IntegralCoherentStates}), implying that the two-point function is given by
\begin{equation}
	\prod_{jm}\int \frac{d^2 \lambda_{jm}}{\pi}e^{(q^m-1)|\lambda_{jm}|^2}e^{\mu\sqrt{\frac{1}{2m}} \lambda_{jm}^* (k_{1j}z_1 ^m + k_{2j}z_2 ^m) }e^{-\mu\sqrt{\frac{1}{2m}}  q^m \lambda_{jm}(k_{1j}z_1 ^{-m} + k_{2j}z_2 ^{-m}) }.
\end{equation}
The integrals can now be carried out by completing the squares, yielding
\begin{equation}\label{TwoPointFunction5}
	\left\langle	: e^{i \vec{k}_1 \cdot \vec{f}(z_1)+i\vec{k}_2 \cdot \vec{f}(z_2) }: \right\rangle = \frac{1}{q^{-\frac{D}{24}}\eta(q)^D} e^{-\mu^2\frac{1}{2}\sum_{m=1}^{\infty} \frac{q^m}{m(1-q^m)}\left[ (\vec{k}_1+\vec{k}_2)^2+\vec{k}_1 \cdot \vec{k}_2 \left( \left(\frac{z_1}{z_2}\right)^m+\left(\frac{z_2}{z_1}\right)^m-2 \right)\right]}.
\end{equation}
The eta-Dedekind function $\eta(q)$ is defined in
\begin{equation}
    \eta(q) = q^{\frac{1}{24}}\prod_{m=1} ^{\infty} (1-q^m).
\end{equation}
Formula (8.1.32) of \cite{GSW} allows one to simplify the argument of the exponential of eq. (\ref{TwoPointFunction5}):
\begin{equation}
	\sum_{n=1} ^{\infty} \frac{1}{n} \frac{x^n}{1-y^n}=-\sum_{m=0} ^{\infty} \log(1-xy^m).
\end{equation}
Then
\begin{align}
	&-\sum_{m=1}^{\infty} \frac{q^m}{m(1-q^m)}\left[ (\vec{k}_1+\vec{k}_2)^2+\vec{k}_1 \cdot \vec{k}_2 \left( \left(\frac{z_1}{z_2}\right)^m+\left(\frac{z_2}{z_1}\right)^m-2 \right)\right] =  \nonumber \\
	&=\vec{k}_1 \cdot \vec{k}_2 \left(\log\left( 1-\frac{z_1}{z_2}\right)+\log\left( 1-\frac{z_2}{z_1}\right)\right)	+\sum_{m=1} ^{\infty} \log(1-q^m)\left[ (\vec{k}_1 + \vec{k}_2)^2-2\vec{k}_1 \cdot \vec{k}_2\right]+ \nonumber\\
	&+\vec{k}_1 \cdot \vec{k}_2 \sum_{m=1} ^{\infty} \left[\log\left( 1-\frac{q^m z_1 }{z_2 }\right)+\log\left( 1-\frac{q^m z_2 }{z_1}\right)\right].
\end{align}
Let us substitute 
\begin{equation}
    \vec{k}_1 = 2\vec{k} ~,~ \vec{k}_2 = -2\vec{k},
\end{equation}
and take a small chemical potential limit.
We use the property that the Riemann zeta function at zero is
\begin{equation}
	\zeta(0)=\sum_{m=1} ^{\infty} 1=-\frac{1}{2}.
\end{equation}
Also, the following limit $q\to 1^-$ is useful:
\begin{equation}
	\sum_{n=1} ^{\infty}\log(1-q^n) \sim \frac{1}{2}\log(2\pi)-\frac{1}{2}\log(1-q)-\frac{\pi^2}{6(1-q)}~.
\end{equation}
Dividing the correlation function by the partition function gives rise to the following result
\begin{equation}\label{TwoPointFunction6}
	\left\langle	: e^{2i \vec{k} \cdot \vec{f}(z_1)-2i\vec{k} \cdot \vec{f}(z_2) }: \right\rangle \to  e^{-2\mu^2 |\vec{k}|^2  \left[  \log\Big(1-\frac{z_2}{z_1}\Big)+\log\Big(1-\frac{z_1}{z_2}\Big) -\log\Big(\frac{2\pi}{1-q}\Big)+\frac{\pi^2}{3(1-q)}\right]}.
\end{equation}
This can be simplified by neglecting the terms $\log\Big(1-\frac{z_2}{z_1}\Big)+\log\Big(1-\frac{z_1}{z_2}\Big)$ in the exponential, in the limit of $q\to 1^{-}$:
\begin{equation}
 q= e^{-2\pi \tau_2} ~,~ (\tau_1=0)
\end{equation}
\begin{equation}\label{TwoPointFunction7}
	\left\langle	: e^{2i \vec{k} \cdot \vec{f}(z_1)-2i\vec{k} \cdot \vec{f}(z_2) }: \right\rangle \to  e^{-|\vec{k}|^2  E},
\end{equation}
where
\begin{equation}
  E= 2\mu^2 \left[  \frac{\pi}{6\tau_2} +\log(\tau_2)\right].
\end{equation}

\section{Details on the ``rotating Green's function''}
\label{Appendix:DetailsRotatingGreenI}

The purpose of this appendix is to calculate
\begin{equation}\label{GswDefinition2}
	G(s,w,\tau) = \sum_{n,m} \frac{1}{|m\tau - n + w|^{2s}}~,
\end{equation}
and isolate the contribution that arises purely from BPS holomorphic oscillators.
Eq.~(\ref{GswDefinition2}) can be rewritten as
\begin{equation}
	G(s,w,\tau) = \sum_{n,m} \frac{1}{((m\tau_1 - n+\text{Re}(w))^2 + (m\tau_2+\text{Im}(w))^2)^{s}}~.
\end{equation}
For $m=0$ and $\text{Im}(w)=0$ we obtain
\begin{align}
	&\big\{G(s,w,\tau),m=0\big\} = \sum_{n\to -\infty} ^{\infty} \frac{1}{(n-\text{Re}(w))^{2s}}  =\zeta(2s,\text{Re}(w))+\zeta(2s,1-\text{Re}(w))~,
\end{align}
where $\zeta(s,a)$ is the Hurwitz zeta function
\begin{equation}
\label{Hurwitz}
	\zeta(s,a)\equiv \sum_{n=0} ^{\infty} \frac{1}{(n+a)^s}~.
\end{equation}
Some useful identities about this function are written below \rcite{ModernAnalysis}, 
\begin{equation}\label{IdentityZeta1}
	\zeta(0,a)=\frac{1}{2}-a ~,
\end{equation}
\begin{equation}\label{IdentityZeta2}
	\frac{\partial}{\partial s} \zeta(s,a)|_{s=0} = \log(\Gamma(a))-\frac{1}{2}\log(2\pi)~.
\end{equation}
Near $s=1$, the diagamma function $\psi(a)$ plays a role:
\begin{equation}\label{IdentityZeta3}
	\zeta(s,a) = \frac{1}{s-1}-\psi(a) + O(s-1)~.
\end{equation}
The reflection formula is
\begin{equation}
\Gamma(z)\Gamma(1-z)=\frac{\pi}{\sin(\pi z)}~,	
\end{equation}
These identities imply that
\begin{align}
	\Big\{G(s=0,w,\tau),m=0\Big\} &=0
\nn\\[.2cm]
\label{ContributionG'(0)}
	\Big\{\frac{\partial}{\partial s}G(s,w,\tau) \big|_{s=0},m=0\Big\} &= 2\log\left(\frac{1}{2\sin(\pi \text{Re}(w)) }\right)~.
\end{align}
For $s\approx 1$:
\begin{equation}\label{G(1)_1}
	\Big\{G(s\approx 1,w,\tau) ,m=0\Big\} =\zeta(2,\text{Re}(w))+\zeta(2,1-\text{Re}(w))~.
\end{equation}
We next set $m\neq0$ and use the identity 
\begin{equation}\label{Identitya^{-s}}
	\frac{1}{a^s} = \frac{\pi^s}{\Gamma(s)} \int_0 ^{\infty} \frac{dt}{t}t^se^{-\pi a t} ~.  
\end{equation}
 Applying this equation to the sum of $G(s)$ for $m\neq 0$, one obtains
\begin{align}\label{ExponentiatingDenominator}
	&\sum_{m\neq0, n}\frac{1}{\left((m\tau_1 - n+\text{Re}(w))^2 + (m\tau_2 + \text{Im}(w))^2\right)^s} =\nonumber\\[.1cm]
	 &\hskip2cm=\frac{\pi^s}{\Gamma(s)} \sum_{n,m\neq 0} \int_0 ^{\infty} \frac{dt}{t} t^s e^{-\pi  \left((m\tau_1 - n+\text{Re}(w))^2 + (m\tau_2 + \text{Im}(w))^2 \right)t}~.
\end{align}
Let us use the Poisson summation formula to re-express the sum
\begin{equation}\label{PoissonResummation}
	\sum_n e^{-\pi (m\tau_1 - n+\text{Re}(w))^2t} = \frac{1}{\sqrt{t}}\sum_{\tilde{n}} e^{-2\pi i  \tilde{n}(m\tau_1+\text{Re}(w)) -\frac{\pi \tilde{n}^2}{t}}~.
\end{equation}
Plugging eq.~(\ref{PoissonResummation}) into eq.~(\ref{ExponentiatingDenominator}) results in
\begin{equation}\label{GoalForKCLformula}
	\frac{\pi^s}{\Gamma(s)} \sum_{\tilde{n},m\neq 0} \int_0 ^{\infty} \frac{dt}{t} t^{s-\frac{1}{2}} e^{-2\pi i \tilde{n} (m \tau_1+\text{Re}(w)) -\pi   (m\tau_2 + \text{Im}(w))^2 t -\frac{\pi \tilde{n}^2}{t}}~.
\end{equation}
Let us compute the $n=0$ contribution to the sum above using the identity in eq.~(\ref{Identitya^{-s}}) for $s\to s-\frac{1}{2}$:
\begin{equation}
\Big\{G(s,n,w)\;,~\tilde{n}=0\Big\}	= \frac{\Gamma(s-\frac{1}{2})\sqrt{\pi}}{\Gamma(s)}\sum_{m\neq0} \frac{1}{(m\tau_2 + \text{Im}(w))^{2\left(s-\frac{1}{2}\right)}}~.
\end{equation}
Suppose $\text{Im}(w)=0$ and use the fact that the definition of the Riemann zeta function is
\begin{equation}
	\zeta(a) \equiv \sum_{n=1} ^{\infty}\frac{1}{n^a}~.
\end{equation}
Then
\begin{equation}
	\Big\{G(s,n,w)\;,~\tilde{n}=0\Big\}	= 2\frac{\Gamma(s-\frac{1}{2})\sqrt{\pi}}{\Gamma(s)\tau_2 ^{2s-1}}\zeta(2s-1)~.
\end{equation}
Near $s\approx 0$ this approximates to $\frac{\pi \tau_2}{3}$. Therefore,
\begin{equation}\label{G'(0)_2}
	\Big\{G(s=0,w,\tau)\;,~\tilde{n}=0\Big\}	= 0 ~~,~~~~ \Big\{\frac{\partial}{\partial s}G(s,w,\tau)\;,~\tilde{n}=0 \Big\}	=\frac{\pi \tau_2}{3}=-\log\left((q\bar{q})^{\frac{1}{12}}\right)~.
\end{equation}
Near $s= 1$, a few manipulations bring about
\begin{equation}\label{G(1)_2}
	\Big\{G(s\approx 1,w,\tau)\;,~\tilde{n}=0 \Big\} = \frac{\pi}{\tau_2 (s-1)}+\frac{2\pi}{\tau_2}\left(\gamma_E-\log(2\sqrt{\tau_2})\right)+O(s-1)~.
\end{equation}
We now move to compute eq.~(\ref{GoalForKCLformula}) for both $m\neq0$ and $\tilde{n}\neq 0$. To this end, we require the formulas
\begin{equation}\label{Ksab}
	K_s (a,b) \equiv \int_0 ^{\infty} e^{-a^2 t - \frac{b^2}{t}}t^s \frac{dt}{t}=\left(\frac{b}{a}\right)^s K_s (ab)~,
\end{equation}
where
\begin{equation}
	K_s (c) \equiv \int_0 ^{\infty} e^{-c\left(t+\frac{1}{t}\right)}t^s \frac{dt}{t}~.
\end{equation}
The right transition in eq.~(\ref{Ksab}) follows from the latter definition and the change of variables $t\to \frac{b}{a}t$. Also, it can be proven that
\begin{equation}\label{K1/2}
	K_{\frac{1}{2}} (c) = \sqrt{\frac{\pi}{c}} e^{-2c}~.
\end{equation}
Back in eq.~(\ref{GoalForKCLformula}),  the contribution for both $m,\tilde{n}\neq0$ is
\begin{equation}\label{IntermediateKronecker}
	\frac{\pi^s}{\Gamma(s)} \sum_{\tilde{n}\neq0,m\neq 0}   e^{-2\pi i \tilde{n}(m \tau_1+\text{Re}(w))} K_{s-\frac{1}{2}}\left(\sqrt{\pi} |m\tau_2 + \text{Im}(w)|,\sqrt{\pi}|\tilde{n}|\right)~.
\end{equation}
We will first set $s=1$ and then $s=0$. For $s=1$ eq.~(\ref{K1/2}) implies
\begin{equation}
	\pi \sum_{\tilde{n}\neq0,m\neq 0}   e^{-2\pi i \tilde{n}(m \tau_1+\text{Re}(w))} \frac{1}{|m\tau_2 + \text{Im}(w)|}e^{-2\pi |\tilde{n}| |m\tau_2 + \text{Im}(w)|}~.
\end{equation}
Let us compute this for $\text{Im}(w)=0$. The pair of sums over $\{\tilde{n},m>0; \tilde{n},m<0\}$ are equal and similarly the sums over $\{\tilde{n}>0~m<0, \tilde{n}<0 ~ m>0\}$ are equal. Also $q^{\tilde{n}m} = e^{2\pi i\tilde{n}m \tau_1 - 2\pi \tilde{n}m \tau_2}$, therefore we obtain
\begin{equation}
		\frac{2\pi}{\tau_2} \sum_{\tilde{n}=1,m=1} ^{\infty} \frac{1}{m}\left(\bar{q}^{\tilde{n}m}e^{-2\pi i \tilde{n} \text{Re}(w)}+q^{\tilde{n}m}e^{2\pi i \tilde{n} \text{Re}(w)}\right)~.
\end{equation} 
Summing over the index $m$ yields
\begin{equation}\label{G(1)_3}
-\frac{2\pi}{\tau_2} \sum_{\tilde{n}=1} ^{\infty} \left( e^{-2\pi i \tilde{n} \text{Re}(w)}\log(1-\bar{q}^{\tilde{n}})+e^{2\pi i \tilde{n} \text{Re}(w)}\log(1-q^{\tilde{n}})\right)~.
\end{equation}
Summing instead over the index $\tilde{n}$ leads to
\begin{equation}
	\frac{2\pi}{\tau_2} \sum_{m=1} ^{\infty} \left( \frac{\bar{q}^m e^{-2\pi i \text{Re}(w)}}{m(1-\bar{q}^m e^{-2\pi i \text{Re}(w)})}+\frac{q^m e^{2\pi i \text{Re}(w)}}{m(1-q^m e^{2\pi i \text{Re}(w)})}\right)~.
\end{equation}
Let us assemble the result for $G(s\approx 1,w,\tau)$ from eqs. (\ref{G(1)_1}),(\ref{G(1)_2}) and (\ref{G(1)_3}). We obtain
\begin{align}
	&G(s\approx 1,w,\tau)= \zeta(2,\text{Re}(w))+\zeta(2,1-\text{Re}(w)) +\frac{\pi}{\tau_2 (s-1)}+\frac{2\pi}{\tau_2}\left(\gamma - \log(2\sqrt{\tau_2})\right)+O(s-1)+\nonumber\\
	&-\frac{2\pi}{\tau_2}\sum_{\tilde{n}=1} ^{\infty} \left[ e^{-2\pi i \tilde{n} \text{Re}(w)} \log(1-\bar{q}^{\tilde{n}}) + e^{2\pi i \tilde{n} \text{Re}(w)} \log(1-q^{\tilde{n}})\right]~.
\end{align}
We now consider
\begin{align}
	&\frac{\tau_2}{2}\left( G(s\approx1,w,\tau )+G(s\approx 1,-w,\tau)\right)=\tau_2\left( \zeta(2,\text{Re}(w))+\zeta(2,1-\text{Re}(w)) \right)+\frac{\pi}{ s-1}+2\pi(\gamma-\log(2\sqrt{\tau_2})) 
 \nonumber\\
	&-\pi \sum_{\tilde{n}=1} ^{\infty} \left[ e^{-2\pi i \tilde{n} \text{Re}(w)} \log(1-q^{\tilde{n}}) + e^{2\pi i \tilde{n} \text{Re}(w)} \log(1-q^{\tilde{n}})+ e^{-2\pi i \tilde{n} \text{Re}(w)} \log(1-\bar{q}^{\tilde{n}}) + e^{2\pi i \tilde{n} \text{Re}(w)} \log(1-\bar{q}^{\tilde{n}})\right]~.
\end{align}
Next, we set $s=0~,~ \text{Im}(w)=0$ in eq.~(\ref{IntermediateKronecker}). Eqs. (\ref{Ksab}) and (\ref{K1/2}) allow one to deduce that the $n,m\neq0$ contributions to $G(s=0,\tau,w)$ sum to
\begin{align}\label{G'(0)_3}
	&\frac{1}{\Gamma(s)} \sum_{\tilde{n}\neq0,m\neq 0}   e^{-2\pi i \tilde{n}(m \tau_1+\text{Re}(w))} \frac{1}{|\tilde{n}|} e^{-2\pi |\tilde{n}| |m| \tau_2}=\frac{2}{\Gamma(s)}  \sum_{\tilde{n},m=1} ^{\infty} \frac{1}{\tilde{n}}\left( q^{\tilde{n}m} e^{2\pi i\tilde{n} \text{Re}(w)}+\bar{q}^{\tilde{n}m} e^{-2\pi i\tilde{n}\text{Re}(w)}\right)
 \nonumber\\
	&\hskip3cm =-\frac{2}{\Gamma(s)} \sum_{m=1} ^{\infty} \left[ \log\left(1-q^m e^{2\pi i \text{Re}(w)}\right)+\log\left(1-\bar{q}^m e^{-2\pi i \text{Re}(w)}\right)\right]~. 
\end{align}
Recall that $\frac{1}{\Gamma(s)}=s+O(s^2)$.
For $s\to0$ if one assembles the result for $G(s\approx 0,w,\tau)$, then
\begin{equation}\label{G0wtau}
	G(0,w,\tau)=0~.
\end{equation}
This equation has an important implication for the modular invariance of the partition function.  This function is copied again below:
\begin{equation}
	Z = \exp\left(\frac{1}{2}\frac{\partial}{\partial p} {\sum_{n,m}}\,'
	\frac{(4\pi^2 \mu^2)^p}{\left[\big|\frac{m-n\tau}{\tau_2}+\omega_{12}\big|^2\right] ^{p} }~\Bigg|_{p=0}+(\omega_{12}\to -\omega_{12})\right)~.
\end{equation}
The numerator can be multiplied by $\frac{1}{\tau_2 ^s}$ without affecting the result, because eq.~(\ref{G0wtau}) guarantees independence on the presence of overall multiplicative factors. 

From eqs. (\ref{ContributionG'(0)}),(\ref{G'(0)_2}) and (\ref{G'(0)_3}), one obtains
\begin{equation}
	\frac{\partial}{\partial s} \left(G(s,w,\tau)+G(s,-w,\tau)\right) |_{s=0} = -2\log\left(\left|\frac{\theta_1 (\text{Re}(w)|\tau)}{\eta(\tau)}\right|^2\right)~.
\end{equation}
Then the partition function reads
\begin{equation}
	Z(\tau,\omega_{12}) = e^{\frac{1}{2}(-2)\log\left(\left|\frac{\theta_1 (\text{Re}(w)|\tau)}{\eta(\tau)}\right|^2\right)}= \left|\frac{\eta(\tau)}{\theta_1(\text{Re}(w)|\tau)} \right|^2~.
\end{equation}
Considering only the holomorphic part of the $\tilde{n},m\neq0$ modes, one obtains
\begin{equation}
	Z(\tau,\omega_{12}) =  \frac{2\sin(\pi \text{Re}(w))\eta(\tau)}{\theta_1(\text{Re}(w)|\tau)}~.
\end{equation}

\section{Two-point function of exponentials - fixed angular potential}
\label{app:rot details}

 This section concerns with the two-point function of exponentials, which enables one to compute expectation values of forms and harmonic functions. 
This quantity admits a path integral expression
\begin{equation}
	\langle e^{ik_1 \cdot X(z_1) + ik_2 \cdot X(z_2)}\rangle = \int DX e^{ik_1 \cdot X(z_1) + ik_2 \cdot X(z_2)} e^{-\frac{1}{4\pi \mu^2} \int d^2 \sigma \left( \partial_a X_j \partial^a X^j +\omega ^2 X_j ^2-i\omega L_{12} \right)}~.
\end{equation}
We focus on a 2D target-space; it is straightforward to generalize the result below for a 4D target-space with two independent angular potentials associated with orthogonal planes. One repeats the steps in eqs. (\ref{XiSum}),(\ref{EigenfunctionsDef}),(\ref{EigenfunctionsExp}),(\ref{Orthonormality}), (\ref{InverseTransformation}) and defines
\begin{equation}
k_{xy}\equiv \frac{k_x + ik_y}{\sqrt{2}}\Rightarrow	k_x = \frac{ k_{xy}+k_{xy}^*}{\sqrt{2}}~,~ k_y = \frac{k_{xy}-k_{xy}^*}{\sqrt{2}i}~.
\end{equation}
It is useful to compute
\begin{align}
	&\vec{k}_1\cdot \vec{X}_{n,m} \Phi_{n,m}(\sigma_1,\sigma_2) + \vec{k}_2 \cdot \vec{X}_{n,m} \Phi_{n,m}(\sigma_1',\sigma_2') =\nonumber\\
	&\hskip2cm= \left(k_{1xy} ^* X_{n,m} ^+ +k_{1xy} X_{n,m} ^-\right) \Phi_{n,m}(\sigma_1,\sigma_2)+\left(k_{2xy} ^* X_{n,m} ^+ +k_{2xy} X_{n,m} ^-\right) \Phi_{n,m}(\sigma_1',\sigma_2')~. \nonumber
\end{align}
Denote
\begin{equation}
	\omega_{\pm,n,m} ^2 = \left| \frac{n\tau-m}{\tau_2}\pm \omega\right|^2~.
\end{equation}
Then the two-point function becomes
\begin{align}
	\left\langle e^{ik_1 \cdot X(z_1) + ik_2 \cdot X(z_2)}\right\rangle &= \prod_{n,m} \int dX^+ _{n,m} d X^- _{n,m} e^{i\left(k_{1xy} ^* X_{n,m} ^+ +k_{1xy} X_{n,m} ^-\right) \Phi_{n,m}(\sigma_1,\sigma_2)}  \nonumber\\
	&\hskip.5cm 
 \times e^{i\left(k_{2xy} ^* X_{n,m} ^+ +k_{2xy} X_{n,m} ^-\right) \Phi_{n,m}(\sigma_1',\sigma_2')} e^{- \frac{\omega_{-,n,m}^2}{4\pi \mu^2} X^+ _{n,m} X^- _{-n,-m}-\frac{\omega_{+,n,m}^2}{4\pi \mu^2} X^- _{n,m} X^+ _{-n,-m}}~.
\end{align}
We perform the following change of variables:
\begin{align}
	\tilde{X}_{n,m} ^+ &= e^{i\text{arg}\left(k_{1xy}^* \Phi_{n,m}(\sigma_1,\sigma_2)+k_{2xy}^* \Phi_{n,m}(\sigma_1',\sigma_2')\right)}X_{n,m} ^+ 
\nn\\[.2cm]
	\tilde{X}_{n,m} ^- &= e^{i\text{arg}\left(k_{1xy} \Phi_{n,m}(\sigma_1,\sigma_2)+k_{2xy} \Phi_{n,m}(\sigma_1',\sigma_2')\right)}X_{n,m} ^- ~. 
\end{align}
Now the contour of the integrals is the real line. 
Using $|\Phi_{n,m}(\sigma_1,\sigma_2)|^2 = \frac{1}{4\pi^2 \tau_2}$, $|k_{1xy}|^2 = \frac{1}{2}\left(k_{1x} ^2 + k_{1y}^2\right)$ and dividing by the partition function, it follows that
\begin{align}
	\left\langle e^{ik_1 \cdot X(z_1) + ik_2 \cdot X(z_2)}\right\rangle &= e^{-\frac{1}{2}(|\vec{k}_1|^2+|\vec{k}_2|^2)\frac{\mu^2\tau_2}{4\pi}\sum_{n,m}\left(\frac{1 }{|n\tau-m+\omega \tau_2 |^2}+\frac{1 }{|n\tau-m-\omega \tau_2 |^2}\right)} 
 \\
	&\hskip.5cm\times e^{-\sum_{n,m}\frac{\pi\mu^2}{\omega_{-,n,m} ^2} \left(k_{1xy}k_{2xy} ^* \Phi_{n,m} ^* (\sigma_1,\sigma_2) \Phi_{n,m}  (\sigma_1',\sigma_2')+k_{1xy}^*k_{2xy}  \Phi_{n,m} (\sigma_1,\sigma_2) \Phi_{n,m}^*  (\sigma_1',\sigma_2') \right)}  \nonumber\\
	&\hskip.5cm\times e^{-\sum_{n,m}\frac{\pi\mu^2}{\omega_{+,n,m} ^2} \left(k_{1xy}^*k_{2xy}  \Phi_{n,m} ^* (\sigma_1,\sigma_2) \Phi_{n,m}  (\sigma_1',\sigma_2')+k_{1xy}k_{2xy}^*  \Phi_{n,m} (\sigma_1,\sigma_2) \Phi_{n,m}^*  (\sigma_1',\sigma_2') \right)}~.\nonumber
\end{align}
The two-point function is equal to
\begin{align}\label{LongTwoPointFunction}
	\left\langle e^{ik_1 \cdot X(z_1) + ik_2 \cdot X(z_2)}\right\rangle &= e^{-\frac{1}{2}(|\vec{k}_1|^2+|\vec{k}_2|^2)\frac{\mu^2\tau_2}{4\pi}\sum_{n,m}\left(\frac{1 }{|n\tau-m+\omega \tau_2 |^2}+\frac{1 }{|n\tau-m-\omega \tau_2 |^2}\right)} \nonumber\\
	&\hskip.5cm
 e^{- k_1 \cdot k_2~\frac{\mu^2 \tau_2}{8\pi}\sum_{n,m}\left[ \frac{e^{in(\sigma_1 - \sigma_1')+i(m-n\tau_1) (\tilde{\sigma}_2-\tilde{\sigma_2}')}}{|n\tau-m-\omega \tau_2|^2}+  \frac{e^{in(\sigma_1 - \sigma_1')+i(m-n\tau_1) (\tilde{\sigma}_2-\tilde{\sigma_2}')}}{|n\tau-m+\omega \tau_2|^2}\right]} \nonumber\\
	&\hskip.5cm 
 e^{- k_1 \cdot k_2~\frac{\mu^2 \tau_2}{8\pi}\sum_{n,m} \left[\frac{e^{in(\sigma_1' - \sigma_1)+i(m-n\tau_1) (\tilde{\sigma}_2'-\tilde{\sigma_2})}}{|n\tau-m-\omega \tau_2|^2} + \frac{e^{in(\sigma_1' - \sigma_1)+i(m-n\tau_1) (\tilde{\sigma}_2'-\tilde{\sigma_2})}}{|n\tau-m+\omega \tau_2|^2}\right] }  \\
	&\hskip.5cm
 e^{i(k_{1y} k_{2x} - k_{1x} k_{2y}) \frac{\mu^2 \tau_2}{8\pi}\sum_{n,m} \frac{e^{in(\sigma_1 - \sigma_1')+i(m-n\tau_1) (\tilde{\sigma}_2-\tilde{\sigma_2}')}-e^{in(\sigma_1' - \sigma_1)+i(m-n\tau_1) (\tilde{\sigma}_2'-\tilde{\sigma_2})}}{|n\tau-m-\omega \tau_2|^2}} \nonumber\\
	&\hskip.5cm
 e^{-i(k_{1y} k_{2x} - k_{1x} k_{2y}) \frac{\mu^2 \tau_2}{8\pi}\sum_{n,m} \frac{e^{in(\sigma_1 - \sigma_1')+i(m-n\tau_1) (\tilde{\sigma}_2-\tilde{\sigma_2}')}-e^{in(\sigma_1' - \sigma_1)+i(m-n\tau_1) (\tilde{\sigma}_2'-\tilde{\sigma}_2)}}{|n\tau-m+\omega \tau_2|^2}}~.\nonumber
\end{align}
Here, 
\begin{equation}
\tilde{\sigma}_{1,2} = \frac{\sigma_{1,2}}{\tau_2}~.	
\end{equation}
We want to calculate
\begin{equation}
	K(\sigma_1,\sigma_2,s,\tau,w)\equiv	\sum_{n,m} \frac{e^{in\sigma_1 + i (m-n\tau_1)\tilde{\sigma}_2}}{\left|n\tau-m-w\right|^{2s}}~,
\end{equation}
in the vicinity of $s=1$.
This equation can be written as
\begin{equation}
K(\sigma_1,\sigma_2,s,\tau,w)=	\sum_{n,m} \frac{e^{in\sigma_1 + i (m-n\tau_1)\tilde{\sigma}_2}}{\left[(n\tau_2 - \text{Im}(w))^2 + (n\tau_1 -m- \text{Re}(w))^2\right]^s}~.
\end{equation}
We first write whether this function exhibits periodicity in the worldsheet coordinates and the $w$-variable. Second, we show that this function times $\tau_2 ^s$ is invariant under modular transformations. Third, we present the result of the calculation prior to doing it.

By construction, $K$ is doubly-periodic in the worldsheet coordinates - it is built from the eigenfunctions of the Laplacian operator which respect these periodic boundary conditions.
\begin{align}
	K(\sigma_1+2\pi,\sigma_2,s,\tau,w) &=	K(\sigma_1,\sigma_2,s,\tau,w)~,~\nonumber\\[.2cm]
 	K(\sigma_1+2\pi\tau_1,\sigma_2+2\pi \tau_2,s,\tau,w) &=	K(\sigma_1,\sigma_2,s,\tau,w)~.
\end{align}
 By taking $w\to w+1$ and defining $m'\equiv m+1$,
\begin{equation}
	K(\sigma_1,\sigma_2,s,\tau,w+1) = \sum_{n,m'} \frac{e^{in\sigma_1 + i (m'-1-n\tau_1)\tilde{\sigma}_2}}{\left|n\tau-m'-w\right|^{2s}}=e^{-i\tilde{\sigma}_2} K(\sigma_1,\sigma_2,s,\tau,w)~.
\end{equation} 
Similarly, for $w\to w+\tau$ and defining $n'\equiv n-1$, 
\begin{equation}
	K(\sigma_1,\sigma_2,s,\tau,w+\tau) = \sum_{n',m} \frac{e^{i(n'+1)\sigma_1 + i (m-(n'+1)\tau_1)\tilde{\sigma}_2}}{\left|n'\tau-m-w\right|^{2s}}=e^{i\sigma_1-i\tau_1 \tilde{\sigma}_2} K(\sigma_1,\sigma_2,s,\tau,w)~.
\end{equation} 
Next, when the T-transformation applies $\tau\to \tau + 1$ then
\begin{equation}
	K(\sigma_1,\sigma_2,s,\tau+1,w)=	\sum_{n,m} \frac{e^{in\sigma_1 + i(m-n\tau_1 -n)\tilde{\sigma}_2}}{|n\tau+n-m-w|^{2s}}=	K(\sigma_1,\sigma_2,s,\tau,w)~.
\end{equation}
The reason for the last transition is the legitimacy of changing the dummy index from $m$ to $m'\equiv m-n$.\\
The S-transformation $\tau\to -\frac{1}{\tau}$ is slightly more complicated because one should also take 
\begin{equation}
\sigma_1+i\sigma_2 \to \frac{\sigma_1+i\sigma_2}{\tau} ~\text{and}~ w\to \frac{w}{\tau}~.	
\end{equation}
It follows that
\begin{equation}
\tau_1 \to \tau_1'\equiv -\frac{\tau_1}{|\tau|^2}~,~ \tau_2 \to \tau_2'\equiv  \frac{\tau_2}{|\tau|^2}	
\end{equation}
and
\begin{equation}
\sigma_1 \to \sigma_1'\equiv \frac{\sigma_1\tau_1 + \sigma_2\tau_2}{|\tau|^2}~,~ \sigma_2 \to \sigma_2 '\equiv \frac{\sigma_2 \tau_1 -\sigma_1 \tau_2}{|\tau|^2}~.	
\end{equation}
It can be checked that
\begin{equation}
	in\sigma_1' + i(m-n\tau_1')\frac{\sigma_2'}{\tau_2 '} = -im\sigma_1 + i(n+m\tau_1)\frac{\sigma_2}{\tau_2}~. 
\end{equation}
The function $K$ is also multiplied by $\tau_2 ^s$ at this stage. This factor naturally arises for $s=1$ in the Green's function as computed in the path integral formulation.
\begin{align}
	&(\tau_2') ^s K\left(\sigma_1',\sigma_2',s,-\frac{1}{\tau},\frac{w}{\tau}\right) =\sum_{n,m} \frac{\frac{\tau_2 ^s}{|\tau|^{2s}}e^{-im\sigma_1+i(n+m\tau_1 )\tilde{\sigma_2}}}{|-\frac{n}{\tau}-m-w|^{2s}}==\sum_{n,m} \frac{\tau_2 ^s e^{i\left(in\sigma_1 + i(m-n\tau_1)\frac{\sigma_2}{\tau_2}\right)}}{\left|n\tau-m -w\right|^{2s}}~.
\end{align} 
In the last transition we changed $m\to-n,n\to m$. Consequently,
\begin{equation}
		(\tau_2') ^s K\left(\sigma_1',\sigma_2',s,-\frac{1}{\tau},\frac{w}{\tau}\right) =	\tau_2 ^s K\left(\sigma_1,\sigma_2,s,\tau,w\right)~.
\end{equation}
We conclude that the function $\tau_2 ^sK\left(\sigma_1,\sigma_2,s,\tau,w\right) $ is modular invariant as well as elliptic in $\sigma_1,\sigma_2$. Nonetheless, it is not strictly elliptic in $w$.\\
Next, we compute
\begin{equation}
	K(\sigma_1,\sigma_2,s,\tau,w)\equiv	\sum_{n,m} \frac{e^{in\sigma_1 + i (m-n\tau_1)\tilde{\sigma}_2}}{\left|n\tau-m-w\right|^{2s}}~.
\end{equation}
The result is that $K(s=1)$ contains terms of the following form:
\begin{align}\label{ResultK12}
	&		
 \frac{\pi}{\tau_2}e^{-i\text{Re}{w}(\tilde{\sigma}_2-\tilde{\sigma}_2 ')}\sum_{n=1} ^{\infty} \frac{1}{n}\left[ \frac{q^n e^{-2\pi i \text{Re}(w)}}{1-q^n e^{-2\pi i \text{Re}(w)}}\left(\frac{z}{z'}\right)^n +\frac{q^n e^{2\pi i \text{Re}(w)}}{1-q^n e^{2\pi i \text{Re}(w)}} \left( \frac{z'}{z}\right)^n\right]~,\nonumber	
\end{align}
where\footnote{The notation $z$ here differs from the notation $z=e^{i\frac{2\pi v}{L}}$ written in the main text.}
\begin{equation}
	z= e^{i(\sigma_1 + i\sigma_2)}~.
\end{equation}
The first contribution we calculate is $n=0,\text{Im}(w)=0$. 
\begin{align}
	\sum_{m\to -\infty} ^{\infty} \frac{e^{im\tilde{\sigma}_2}}{(m+\text{Re}(w))^{2s}} &= \sum_{m\to -\infty} ^{-1} \frac{e^{im\tilde{\sigma}_2}}{(m+\text{Re}(w))^{2s}}+\sum_{m=0} ^{\infty} \frac{e^{im\tilde{\sigma}_2}}{(m+\text{Re}(w))^{2s}} \nonumber\\[.2cm]
	&=\sum_{l=0} ^{\infty} \frac{e^{i(-l-1)\tilde{\sigma}_2}}{(l+1-\text{Re}(w))^{2s}}+\sum_{m=0} ^{\infty} \frac{e^{im\tilde{\sigma}_2}}{(m+\text{Re}(w))^{2s}}~.
\end{align}
Now, the Lerch zeta function is defined in
\begin{equation}
	L(\lambda,s,\alpha)\equiv \sum_{m=0} ^{\infty}\frac{ e^{2\pi i m\lambda}}{(m+\alpha)^s}~.
\end{equation}
A related function is the ``Lerch transcendent'', defined in
\begin{equation}
	\Phi(z,s,a)\equiv \sum_{n=0} ^{\infty} \frac{z^n}{\left(n+a\right)^s}~.
\end{equation}
This allows one to extract a contribution to the Green's function:
\begin{equation}
	\{K(\sigma_1,\sigma_2,\tau,w)~,~n=0\} = L\left( \frac{\tilde{\sigma}_2}{2\pi},2s,\text{Re}(w)\right)+ e^{-i\tilde{\sigma}_2} L\left( -\frac{\tilde{\sigma}_2}{2\pi},2s,1-\text{Re}(w)\right)~.
\end{equation}
The Lerch transcendent satisfies the following identity, which allows for the permutation of the arguments of the function:
\begin{equation}
	\Phi\left( e^{2\pi i x},1-s,a\right) = \frac{\Gamma(s)}{(2\pi)^s}\left( e^{i\pi \left( \frac{s}{2}-2ax\right)}\Phi\left(e^{-2\pi i a },s,x\right)+e^{i\pi \left(2a(1-x)-\frac{s}{2}\right)}\Phi\left( e^{2\pi i a},s,1-x\right)\right)~.
\end{equation}
One can set $x=\text{Re}(w),s\to2s,a= -\frac{\tilde{\sigma}_2}{2\pi}$ and obtain
\begin{align}
	&\frac{(2\pi)^{2s}}{\Gamma(2s)}L\left(\text{Re}(w),1-2s,-\frac{\tilde{\sigma}_2}{2\pi}\right)=e^{i\tilde{\sigma}_2 \text{Re}(w)}=\left(e^{i\pi s}L\left( \frac{\tilde{\sigma}_2}{2\pi},2s,\text{Re}(w)\right)+e^{-i \pi s-i\tilde{\sigma}_2}L\left(-\frac{\tilde{\sigma}_2}{2\pi},2s,1-\text{Re}(w)\right)\right)~.
\end{align}
Therefore,
\begin{align}
	\{K(\sigma_1,\sigma_2,\tau,w)~,~n=0\} &= -4\pi^2e^{-i\tilde{\sigma}_2 \text{Re}(w)} L\left( \text{Re}(w),-1,-\frac{\tilde{\sigma}_2}{2\pi} \right)
 \nonumber\\
 &=-4\pi^2e^{-i\tilde{\sigma}_2 \text{Re}(w)} \left(\frac{e^{2\pi i \text{Re}(w)}}{\left(1-e^{2\pi i \text{Re}(w)}\right)^2}-\frac{\frac{\tilde{\sigma}_2}{2\pi}}{1-e^{2\pi i \text{Re}(w)}}\right)~.
\end{align}
Below we consider $n\neq0$ and apply the integral representation of the power function
\begin{equation}
	\frac{1}{a^s} = \frac{\pi^s}{\Gamma(s)} \int_0 ^{\infty} \frac{dt}{t}t^se^{-\pi a t}~,
\end{equation}
 in
\begin{align}
	&K(\sigma_1,\sigma_2,\tau,w) = \frac{\pi^s}{\Gamma(s)} \sum_{n,m} e^{i n\sigma_1 + i (m-n\tau_1)\tilde{\sigma}_2} \int_0 ^{\infty} \frac{dt}{t}t^s e^{-\pi t \left[ (-n\tau_2 + \text{Im}(w))^2 + (m-n\tau_1 + \text{Re}(w))^2\right]}~.
\end{align}
Note that the total argument in the exponential can be written as
\begin{align}
	&i n\sigma_1 + i (m-n\tau_1)\tilde{\sigma}_2-\pi t \left[ (-n\tau_2 + \text{Im}(w))^2 + (m-n\tau_1 + \text{Re}(w))^2\right] =
 \nonumber\\
	\hskip1cm &= -\pi t\left[ m-n\tau_1 + \text{Re}(w) - \frac{i \tilde{\sigma}_2}{2\pi t}\right]^2 -\pi t \left[ -n\tau_2 +\text{Im}(w) +\frac{i\tilde{\sigma}_1}{2\pi t}\right]^2 -\frac{\tilde{\sigma}_1 ^2 +\tilde{\sigma}_2 ^2 }{4\pi t} - i\text{Re}(w) \tilde{\sigma}_2 + i \text{Im} (w) \tilde{\sigma}_1 ~.
\end{align}
We next use the Poisson summation formula for the index $m$:
\begin{equation}
	\sum_m e^{-\pi t \left[ m-n\tau_1 + \text{Re}(w)-\frac{i \tilde{\sigma}_2}{2\pi t}\right]^2} = \frac{1}{\sqrt{t}} \sum_{\tilde{m}} e^{-\frac{\pi \tilde{m}^2}{t}+2\pi i \tilde{m} \left( n\tau_1 - \text{Re}(w)+\frac{i \tilde{\sigma}_2}{2\pi t}\right)} ~.
\end{equation}
It follows that
\begin{align}
	&K(\sigma_1,\sigma_2,\tau,w) =e^{i\left(- \text{Re}(w) \tilde{\sigma}_2 + \text{Im}(w) \tilde{\sigma}_1\right)-\frac{\tilde{\sigma}_1 ^2 + \tilde{\sigma}_2 ^2}{4\pi t}} \frac{\pi^s}{\Gamma(s)} \sum_{n,m}  \int_0 ^{\infty} \frac{dt}{t}t^{s-\frac{1}{2}}\nonumber\\
	&\times e^{-\pi t \left(-n\tau_2 + \text{Im}(w) + \frac{i\tilde{\sigma}_1}{2\pi t}\right)^2 }e^{-\frac{\pi \tilde{m}^2}{t}+2\pi i \tilde{m} \left( n\tau_1 - \text{Re}(w)+\frac{i \tilde{\sigma}_2}{2\pi t}\right)} ~.
\end{align}
In the above argument of the exponential, the quadratic term proportional to $\sigma_1^2$  cancels.  
Then
\begin{align}\label{GreenIntermediate}
	K(\sigma_1,\sigma_2,\tau,w) &= 
 e^{i\left(- \text{Re}(w) \tilde{\sigma}_2 + \text{Im}(w) \tilde{\sigma}_1\right)}\frac{\pi^s}{\Gamma(s)} \sum_{n,m}  e^{-i\tilde{\sigma}_1 \left(-n\tau_2 + \text{Im}(w)\right)} 
 \nonumber\\
	&\hskip1cm\times \int_0 ^{\infty} \frac{dt}{t}t^{s-\frac{1}{2}} e^{-\pi t (-n\tau_2 + \text{Im}(w))^2}e^{-\frac{\pi \left(\tilde{m}+\frac{\tilde{\sigma}_2}{2\pi}\right)^2}{t}+2\pi i \tilde{m} \left( n\tau_1 - \text{Re}(w)\right)} ~.
\end{align} 
We require the following integral formula 
\begin{equation}\label{K1/2same}
\int_0 ^{\infty} \frac{dt}{\sqrt{t}} e^{-a^2 t -\frac{b^2}{t}} = \frac{\sqrt{\pi}}{a}e^{-2ab} ~.
\end{equation}
In general, we will take $a=\sqrt{\pi}|-n\tau_2 +\text{Im}(w)|$ and $b=\sqrt{\pi} |\tilde{m} + \frac{\tilde{\sigma}_2}{2\pi}|$. However, we start by setting $s=1,\tilde{m}=0,\text{Im}(w)=0$. The formulas above with $a=\sqrt{\pi}|n|\tau_2$ and $b=\frac{|\tilde{\sigma}_2|}{2\sqrt{\pi}}$ imply 
\begin{align}
	\Big\{K(\sigma_1,\sigma_2,\tau,\omega) ~,~ \tilde{m}=0\Big\} &=\frac{\pi}{\tau_2} e^{-i\text{Re}(w) \tilde{\sigma}_2} \sum_{n\neq0} \frac{1}{|n|} e^{-|n|\tau_2 |\tilde{\sigma}_2| + i \tilde{\sigma}_1 n \tau_2 }\nonumber\\
	&=-	\frac{\pi}{\tau_2} e^{-i\text{Re}(w) \tilde{\sigma}_2}\left[ \log\left(1-e^{-i(\sigma_1 -i \sigma_2)}\right)+\log\left( 1- e^{i(\sigma_1 +i \sigma_2)}\right)\right] ~.
\end{align}
This displays a logarithmic singularity when $\sigma_1,\sigma_2 \to0$.\\
The third contribution comes from both $n\neq0,\tilde{m}\neq0$. In this case, for $s=1,\text{Im}(w)=0$ eqs. (\ref{GreenIntermediate}) and (\ref{K1/2same}) lead to 
\begin{align}
	&\frac{\pi}{\tau_2} e^{-i\text{Re}{w}\tilde{\sigma}_2}\sum_{n\neq0, \tilde{m}\neq0} \frac{1}{|n|}e^{-2\pi \tau_2 |n| |\tilde{m}+\frac{\tilde{\sigma}_2}{2\pi}|+in\sigma_1 +2\pi i n\tilde{m} \tau_1 - 2\pi i \tilde{m} \text{Re}(w)}
 \nonumber\\
	&\hskip1cm =\frac{\pi}{\tau_2} e^{-i\text{Re}{w}\tilde{\sigma}_2}\sum_{n=1} ^{\infty} \sum_{\tilde{m}\neq0} \frac{1}{n} e^{-2\pi \tau_2 n |\tilde{m} + \frac{\tilde{\sigma}_2}{2\pi}|-2\pi i \tilde{m}\text{Re}(w)} \left( e^{in\sigma_1 + 2\pi i n \tilde{m}\tau_1}+e^{-in\sigma_1 - 2\pi i n \tilde{m}\tau_1}\right)
 \nonumber\\[.2cm]
	&\hskip1cm  =\frac{\pi}{\tau_2} e^{-i\text{Re}{w}\tilde{\sigma}_2}\sum_{n=1} ^{\infty}  \sum_{\tilde{m}=1} ^{\infty}\frac{1}{n} \left[ e^{-2\pi \tau_2 n(\tilde{m}+\frac{\tilde{\sigma}_2}{2\pi})-2\pi i \tilde{m}\text{Re}(w)} \left( e^{in\sigma_1 + 2\pi i n \tilde{m}\tau_1}+e^{-in\sigma_1 - 2\pi i n \tilde{m}\tau_1}\right)\right.\nonumber\\ 
	&\hskip2cm \left. + e^{-2\pi \tau_2 n(\tilde{m}-\frac{\tilde{\sigma}_2}{2\pi})+2\pi i \tilde{m}\text{Re}(w)} \left( e^{in\sigma_1 - 2\pi i n \tilde{m}\tau_1}+e^{-in\sigma_1 + 2\pi i n \tilde{m}\tau_1}\right) \right]
 \nonumber\\[.2cm]
	&\hskip1cm =\frac{\pi}{\tau_2} e^{-i\text{Re}{w}\tilde{\sigma}_2}\sum_{n=1} ^{\infty} \sum_{\tilde{m}=1} ^{\infty}  \frac{1}{n} e^{-2\pi i \text{Re}(w)\tilde{m}}\left[q^{n\tilde{m}} e^{in(\sigma_1 + i\sigma_2)}+\bar{q} ^{n\tilde{m}} e^{-in (\sigma_1 - i\sigma_2)}\right]
 \nonumber\\
	&\hskip2cm + \sum_{n=1} ^{\infty} \sum_{\tilde{m}=1} ^{\infty}  \frac{1}{n} e^{2\pi i \text{Re}(w)\tilde{m}}\left[q^{n\tilde{m}} e^{-in(\sigma_1 + i\sigma_2)}+\bar{q} ^{n\tilde{m}} e^{in (\sigma_1 - i\sigma_2)}\right] ~.
\end{align}
Summing over $\tilde{m}$ yields
\begin{align}
	&e^{-i\text{Re}{w}\tilde{\sigma}_2}\frac{\pi}{\tau_2}\sum_{n=1} ^{\infty} \frac{1}{n}\left[ \frac{q^n e^{-2\pi i \text{Re}(w)+in(\sigma_1+i\sigma_2)}}{1-q^n e^{-2\pi i \text{Re}(w)}} +\frac{\bar{q}^n e^{-2\pi i \text{Re}(w)-in(\sigma_1-i\sigma_2)}}{1-\bar{q}^n e^{-2\pi i \text{Re}(w)}}+\right. \nonumber\\
	&\hskip4cm\left.  +\frac{q^n e^{2\pi i \text{Re}(w)-in(\sigma_1+i\sigma_2)}}{1-q^n e^{2\pi i \text{Re}(w)}} +\frac{\bar{q}^n e^{2\pi i \text{Re}(w)+in(\sigma_1-i\sigma_2)}}{1-\bar{q}^n e^{2\pi i \text{Re}(w)}}\right] ~.
\end{align}
In the above derivation we can replace $\sigma_1 \to \sigma_1 - \sigma_1',\sigma_2\to \sigma_2-\sigma_2 '$ and define
\begin{equation}
	z= e^{i(\sigma_1+ i \sigma_2)} ~~,~~~~ z'  = e^{i(\sigma_1'+i\sigma_2')}~.
\end{equation}
Then the holomorphic part of the expression for the third contribution above reads
\begin{equation}
	\frac{\pi}{\tau_2}e^{-i\text{Re}{w}(\tilde{\sigma}_2-\tilde{\sigma}_2')}\sum_{n=1} ^{\infty} \frac{1}{n}\left[ \frac{q^n e^{-2\pi i \text{Re}(w)}}{1-q^n e^{-2\pi i \text{Re}(w)}}\left(\frac{z}{z'}\right)^n +\frac{q^n e^{2\pi i \text{Re}(w)}}{1-q^n e^{2\pi i \text{Re}(w)}} \left( \frac{z'}{z}\right)^n\right]~.
\end{equation}
The conclusion is that the holomorphic oscillators contribution to the Green's function is 
\begin{align} 
 \frac{\pi}{\tau_2}\sum_{n=1} ^{\infty} \frac{1}{n}\left[ \frac{q^n e^{-2\pi i \text{Re}(w)}}{1-q^n e^{-2\pi i \text{Re}(w)}}z^n +\frac{q^n e^{2\pi i \text{Re}(w)}}{1-q^n e^{2\pi i \text{Re}(w)}} z^{-n}\right]~.
\end{align} 
The two-point function is proportional to a purely holomorphic factor relevant for the physics of half-BPS states:
 \begin{align}\label{Rotating2PointFunctionApp}
 	&\left\langle e^{ik_1 \cdot X(z_1) + ik_2 \cdot X(z_2)}\right\rangle \propto e^{-\frac{1}{2}(|\vec{k}_1|^2+|\vec{k}_2|^2)G^r(0,\omega,\tau)-\vec{k}_1 \cdot \vec{k}_2 G(z_1,z_2,\omega,\tau)-i\vec{k}_1 \wedge \vec{k}_2 \mathcal{G}(z_1,z_2,\omega,\tau)}~,\nonumber
 \end{align}
where
\begin{align}
	G^r (0,\omega,\tau)&\equiv \frac{\mu^2}{2}\sum_{n=1} ^{\infty} \frac{1}{n} \left( \frac{q^n e^{2\pi i\omega }}{1-q^n e^{2\pi i \omega}}+\frac{q^n e^{-2\pi i\omega }}{1-q^n e^{-2\pi i \omega}}\right)
\\[.2cm]
	G(z_1,z_2,\omega,\tau) &= \frac{\mu^2}{4}\sum_{n=1} ^{\infty} \frac{1}{n} \left( \frac{q^n e^{2\pi i\omega }}{1-q^n e^{2\pi i \omega}}+\frac{q^n e^{-2\pi i\omega }}{1-q^n e^{-2\pi i \omega}}\right)\left( \left( \frac{z_1}{z_2}\right)^n + \left(\frac{z_2}{z_1}\right)^n\right)~.
\end{align}
This equation reduces to the previous one for equal points on the torus $z_1=z_2$. 
\begin{equation}
	\vec{k}_1 \wedge \vec{k}_2 \equiv k_{1x}k_{2y} - k_{1y} k_{2x}~.
\end{equation}
When $\vec{k}_1 || \vec{k}_2$ this vanishes, in particular for $\vec{k}_1 = \pm\vec{k}_2$.
\begin{equation}
	\mathcal{G}(z_1,z_2,\omega,\tau)=\frac{\mu^2}{4}\sum_{n=1} ^{\infty} \frac{1}{n}\left( \frac{q^n e^{2\pi i \omega}}{1-q^n e^{2\pi i \omega}}-\frac{q^n e^{-2\pi i \omega}}{1-q^n e^{-2\pi i \omega}}\right)\left[ \left( \frac{z_2}{z_1}\right)^n - \left( \frac{z_1}{z_2}\right)^n\right]~.
\end{equation}

\newpage

\bibliographystyle{JHEP}      

\bibliography{fivebranes}


\end{document}

